\documentclass[12pt]{article}
\usepackage{amsmath}
\usepackage{graphicx}
\usepackage{natbib}
\usepackage{url} 
\usepackage{array,epsfig,rotating} 
\usepackage{url} 
\usepackage{algcompatible}
\usepackage[]{hyperref}  
\usepackage{sectsty, secdot}
\usepackage{rotating}
\usepackage{xparse}
\usepackage{mathtools}
\usepackage{centernot}
\usepackage{epsfig}
\usepackage{amssymb}
\usepackage{enumerate}
\usepackage{verbatim}
\usepackage{multicol}
\usepackage[english]{babel}
\usepackage{color}
\usepackage{float}
\usepackage{hyperref}
\usepackage{tabularx}
\hypersetup{
     colorlinks=true,
     linkcolor=blue,
     citecolor=blue,
     filecolor=blue,
     urlcolor=blue,
}

\usepackage{mathrsfs}
\usepackage[skip=0pt]{caption}
\usepackage{pdflscape}
\usepackage{xcolor}
\usepackage{algorithm}
\usepackage{amsthm}
\usepackage{multirow}
\newcommand{\blind}{0}

\newcommand{\bI}{\mathbf{I}}

\newcommand{\bX}{\mathbf{X}}
\newcommand{\bY}{\mathbf{Y}}
\newcommand{\bZ}{\mathbf{Z}}

\newcommand{\bF}{\mathbf{F}}
\newcommand{\bG}{\mathbf{G}}

\newcommand{\bU}{\mathbf{U}}
\newcommand{\bV}{\mathbf{V}}

\newcommand{\bH}{\mathbf{H}}
\newcommand{\bx}{\mathbf{x}}
\newcommand{\bz}{\mathbf{z}}

\newcommand{\bu}{\mathbf{u}}

\newcommand{\bv}{\mathbf{v}}

\newcommand{\bmu}{\boldsymbol{\mu}}

\newcommand{\bSigma}{\boldsymbol{\Sigma}}
\newcommand{\0}{\mathbf{0}}

\newcommand*{\QEDB}{\hfill\ensuremath{\square}}
\DeclareRobustCommand{\rchi}{{\mathpalette\irchi\relax}}
\newcommand{\irchi}[2]{\raisebox{\depth}{$#1\chi$}} 
\newcommand{\argmin}{\operatornamewithlimits{arg\,min}}
\newcommand{\argmax}{\operatornamewithlimits{arg\,max}}

\newtheorem{theorem}{Theorem}
\newtheorem{lemma}{Lemma}
\newtheorem{corollary}{Corollary}

\newtheorem{example}{Example}
\newtheorem{remark}{Remark}

\begin{document}

\theoremstyle{definition}
\def\spacingset#1{\renewcommand{\baselinestretch}%
{#1}\small\normalsize} \spacingset{1}


\if0\blind
{
  \title{\bf On Exact Feature Screening in Ultrahigh-dimensional Binary Classification \vspace*{0.1in}}

  \author{
    Sarbojit Roy \\
    \small Computer, Electrical, and Mathematical Sciences and Engineering Division,\\
    \small King Abdullah University of Science and Technology, Thuwal - 23955, Saudi Arabia. \\
	and \\
    Soham Sarkar \\
    \small Theoretical Statistics and Mathematics Unit, Indian Statistical Institute, Delhi - 110016, India. \\
    and \\
    Subhajit Dutta \\
    \small Department of Mathematics and Statistics, IIT Kanpur, Kanpur - 208016, India. \\
  	and \\
    Anil K. Ghosh \\
    \small Theoretical Statistics and Mathematics Unit, Indian Statistical Institute, Kolkata - 700108, India. \\
}
  \maketitle
} \fi

\if1\blind
{
  \bigskip
  \bigskip
  \bigskip
  \begin{center}
    {\LARGE\bf Title}
\end{center}
  \medskip
} \fi

\bigskip
\begin{abstract}
\vspace*{0.05in}
We propose a new model-free feature screening method based on energy distances for ultrahigh-dimensional binary classification problems. 
With a high probability, the proposed method retains only relevant features after discarding all the noise variables. 
The proposed screening method is also extended to identify pairs of variables that are marginally undetectable but have differences in their joint distributions. 
Finally, we build a classifier that maintains coherence between the proposed feature selection criteria and discrimination method and also establish its risk consistency.
An extensive numerical study with simulated and real benchmark data sets shows clear and convincing advantages of our proposed method over 
the state-of-the-art methods.
\vspace*{0.1in}
\end{abstract}

\noindent%
\vfill

\newpage
\spacingset{1.5} 
\section{Introduction}\label{intro}

In a high-dimensional classification problem, the presence of a large number of irrelevant covariates (say, noise variables) usually deteriorates the performance of any classifier. Identifying relevant covariates (or, features), and subsequently, discarding noise often yield improved classification accuracy.
Thus, feature screening continues to be an active area of research in classification problems (see, e.g., \cite{fan2008high,pan2016ultrahigh} for some model-based approaches). 

Existing screening methods are primarily comprised of two main steps. The {\it first step} is to rank the covariates according to their importance in predicting the response. A model-free feature screening method was developed by \cite{zhu2011model}, where the conditional density of the response given a component variable was used to measure its importance. 
For binary classification problems, \cite{mai2013kolmogorov} introduced Kolmogorov filtering (referred to as KF) which uses the two-sample Kolmogorov-Smirnov statistic on the one-dimensional marginals to rank the covariates. A weighted average of the Cramer-von Mises distance between the conditional distribution function of a variable given the response and its unconditional distribution was proposed by \cite{cui2015model} to rank the covariates. Using this weighted average and the idea of sure independent screening (SIS) developed by \cite{fan2008sure}, \cite{cui2015model} proposed the screening method MV-SIS.
\cite{cheng2017robust} developed a robust screening method (referred to as RRS) that uses the difference between the conditional rank of the covariates given the response and their unconditional ranks. VR-SIS was proposed by \cite{MR3852003}, where the authors used the ratio of the variance of a covariate conditioned on response and the marginal variance as the marginal utility of the covariate. 
Using the idea of the Anderson-Darling test, the MV-SIS approach was modified in \cite{MR3922199}. Covariates were ranked based on the classification accuracy of marginal classifiers (referred to as MCS) by \cite{sheng2020model}. Recently, \cite{MR4359625} developed a new class of non-parametric test statistics, namely, the maximum adjusted chi-squared (MAC) statistic for the two-sample testing 
problem, and proposed a variable screening procedure for binary classification problems using this MAC statistic. 
All these screening methods are model-free and possess the {\it sure screening property} \citep[introduced by][]{fan2008sure} in ultrahigh dimensions, viz., when the number of variables $d=d_n$ satisfies $\log d_n = O(n^\beta)$ for some $\beta>0$ and $n \in \mathbb{N}$. In other words, they retain all the features in the screened set with probability tending to one as the sample size tends to infinity, and the dimension is allowed to grow exponentially with the sample size. 

After ranking covariates, the {\it second step} is to select the first $\tilde{n}$ of the ranked covariates, where $\tilde{n}$ is typically set to be $[n/\log n]$ (see \cite{fan2008sure}). Here, $[x]$ denotes the greatest integer less than, or equal to $x \in \mathbb{R}$. Now, if the number of relevant features is strictly smaller than the number of selected components (i.e., $\tilde{n}$), then all these screening methods will inevitably include some noise components.~We~illustrate~this~using an~example.

\begin{example}\label{ex1}
Let $(X_1,\ldots,X_{d_n}) \sim N_{d_n}(\0,\bI)$ and $(Y_1,\ldots,Y_{d_n}) \sim N_{d_n}(\bmu,\bI)$, where ${\bf 0}$ and $\bI$ denote the null vector and the identity matrix, respectively, and $\bmu=(1,1,1,1,0,\ldots,0)^\top$. Here, $N_{d_n}(\bmu,\bSigma)$ denotes the $d_n$-dimensional normal distribution with mean vector $\bmu$ and covariance matrix $\bSigma$.
\end{example}

In this example, only the first four covariates are relevant for classification.
Now, if we have $200$ observations (i.e., $n=200$), then all the aforementioned methods will select $\tilde n=37$ covariates, irrespective of $d$. Eventually, the screened set will contain at least $37-4=33$ noise components. Clearly, the accumulation of noise in the screened set will have detrimental effects on the classification accuracy. The problem becomes even more severe when the sample size increases since $\tilde n$ is an increasing function of $n$. 

A second major limitation of most of the existing screening methods is that they can only detect signals that arise from differences in marginal distributions, but are completely useless if the marginals are identical. We demonstrate this using a second example.

\vspace*{0.1in}
\begin{example}\label{ex2}
Suppose that $(X_{1},X_{2})$, $(X_{3},X_{4}) \stackrel{iid}{\sim} N_2(\0,\bSigma_1)$ with $\bSigma_1=[1,0.9;0.9,1]$ and $(Y_{1},Y_{2})$, $(Y_{3},Y_{4}) \stackrel{iid}{\sim} N_2(\0,\bSigma_2)$ with $\bSigma_2=[1,-0.9;-0.9,1]$, while $X_{5},\ldots, X_{d_n}$ and $Y_{5},\ldots, Y_{d_n}$ are iid $N(0,1)$. Here, $(X_1,\ldots,X_4)$, $ (X_5,\ldots,X_{d_n})$, $(Y_1,\ldots,Y_4)$ and $(Y_5,\ldots,Y_{d_n})$ are mutually independent and `iid' stands for independent and identically distributed.
\end{example}

In this example, the pairs $\{1,2\}$ and $\{3,4\}$ contain signal through their bivariate distributions. But, the individual components are marginally undetectable since all the one-dimensional marginals of the two competing distributions are $N(0,1)$. 
If we use any of the existing screening methods, it will select $\tilde{n}$ component variables (just like Example \ref{ex1}), which are all useless for classification. 
There are some popular methods in the literature like group LASSO \citep[see, e.g.,][]{meier2008group} and its sparse version \citep[see, e.g.,][]{simon2013sparse} that can capture information from differences in joint distributions. More recently, \cite{MR4517888} proposed a grouped feature screening method based on the Gini impurity. However, these methods require apriori knowledge about plausible pairs and selecting the relevant ones out of those, and are not suitable in practice if the information on the pairs is unknown. 
To obtain paired features, \cite{MR3852003} extended the VR-SIS method to select two covariates as a pair if their product yields a large value of their proposed VR index. Similar to their marginal screening method, the proposed modification also assumes the existence of second-order moments for the covariates, and hence, lacks robustness. \cite{MR4359625} generalized their MAC based filtering method for screening features that are marginally undetectable, but have discriminatory information in their joint distributions. But, the performance of this method depends on a targeted false positive rate and this value needs to be tuned in practice.

After screening, our aim is to classify observations based on the screened features. 
In terms of compatibility, the criterion used for finding relevant features should also be reflected in the choice for the discriminant. For instance, suppose that the support vector machine with a radial basis kernel (say, SVMRBF) is used as the marginal classifier for selecting the top $\tilde{n}$ variables. 
Since SVMRBF selected variables that contain discriminatory information between the competing populations in $\mathbb{R}^{d_n}$, it will not be appropriate to use the SVM classifier with a linear kernel (say, SVMLIN) on the reduced space $\mathbb{R}^{\tilde{n}}$. 
Among existing methods, only MCS possesses this congruity between the criteria for feature selection and discrimination. 
Other methods only specify a screening procedure, but leave the choice of the classifier to the user.
In this article, we propose a model-free screening method for ultrahigh-dimensional binary classification problems. 
(a) With high probability, the proposed method retains only relevant features after discarding all the noises. (b) Next, we have extended the method to retain paired along with marginal features. The pairs are screened in a way such that the two classes have maximum separation in terms of their energy distance. 
(c) Lastly, unlike almost all existing methods, there is a coherence between the methodology used for constructing the screening set and the proposed classification rule.


The rest of this article is organized as follows. Section \ref{conc_en} presents the proposed marginal screening method and theoretical results related to the consistency of the screened set. 
In Section \ref{group}, a further generalization of our method is developed that detects paired features by identifying differences between joint distributions of the bivariate components. 
Comparative performance of the existing and proposed algorithms is demonstrated using a variety of simulated data sets in Section \ref{numstud}. We propose a classifier that is coherent with the screening method in Section \ref{class}, and discuss related consistency properties.
Our classification method is compared numerically with several popular classifiers on numerous simulated as well as real benchmark data sets in Sections \ref{numstud_class} and \ref{real}, respectively.
All proofs and mathematical details are provided in a supplementary file, which also contains algorithms of the proposed screening methods and some additional material. 

\section{Marginal Screening Based on Energy Distances} \label{conc_en}

Suppose that $\bF$ and $\bG$ are two absolutely continuous distribution functions (dfs) on $\mathbb{R}^{d_n}$. Let $\bX=(X_1,\ldots,X_{d_n})^\top\sim\bF$ and $\bY=(Y_1,\ldots,Y_{d_n})^\top\sim\bG$ with $X_{k}\sim F_k$ and $Y_{k}\sim G_k$ for $1\le  k\le d_n$. The covariate $X_k$ can \emph{marginally} discriminate between $\bF$ and $\bG$ if $F_k \neq G_k$. Consequently, the set of marginal signals is defined as
\begin{align}\label{defS}
S_n = \{k: F_{k}\neq G_{k}\ \mbox{ for } 1 \leq k \leq d_n\}.
\end{align}
We denote its cardinality by $s_n:=|S_n|$; the number of noise variables 
is $t_n := d_n - s_n$.

Now, we use energy distance between the one-dimensional marginals of $\bF$ and $\bG$ to present an equivalent definition of $S_n$. Suppose that $\bX_1$, $\bX_2$ and $\bY_1$, $\bY_2$ are iid copies of $\bX$ and $\bY$, respectively. Then, the energy distance between $F_{k}$ and $G_{k}$ is defined as
\begin{align}\label{endef}
\mathcal{E}_k = 2E[\gamma(|X_{1k}-Y_{1k}|^2)]-E[\gamma(|X_{1k}-X_{2k}|^2)]-E[\gamma(|Y_{1k}-Y_{2k}|^2)]
\end{align}
for $1 \leq k \leq d_n$. Here, $\gamma$ is a continuous, monotonically increasing function from $\mathbb{R}^+$ to $\mathbb{R}^+$ such that $\gamma$ has non-constant completely monotone derivative \citep{MR0270403} and $\gamma(0)=0$. Some popular choices of the function $\gamma(t)$ are $1-\exp(-t)$, $\log(1+t)$ and $\sqrt{t}$ for $t\geq 0$. It is well known that $\mathcal{E}_k=0$ iff $F_{k}= G_{k}$ \citep[see, e.g.,][]{szekely2005new,BF10}.
Thus, energy distances corresponding to the elements of $S_n$ are strictly positive, whereas they are zero for the noise components. This allows us to express $S_n$ as
\begin{align}\label{defSen}
S_n =\{k: \mathcal{E}_k>0 \mbox{ for } 1 \leq k \leq d_n\},
\end{align}
and the screening problem reduces to identifying the covariates with positive energy distance.
Let us arrange the energy distances $\{\mathcal{E}_k : 1 \leq k \leq d_n\}$ in increasing order of magnitude. Clearly, the smallest $t_n$ energy distances will correspond to the collection of noise variables and will all be equal to zero. In other words, we have
\begin{equation} \label{energy_ordered_popln}
0=\mathcal{E}_{(1)}=\cdots = \mathcal{E}_{(t_n)}<\mathcal{E}_{(t_n+1)}\leq \cdots\leq \mathcal{E}_{(d_n)}.
\end{equation}
This now gives us yet another equivalent representation of $S_n$:
\begin{align}\label{defS_alt}
S_n=\{k: \mathcal{E}_k \geq \mathcal{E}_{(t_n+1)} \mbox{ for } 1 \leq k \leq d_n\},
\end{align}
where $\mathcal{E}_{(t_n+1)}$ denotes the minimum energy among the signals, i.e., $\mathcal{E}_{(t_n+1)} = \min_{k\in S_n}\mathcal{E}_{k}$. A key observation here is that the ratio $\mathcal{E}_{(t_n+1)}/\mathcal{E}_{(t_n)} = \infty$, while $\mathcal{E}_{(k+1)}/\mathcal{E}_{(k)} < \infty$ for $(t_n+1) \leq k \leq (d_n-1)$.

In practice, both $s_n$ and $S_n$ are unknown and our objective is to estimate these quantities. We begin by estimating the energy distances. Assume $\min\{n_1, n_2\} \geq 2$. 
For random samples $\rchi =\{\bX_1,\ldots, \bX_{n_1}\}$ and $\mathcal{Y} =\{\bY_1,\ldots, \bY_{n_2}\}$ drawn from the dfs $\bF$ and $\bG$, respectively, an unbiased estimator of $\mathcal{E}_k$ is given by the \emph{sample energy distance}:
\begin{align} \label{smplenergy} \nonumber
\widehat{\mathcal{E}}_{k} & = \frac{2}{n_1n_2} \sum\limits_{m_1=1}^{n_1} \sum\limits_{m_2=1}^{n_2} \gamma(|X_{m_1k}-Y_{m_2k}|^2) - \frac{1}{{n_1\choose 2}} \mathop{\sum \sum} \limits_{1 \leq m_1< m_2 \leq n_1} \gamma(|X_{m_1k}-X_{m_2k}|^2) \\
&\kern20ex - \frac{1}{{n_2\choose 2}} \mathop{\sum \sum} \limits_{1 \leq m_1< m_2 \leq n_2} \gamma(|Y_{m_1k}-Y_{m_2k}|^2)
\end{align}
for $1 \leq k \leq d_n$. 
It is known that $\widehat{\mathcal E}_k$ is a consistent estimator of $\mathcal E_k$. Thus, if we arrange the sample energy distances in increasing order, following \eqref{energy_ordered_popln}, one expects the first $t_n$ values to correspond to elements of the noise set, while the complementary set should be the signal set. Using this idea, we first construct an estimate of $s_n$, the number of signals.

For absolutely continuous dfs $F_{k}$ and $G_{k}$, we have 
$\widehat{\mathcal{E}}_{k}>0$ almost surely for $1\leq k\leq d_n$. So, the ratio $\widehat{R}_k=\widehat{\mathcal{E}}_{(k+1)}/\widehat{\mathcal{E}}_{(k)}$ is well-defined with probability $1$ for $1 \leq k \leq (d_n-1)$.
Define $R_k$ (the population counterpart of $\widehat{R}_k$) as $\mathcal{E}_{(k+1)} /\mathcal{E}_{(k)}$ for $t_n \leq k \leq (d_n-1)$. In view of the fact that ${R}_{t_n}=\mathcal{E}_{(t_n+1)}/\mathcal{E}_{(t_n)}=\infty$, we expect $\widehat{R}_{t_n}$ to take a significantly large value when compared with the entire sequence $\{\widehat{R}_k: 1 \leq k \leq (d_n-1)\}$. 
Detecting this jump yields the following estimator of the number of signals:
\begin{align}\label{prop_est_sn}
\widehat{t}_n = & \argmax_{1 \leq k \leq (d_n-1)} \widehat{R}_k \qquad\text{ and  }\qquad \widehat{s}_n = d_n - \widehat{t}_n.
\end{align}
After obtaining $\widehat s_n$, we simply define the screened set to consist of the covariates corresponding to the $\widehat s_n$ largest $\widehat{\mathcal E}_k$ values. In other words, we estimate the signal set $S_n$ as:
\begin{align}\label{prop_est}
\widehat{S}_n &= \{k: \widehat{\mathcal{E}}_{k} \geq \widehat{\mathcal{E}}_{(\widehat{t}_n + 1)} \mbox{ for } 1 \leq k \leq d_n \}.
\end{align}
Our proposed screening method is based on the idea of marginal differences, hence, we refer to it as {\it marginal screening} (MarS).

Note that energy based methods like the distance correlation (dCor) \citep[see][]{MR2382665} and the Hilbert-Schmidt Independence Criterion (HSIC) \citep[see][]{gretton2005measuring} have been used for developing variable screening methods in classification \citep[see, e.g.,][]{song2012feature,balasubramanian2013ultrahigh}. These methods typically use tests of independence between the feature and the response to decide the relevance of a feature, whereas here, we use two-sample energy statistics to address this problem.


\subsection{Screening Property of MarS} \label{Screen_Mars}

A screening method is said to possess {\it sure screening property} (SSP) if 
the estimated set $\widehat{S}_n$ contains the true signal set $S_n$ with probability tending to one as the sample size increases, i.e., $P[S_n\subseteq \widehat{S}_n]\to 1$ as $n \to \infty$. 
The proposed method MarS possesses SSP under some regularity conditions.
Recall that we work in the ultrahigh-dimensional regime, where $\log d_n = O(n^\beta)$ for some $0 \leq  \beta < 1$. Consider the following assumptions:
\begin{enumerate}
 \item[A1.] There exists a constant $0 < \alpha_1 < (1-\beta)/2$ such that
 \begin{enumerate}[1.]
 \item $1/\mathcal{E}_{(t_n+1)}= o(n^{\alpha_1})$ and
 \item $\max_{(t_n+1) \leq k \leq (d_n-1)} R_k = o(n^{\alpha_1} \mathcal{E}_{(t_n+1)})$.
 \end{enumerate}
\end{enumerate}

\noindent
Assumption A1.1 provides a lower bound on the {rate of the} minimum energy distance $\mathcal{E}_{(t_n+1)} = \min_{k\in S_n} \mathcal{E}_k$ in the signal set $S_n$. It is easy to see that the assumption can be restated as $n^{\alpha_1} \mathcal{E}_{(t_n + 1)} \to \infty$ as $n \to \infty$ for some $0 < \alpha_1 < (1-\beta)/2$, which is equivalent to the existence of $0 \leq \tau <1/2$ such that $n^{\tau}\mathcal{E}_{(t_n+1)}$ is greater than some fixed constant. 
This is a common assumption in the variable screening literature and it readily holds if the energy distances between marginals are larger than some fixed constant. However, we also allow the minimum energy distance to go to zero, albeit at an appropriate rate. 
Assumption A1.2, on the other hand, states that the maximum value of $R_k = \mathcal E_{(k+1)}/\mathcal E_{(k)}$ for $(t_n+1) \leq k \leq (d_n-1)$ cannot be too large compared to the minimum energy distance $\mathcal E_{(t_n+1)}$. This is equivalent to requiring the difference between the successive energy distances in $S_n$ to be controlled. In other words, we put an upper bound on the rate of the relative growth of the energy distances in $S_n$ (also see Remark \ref{iid_C} in Appendix \ref{Appendix_C}). 
We now state the first theoretical result, which establishes SSP of the proposed method MarS. 

\vspace*{-0.05in}
\begin{theorem} \label{mainthm}
If assumption {\rm A1} is satisfied and $\gamma$ is a bounded function, then there exists a constant $b_1>0$ (not depending on $n$) such that
\[
P[S_n\subseteq \widehat{S}_n]\geq 1- O\big (e^{-b_1 \{n^{1-2\alpha}-n^\beta\}}\big) \quad\text{for all }\, 0< \alpha_1\le \alpha<(1-\beta)/2.
\]
As a consequence, we have $P[S_n \subseteq \widehat{S}_n] \to 1$ as $n \to \infty$.
\end{theorem}
\vspace*{-0.05in}

\noindent
Among the popular choices of $\gamma$, the function  $\gamma(t)=1-\exp(-t)$ for $t\ge 0$ is {\it bounded}.

\begin{remark} \label{R_mainthm}
The boundedness of $\gamma$ in the theorem is sufficient, but not necessary. For instance, if the univariate marginals of $\bF$ and $\bG$ are Gaussian and $\gamma$ is $L-$Lipschitz continuous for some $L>0$ (e.g., $\gamma(t)=\log(1+t)$ for $t\geq 0$ with $L=1$), then Theorem \ref{mainthm} holds (see Lemmas \ref{U_expbound_suff_Gaussian} and \ref{U_expbound_lip} in Appendix \ref{Appendix_A} for more details).
Moreover, if the components of $\bX\sim\bF$ and $\bY\sim\bG$ are bounded and $\gamma$ is continuous, then Theorem \ref{mainthm} remains valid.
\end{remark}

Recall that we had used the ratios $\widehat{R}_k$ for $1\leq k\leq (d_n-1)$ to construct the set $\widehat{S}_n$. Now, if $\widehat{R}_k$ attains the largest value for some $k < t_n$, then we are bound to include some noise variables in $\widehat{S}_n$. 
Under an additional condition on the noise variables, we will prove that $\widehat{S}_n$ not only retains all the signals but also disposes of all the noise components. Such a property is referred to as the {\it exact screening property} (ESP).
To be precise, a screening method possesses ESP if the set of selected signals $\widehat{S}_n$ satisfies $P[\widehat{S}_n = S_n] \to 1$ as $n \to\infty$. In other words, it perfectly estimates the true signal set with probability tending to one. Clearly, ESP is a stronger property than SSP. 

Now, consider the set $\mathcal{N}_n=\{\widehat{\mathcal{E}}_{k}:k\in S^c_n\}$ which is the collection of estimated energy distances corresponding to the noise variables. 
For notational ease, let $N_{k}$ denote the $k$-th element and $N_{(k)}$ be the $k$-th minimum in the set $\mathcal{N}_n$ for $1\leq k\leq t_n$. Now, consider the following assumption:
\begin{enumerate}
\item[A2.] There exists a constant $0 < \alpha_2 <(1-\beta)/2$ such that
\[
\max_{1\leq k\leq (t_n-1)}N_{(k+1)}/N_{(k)}=o_{P}( n^{\alpha_2} \mathcal{E}_{(t_n+1)}).
\]
\end{enumerate}

\begin{theorem}\label{mainthm_E}
If assumptions {\rm A1} and {\rm A2} are satisfied for $0<\alpha_1<(1-\beta)/2$ and $0<\alpha_2<(1-\beta)/2$, respectively, and $\gamma$ is a bounded function, then there exists $b_2>0$ such that 
\[
P[\widehat{S}_n = S_n] \geq 1-O\big (e^{-b_2\{ n^{1-2\alpha}-n^\beta\}}\big) \quad\text{for all }\, 0 < \max\{\alpha_1,\alpha_2\}\le \alpha<(1-\beta)/2.
\]
As a consequence, $P[\widehat{S}_n = S_n]\to 1$ as $n\to\infty$.
\end{theorem}

\begin{remark}
Theorem \ref{mainthm_E} can be proved under weaker versions of assumptions A1.2 and A2, with the `little $o$' being replaced by `big $O$'
(see Remark \ref{relaxed_conditions} of Appendix \ref{Appendix_A} for details).
\end{remark}

Assumption A2 holds the key to proving ESP of MarS, and we first discuss this condition in detail here.
Fix $1\leq k\leq t_n$. If $F_k=G_k$, then \cite{szekely2005new} showed that $n_1 n_2 N_k/n$ converges in distribution to a non-degenerate random variable with some limiting (as $n \to \infty$) df (say, ${H}$) on $\mathbb{R}^{+}$.
So, we can think of $n_1 n_2 N_{(1)}/n, \ldots, n_1 n_2 N_{(t_n)}/n$ as an ordered random sample of size $t_n$ from ${H}$ for large $n$. It is now clear that assumption A2 imposes a condition on the growth of the ratios of these ordered random variables, and allows these ratios to increase at a rate slower than $n^{\alpha_1} \mathcal{E}_{(t_n+1)}$ for some $0<\alpha_1<(1-\beta)/2$. By assumption A1.1, we already have $n^{\alpha_1} \mathcal{E}_{(t_n+1)} \to \infty$ as $n \to \infty$. Using the results of \cite{balakrishnan2008asymptotic} on the ratios of consecutive order statistics, one can show that
$N_{(k+1)}/N_{(k)} = O_P(1)$ for any $1 \leq k \leq d_n$ (under appropriate regularity conditions on the df ${H}$). Further, if $L_n=\max_{1\leq k\leq (t_n-1)} N_{(k+1)}/N_{(k)}$ is $O_{P}(1)$, then A2 holds. 
In general, however, mathematical verification of this assumption is admittedly difficult.

\vspace*{0.2in}
\begin{figure}[htp]
	\begin{center}
		\includegraphics[width = 0.9\linewidth]{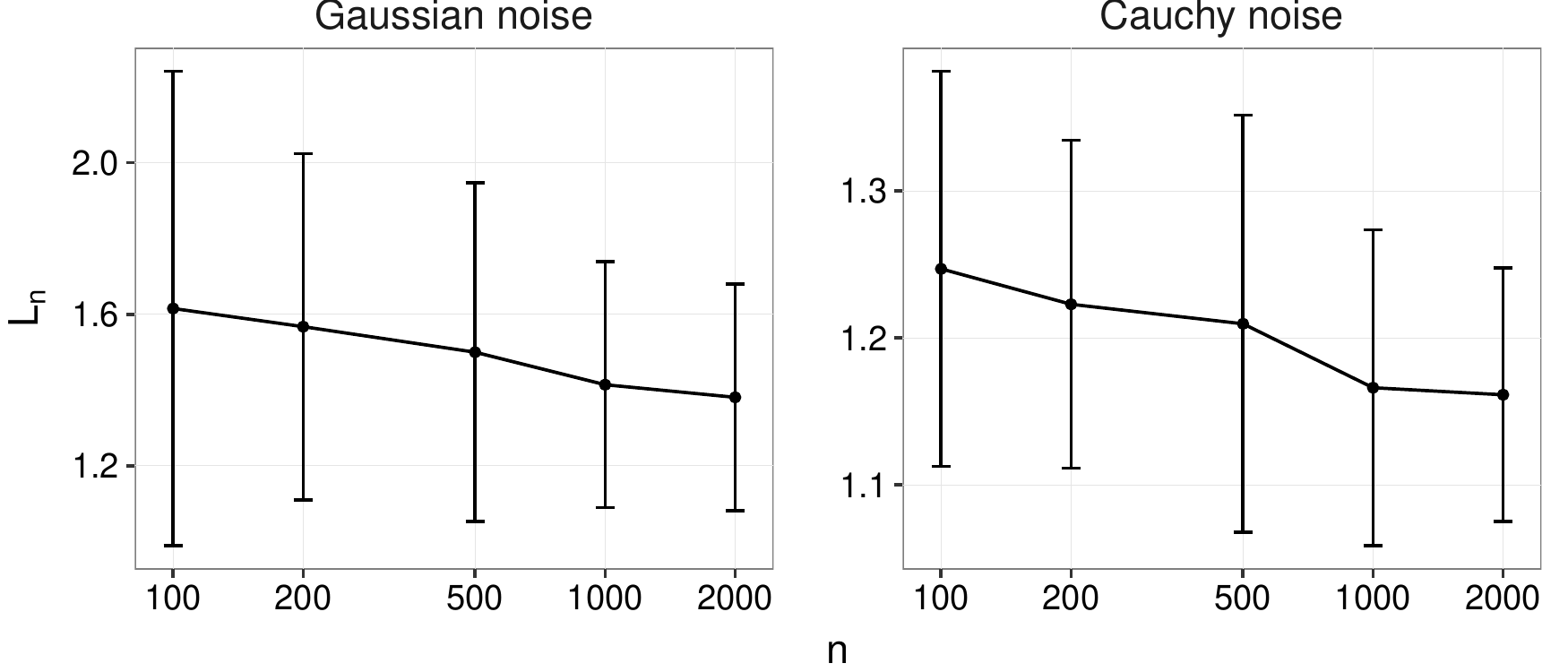}
	\end{center}
	\caption{\it Plot of $\displaystyle \frac{1}{100} \sum_{i=1}^{100} L^{(i)}_n$ and its standard error when the noise variables follow the standard normal distribution (the left panel) and the standard Cauchy distribution (the right~panel) for increasing values of $n$ (in logarithmic scale).}
	\label{heur}
\end{figure}

We now check the validity of assumption A2 numerically. Let us consider a noise set $S^c_n$ with cardinality $t_n=[\exp{(25 n^{1/4})}]$, where the components are iid $N(0,1)$. 
For a fixed $n$, let $N^{(i)}_{(1)},\ldots,N^{(i)}_{(t_n)}$ denote the ordered sample energy distances for the $i$-th replicate with $1\leq i\leq 100$. We compute $L^{(i)}_n=\max\limits_{1\leq k\leq (t_n-1)} {N^{(i)}_{(k+1)}}/{N^{(i)}_{(k)}}$ for $1\leq i\leq 100$.
The average of these values along with their standard errors are plotted against increasing $n$ in Figure \ref{heur}. This figure clearly shows a decreasing trend 
as $n$ increases, indicating that $L_n$ is bounded in probability. We observe a similar phenomenon when the experiment was repeated with the standard Cauchy distribution (say, $C(0,1)$) as the noise distribution.

\begin{figure}[htp]
		\centering
		\begin{tabular}{cc}
 			\multicolumn{1}{c}{\hspace{3.5cm}Example 1}& \multicolumn{1}{c}{\hspace{3.5cm}Example 2}\\
            \multicolumn{2}{c}{\includegraphics[scale = 0.85] 
            {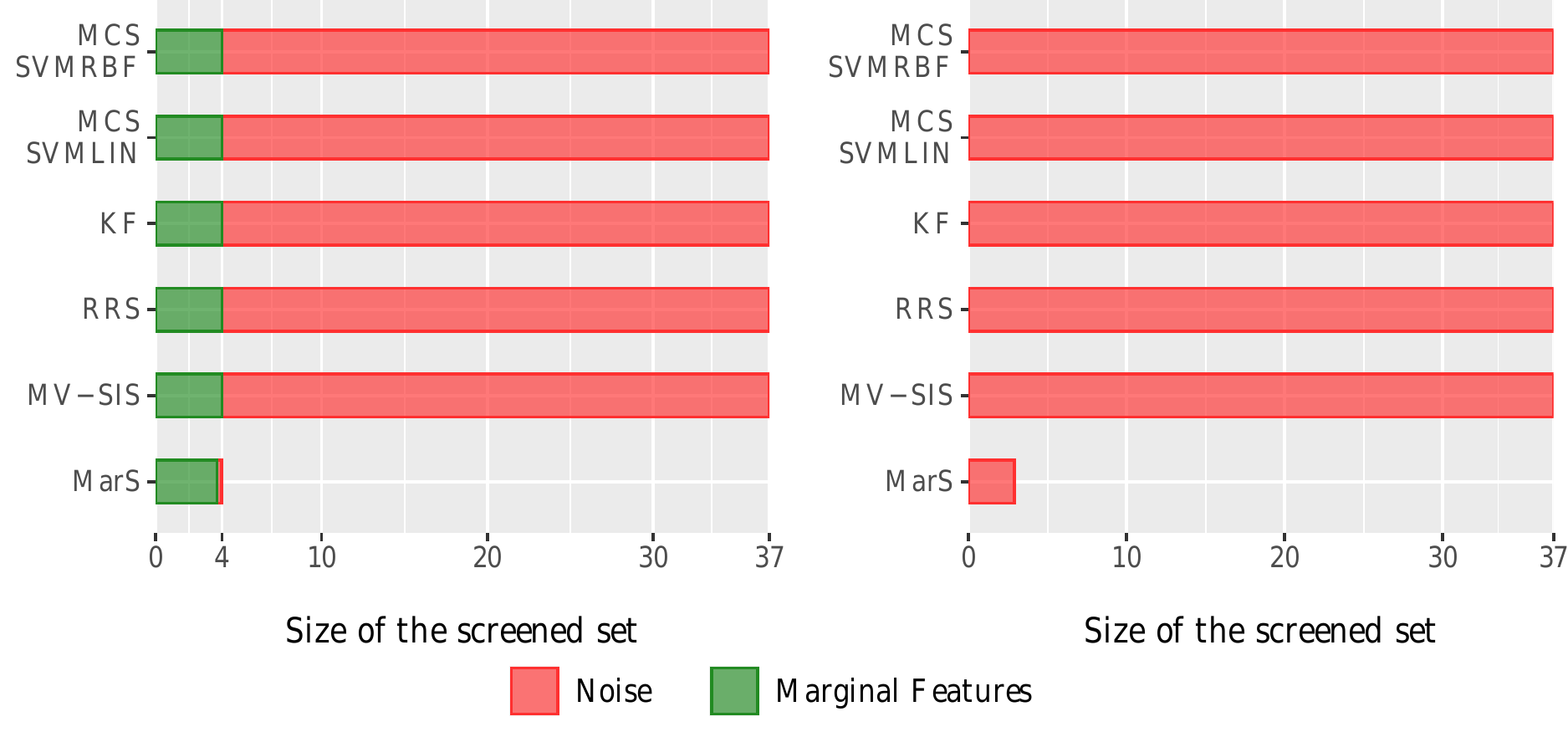}}
		\end{tabular}
		\caption{\it Bar plots indicating the average number of features selected by competing methods as well as the MarS algorithm in Examples \ref{ex1} and \ref{ex2} over $100$ simulation runs with $n_1=n_2=100$ and $d_n=1000$. MarS was implemented with $\gamma(t) = 1 - \exp{(-t)}$ for $t \ge 0$. \vspace*{-0.1in}}
		\label{fig2}
\end{figure}

Let us now revisit Examples \ref{ex1} and \ref{ex2} introduced in Section \ref{intro}. 
In Figure~\ref{fig2}, we show the performance of different screening methods in these examples with $n=200$ ($100$ observations from each class) and $d=1000$ over $100$ simulation runs. The left panel of Figure \ref{fig2} shows that 
MarS retained all the relevant features for most of the simulation runs. 
However, this was not the case for Example \ref{ex2}. The univariate marginals are all equal (viz., $F_{k}=G_{k} \equiv N(0,1)$ for all $1 \leq k \leq d_n$) in this example. 
As a result, MarS as well as all competing methods fail to identify the signal set.
To circumvent this problem, next, we propose a method that is capable of detecting differences in the joint distribution of pairs.

\section{Screening of Paired Features} \label{group}

Let $\bF_{\{i,j\}}$ and $\bG_{\{i,j\}}$ denote the joint distributions of $\bX_{\{i,j\}}=(X_i,X_j)^\top$ and $\bY_{\{i,j\}}=(Y_i,Y_j)^\top$, respectively, for $1\leq i<j\leq d_n$. We define $\{i,j\}$ to be a {\it paired feature} if $F_i=G_i$ and $F_j=G_j$, but $\bF_{\{i,j\}} \neq \bG_{\{i,j\}}$. In other words, we have no discriminatory information in the marginal components $\{i\}$ and $\{j\}$, but only in the joint distribution through the pair $\{i,j\}$ for $1\leq i<j\leq d_n$. If $\bX_1$, $\bX_2$ and $\bY_1$, $\bY_2$ are iid copies of $\bX$ and $\bY$, respectively, then the energy distance between $\bF_{\{i,j\}}$ and $\bG_{\{i,j\}}$ is given by

\noindent
\begin{align}
 \mathcal{E}_{\{i,j\}} = &\ 2{\rm E}\left [\gamma\left(\frac{1}{2}\|\bX_{1,\{i,j\}}-\bY_{1,\{i,j\}}\|^2\right)\right ]-{\rm E}\left [\gamma\left(\frac{1}{2}\|\bX_{1,\{i,j\}}-\bX_{2,\{i,j\}}\|^2\right)\right ]\nonumber \\
 &\kern0.5ex-{\rm E}\left [\gamma\left(\frac{1}{2}\|\bY_{1,\{i,j\}}-\bY_{2,\{i,j\}}\|^2\right)\right ]
\end{align}
for $1\leq i<j\leq d_n$. Here, $\|\cdot\|$ denotes the Euclidean norm on $\mathbb{R}^2$. We have $\mathcal{E}_{\{i,j\}}=0$ iff $\bF_{\{i,j\}}=\bG_{\{i,j\}}$ for a pair $\{i,j\}$. As in Section \ref{conc_en}, one may use $\mathcal{E}_{\{i,j\}}$ to conclude whether a pair $\{i,j\}$ contributes to the signal, or not.
In Example \ref{ex2}, the pairs $\{1,2\}$, $\{3,4\}$ are the only signals. So, $\mathcal{E}_{\{1,2\}}$ and $\mathcal{E}_{\{3,4\}}$ are positive, while $\mathcal{E}_{\{i,j\}}=0$ for all $\{i,j\} \neq \{1,2\},\{3,4\}$.
Using these facts from Example \ref{ex2}, we now develop the idea of screening paired signals. 

We start by assuming $d_n$ to be even. If not, we can make it even by adding an independently distributed noise term (e.g., a $N(0,1)$ variate). 
Let $\mathcal{P}_n$ denote the collection of all possible disjoint pairs which form a partition of $\{1,\ldots, d_n\}$, and 
define $\tilde{d}_n=d_n/2$. For a given partition $P_n=\big \{\{i_1,j_1\},\ldots, \{i_{\tilde{d}_n}, j_{\tilde{d}_n}\}\big\}\in \mathcal{P}_n$, define
$\mathcal{E}(P_n) = \sum_{\{i,j\}\in P_n} \mathcal{E}_{\{i,j\}}$.
In Example \ref{ex2}, $\mathcal{E}(P_n)$ can take four possible values, namely, $\mathcal{E}_{\{1,2\}} + \mathcal{E}_{\{3,4\}}$ if both $\{1,2\}$ and $\{3,4\} \in P_n$, $\mathcal{E}_{\{1,2\}}$ if only $\{1,2\} \in P_n$, $\mathcal{E}_{\{3,4\}}$ if only $\{3,4\} \in P_n$ and $0$ otherwise. 
Clearly, the maximum value that $\mathcal{E}(P_n)$ can attain is $\mathcal{E}_{\{1,2\}}+\mathcal{E}_{\{3,4\}}$, and it is achieved when the partition $P_n$ contains both the pairs $\{1,2\}$ and $\{3,4\}$.
So, maximizing $\mathcal{E}(P_n)$ over the set of all disjoint pairs $\mathcal{P}_n$ yields the set of paired signals. 
We now formalize this idea below.

Among the $\tilde{d}_n$ paired components, suppose that we have signal only in $s_n (< \tilde{d}_n)$ paired features.
In other words, let $i_1,\ldots, i_{s_n}$ and $j_1,\ldots, j_{s_n}$ be distinct integers in $\{1,\ldots,d_n\}$ such that $F_{k}=G_{k}$ for all $1\leq k\leq d_n$, but $\bF_{\{i_k,j_k\}} \neq \bG_{\{i_k,j_k\}}$ for $1\leq k\leq  s_n$. This now implies that $\{i_1,j_1\}, \ldots, \{i_{s_n},j_{s_n}\}$ are the paired signals, while the rest of the components are noise.
In this case, it clearly holds that
\begin{align}\label{Ptheta}
\mathcal{E}(P_n) = \sum_{\{i,j\} \in P_n} \mathcal{E}_{\{i,j\}}= \sum\limits_{k=1}^{s_n} \mathcal{E}_{\{i_k,j_k\}} \ \mathbb{I}[\{i_k,j_k\} \in P_n]
\end{align}
with $\mathbb{I}[\cdot]$ denoting the indicator function.
The next result gives us a set of sufficient conditions under which
$\mathcal{E}(P_n)$ is maximized iff $P_n$ contains all the paired signals. 

Define
\begin{equation}\label{pairedSn}
S_n=\ \big \{\{i_1,j_1\}, \ldots, \{i_{s_n},j_{s_n}\}\big \} \text{ and } S_n^c=\ \{1,\ldots,d_n\} \setminus \{i_1,\ldots, i_{s_n},j_1,\ldots, j_{s_n}\}.
\end{equation}
These can be viewed as the set of paired signals and noise components, respectively.
\begin{lemma} \label{joint_indep}
Suppose that $\bX_{\{i,j\}}$ and $\bX_{\{i^\prime,j^\prime\}}$ are mutually independent for $\{i,j\},\{i^\prime,j^\prime\}\in S_n$ with $i\neq i^\prime$ and $j\neq j^\prime$. 
Also, let $\bX_{\{i,j\}}$ and $X_l$ be mutually independent for any $\{i,j\}\in S_n$ and $l\in S^c_n$.
Then, $\max\limits_{P_n \in \mathcal{P}_n} \mathcal{E}(P_n) = \sum \limits_{\{i,j\}\in S_n} \mathcal{E}_{\{i,j\}} = \sum_{k=1}^{s_n} \mathcal{E}_{\{i_k,j_k\}}$ and the maximum is attained iff $S_n \subseteq P_n$ for $P_n \in \mathcal{P}_n$.
\end{lemma}

\noindent
This formulation allows us to transform the problem of paired-feature screening into a maximization problem,
with a nice interpretation from the graph-theoretic point of view. Let $G=(V,E)$ be an undirected graph with vertex set $V=\{1,\ldots, d_n\}$
and ${\mathcal{E}}_{\{i,j\}}$ denote the weight of the edge between the $i$-th and $j$-th nodes for $1 \leq i < j \leq d_n$. Under this setting, maximizing ${\mathcal{E}}(P_n)$ w.r.t $P_n$ is equivalent to maximizing the sum of pairwise edge weights, where no two edges share the same node.
This is the same as minimizing $-\sum_{\{i,j\} \in P_n} \mathcal E_{\{i,j\}}$, or equivalently, $\sum_{\{i,j\} \in P_n} \big(M-\mathcal E_{\{i,j\}}\big)$ for a constant $M>\max_{\{i,j\} \in P_n} \mathcal E_{\{i,j\}}$. This essentially leads us to {an optimal} non-bipartite (NBP) matching problem of a graph \citep[see, e.g.,][]{derigs1988solving} with the weight of the $(i,j)$-th edge being $M-{\mathcal{E}}_{\{i,j\}}$ for $1 \leq i < j \leq d_n$. 
Note that $\argmax_{P_n \in \mathcal{P}_n} \mathcal{E}(P_n)$ may not be unique. However,
once we obtain a partition solving the NBP matching problem, Lemma \ref{joint_indep} ensures that it contains all the signals and the remaining pairs are noise. 

\begin{remark}
Although we prove Lemma \ref{joint_indep} under the condition of independence of signal and noise variates, this is not necessary for the implementation of PairS. In practice, we maximize the criterion \eqref{Ptheta}, which gives us the pairs containing the maximum information for classification.
\end{remark}

Our goal is now to discard the noise pairs, and we adopt the same strategy as in the case of marginal signals.
First of all, we define the empirical estimator of $\mathcal E_{\{i,j\}}$ based on the training sample as
\begin{align}\label{smplenergy_2} \nonumber
\widehat{\mathcal{E}}_{\{i,j\}} & =  \frac{2}{n_1 n_2} \sum_{m_1=1}^{n_1} \sum_{m_2=1}^{n_2} \gamma \bigg (\frac{1}{2} \|\bX_{m_1,\{i,j\}} - \bY_{m_2,\{i,j\}}\|^2 \bigg) \nonumber \\
& \kern5ex - \frac{1}{{n_1\choose 2}} \mathop{\sum \sum} \limits_{1 \leq m_1< m_2 \leq n_1} \gamma \bigg(\frac{1}{2} \|\bX_{m_1,\{i,j\}} - \bX_{m_2,\{i,j\}}\|^2 \bigg) \nonumber\\
& \kern5ex - \frac{1}{{n_2 \choose 2}} \mathop{\sum \sum} \limits_{1 \leq m_1< m_2 \leq n_2} \gamma \bigg (\frac{1}{2} \|\bY_{m_1,\{i,j\}} - \bY_{m_2,\{i,j\}}\|^2 \bigg )
\end{align}
for $1 \leq i \neq j \leq d_n$. 
We solve the {optimal} NBP matching by maximizing the following: 
\begin{align}\label{Ptheta_sample}
\widehat{\mathcal{E}}(P_n) = \sum_{\{i,j\}\in P_n} \widehat{\mathcal{E}}_{\{i,j\}} \text{ with respect to } P_n \in \mathcal{P}_n.
\end{align}
Let $\widehat{P}_n = \big \{\{i_1,j_1\}, \ldots, \{i_{\tilde{d}_n},j_{\tilde{d}_n}\}\big \}$ denote a maximizer of \eqref{Ptheta_sample}. 
To reduce the notational burden, we denote $\widehat{\mathcal{E}}_{\{i_k,j_k\}}$ simply by $\widehat{\mathcal{E}}_k$ for $1\leq k\leq \tilde{d}_n$ and denote their ordered values as $\widehat{\mathcal{E}}_{(1)} \leq \widehat{\mathcal{E}}_{(2)} \leq \cdots \leq \widehat{\mathcal{E}}_{(\tilde{d}_n)}$.
Following the formulation of the MarS algorithm (cf.~\eqref{prop_est_sn} and \eqref{prop_est}), we define
\begin{align}\label{prop_est_sn_g}
\widehat{t}_n & = \argmax_{1 \leq k \leq (\tilde{d}_n-1)} \widehat{R}_k,\ \widehat{s}_n = \tilde{d}_n - \widehat{t}_n 
\mbox{ and } \widehat{S}_n = \big \{\{i_k,j_k\}: \widehat{\mathcal{E}}_k \geq \widehat{\mathcal{E}}_{(\widehat{t}_n + 1)} \text{ for } 1 \leq k \leq \tilde{d}_n\} \big \}.
\end{align}

\noindent
This screening method identifies differences between the joint distributions of pairs, hence, we refer to it as {\it paired screening} (or, PairS). Based on our discussion above, it is clear that PairS will possess SSP/ESP under conditions similar to those used in Theorems \ref{mainthm} and \ref{mainthm_E}.

\begin{figure}[H]
		\centering
		\begin{tabular}{cc}
			\multicolumn{1}{c}{\hspace{3.5cm}Example 2} & \multicolumn{1}{c}{\hspace{3.5cm}Example 3}\\
                \multicolumn{2}{c}{\includegraphics[scale = 0.8]
                {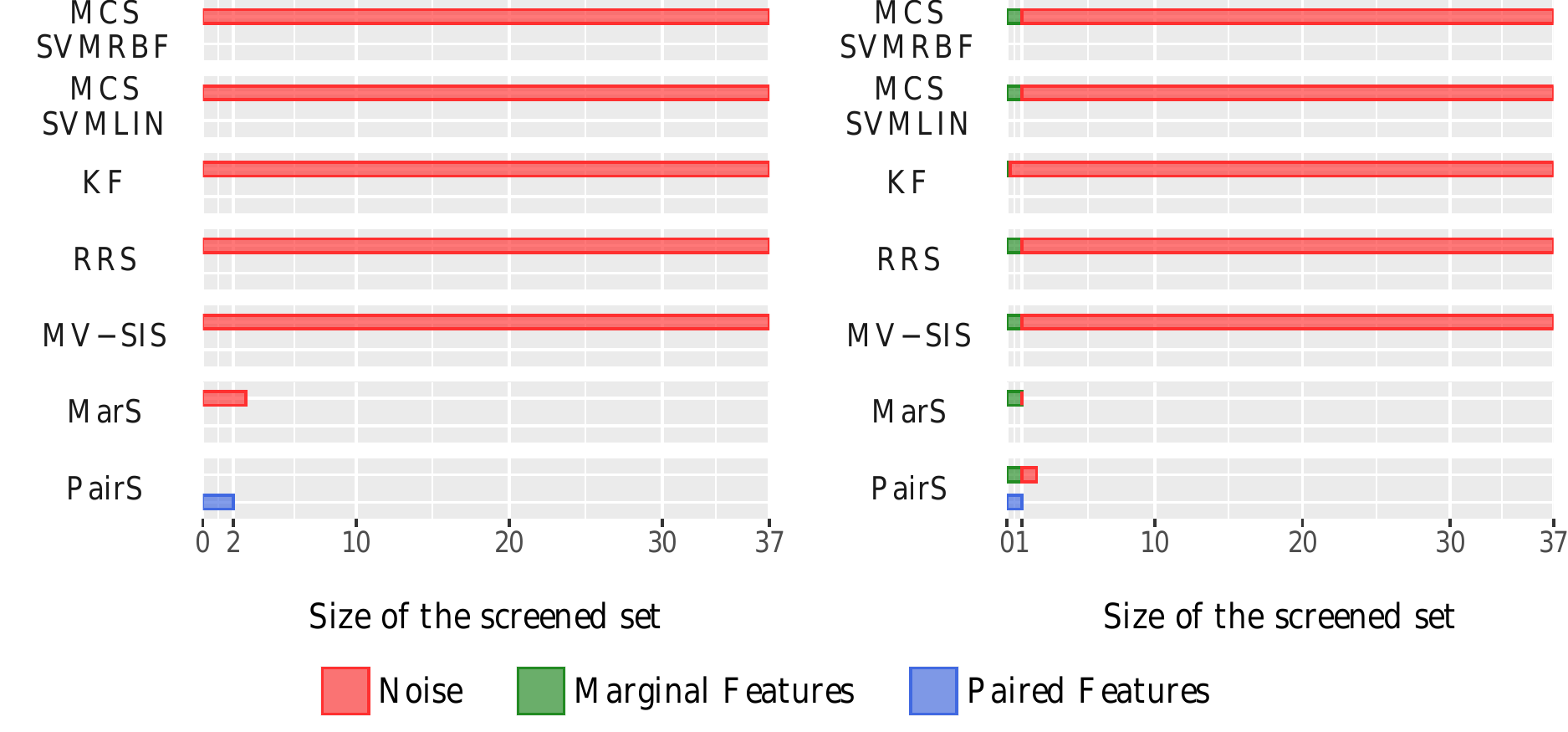}}
		\end{tabular}
		\caption{\it Bar plots indicating the average number of variables selected by competing methods as well as the MarS and PairS algorithms in Examples \ref{ex2} and  \ref{ex3} over $100$ simulation runs. Our proposed algorithms were implemented with $\gamma(t) = 1 - \exp{(-t)}$ for $t\geq 0$.}
		\label{fig3}
\end{figure}

Recall Example \ref{ex2}. The left panel in Figure \ref{fig3} shows that PairS successfully screened both the pairs $\{1,2\}$ and $\{3,4\}$ in this example. Moreover, PairS did not select any noise component.
The existing methods selected $\tilde{n}=37$ components that contain no discriminatory information in their marginals and are clearly incapable of retaining the paired features.
However, in the presence of a marginal feature, the NBP matching algorithm will forcefully couple the marginal component with some other (possibly a noise) component.
We now introduce a third example to demonstrate this limitation of the PairS algorithm.

\begin{example}\label{ex3}
$\bX_{\{1,2\}} \sim N_2(\0,\bSigma_1)$ with $\bSigma_1=[1,0.9;0.9,1]$, $X_3 \sim N(1,1)$ and $\bY_{\{1,2\}} \sim N_2(\0,\bSigma_2)$ with $\bSigma_2=[1,-0.9;-0.9,1]$, while $X_{4},\ldots,X_{d_n}$ and $Y_3,\ldots,Y_{d_n}$ are iid $N(0,1)$.
\end{example}

\noindent
In Example \ref{ex3}, 
the pair $\{1,2\}$ contributes to the signal set through its joint (bivariate) distribution, while the third component contributes through the marginal distribution. The remaining $(d_n-3)$ components all correspond to noise variables.

The right panel of Figure \ref{fig3} shows that in Example \ref{ex3}, competing methods
successfully captured the marginal signal, but the remaining $36$  components are all noises. The PairS method retained both $\{1,2\}$ as well as $\{3\}$, but the marginal signal $\{3\}$ brought an additional noise with it. 
To summarize, neither the existing nor the proposed methods could retain both $\{1,2\}$ and $\{3\}$ as signals, and dispose of the remaining components as noise.

\subsection{Screening of Mixed Features}\label{mixed}

{The main issue with the PairS algorithm is that it always detects pairs of features. However, some of the screened pairs may contain a marginal feature and a noise, or two marginal features. Here we modify the PairS algorithm to circumvent this issue.}
Consider $\widehat{P}_n$, the partition obtained by solving the NBP matching problem. After screening significant pairs using the PairS algorithm, we obtain the estimated set $\widehat{S}_n$ given by \eqref{prop_est_sn_g}. Without loss of generality, let us denote the screened pairs as $\{i_1,j_1\},\ldots, \{i_{\widehat{s}_n},j_{\widehat{s}_n}\}$. For each of these screened pairs, there are three possibilities: (i) {one component is a marginal signal, while the other is a noise}; (ii) {both the components are marginal signals}; 
and (iii) {the individual components have no marginal signal, together they form a paired signal}. More rigorously, for a screened pair $\{i_k,j_k\}\in \widehat{S}_n$ for $1 \leq k \leq \widehat{s}_n$, we have the following four possibilities:

\begin{enumerate}
 \item $i_k$ is a marginal signal, $j_k$ is a noise, i.e., $F_{i_k}\neq G_{i_k}$ and $F_{j_k}= G_{j_k}$,
 \item $j_k$ is a marginal signal, $i_k$ is a noise, i.e., $F_{i_k} = G_{i_k}$ and $F_{j_k}\neq G_{j_k}$,
 \item both $i_k$ and $j_k$ are marginal signals, i.e., $F_{i_k} \neq G_{i_k}$ and $F_{j_k}\neq G_{j_k}$,
 \item $i_k$ and $j_k$ do not have marginal signals, but taken together they constitute a paired signal, i.e., $F_{i_k} = G_{i_k}$ and $F_{j_k} = G_{j_k}$, but $\bF_{\{i_k,j_k\}} \neq \bG_{\{i_k,j_k\}}$.
\end{enumerate}
Among these four possibilities, we identify the most likely case by {identifying which among the following four mutually exhaustive null hypotheses is the most plausible}:
\begin{equation*}
H^k_{1,0} : F_{i_k}=G_{i_k}, H^k_{2,0} : F_{j_k}=G_{j_k}, H^k_{3,0} : F_{i_k}=G_{i_k},F_{j_k}=G_{j_k} \text{ and } H^k_{4,0} :  \bF_{\{i_k,j_k\}}=\bG_{\{i_k,j_k\}}.
\end{equation*}
Using energy distances, these hypotheses can be equivalently expressed as follows:
\begin{align*}
H^k_{1,0} : \mathcal{E}_{i_k}=0,\
\ H^k_{2,0} : \mathcal{E}_{j_k}=0,\
\ H^k_{3,0} :
\mathcal{E}_{i_k}+\mathcal{E}_{j_k}=0 \text{ and }
\ H^k_{4,0} :  \mathcal{E}_{\{i_k,j_k\}}=0.
\end{align*}

Although we have formulated the possibilities listed above as testing problems, it is not our aim to perform formal hypothesis testing. Instead, our goal is to check which of the four possible scenarios is the most plausible, which we achive via comparing the corresponding $p$-values. In particular, we carry out the four hypotheses tests mentioned above using the energy distance-based testing procedure developed by \cite{szekely2004testing}. The test statistics of interest are $\widehat{\mathcal{E}}_{i_k}$, $\widehat{\mathcal{E}}_{j_k}$, $\widehat{\mathcal{E}}_{i_k}+\widehat{\mathcal{E}}_{j_k}$ and $\widehat{\mathcal{E}}_{\{i_k,j_k\}}$, respectively, for $1\leq k\leq \widehat{s}_n$. Let the corresponding $p$-values be $p_1$, $p_2$, $p_3$ and $p_4$, respectively. 
The minimum among these four $p$-values points towards the most likely null hypothesis that is to be rejected. 
For instance, if $\min_{i} p_i = p_1$, then we have the strongest evidence for $F_{i_k}\neq G_{i_k}$, but $F_{j_k} = G_{j_k}$. In this case, we retain $i_k$ in the screened set as a singleton while discarding the $j_k$-th component as a noise.
The case when $\min_i p_i = p_2$ is similar, and the signal and noise indices are essentially swapped. If $\min_i p_i = p_3$, then scenario 3 has the strongest evidence, i.e., both $i_k$ and $j_k$ are screened
as marginal signals. Finally, if $\min_i p_i = p_4$, then we retain {$\{i_k,j_k\}$ as a paired signal}. We repeat this procedure for each of the selected pairs, and without loss of generality, we again denote the updated screened set as $\widehat{S}_n$. 
Since this modified method is capable of screening \emph{mixed features}, i.e., both marginal as well as paired features, we refer to it as {\it mixed screening} (or, MixS). 
\section{Numerical Studies} \label{numstud}

Using a diverse set of examples, we now study the performance of our proposed screening algorithms, viz., MarS and MixS (we exclude PairS due to its limitations and the subsequent introduction of MixS) and compare them with some existing methods. We have already discussed Examples \ref{ex1}-\ref{ex3} in earlier sections. 
Here, we introduce five new examples.
In Examples \ref{ex1}--\ref{ex4} and \ref{ex7}--\ref{ex8}, the noise variables are iid $N(0,1)$. For Examples \ref{ex5} and  \ref{ex6}, the noise variables are iid\ $C(0,1)$. 
The behaviour of MarS and MixS is studied for the three different choices of the $\gamma$ function mentioned after \eqref{endef}, viz., $\gamma_1(t)=1-\exp (-t)$, $\gamma_2(t) = \log(1+t)$ and $\gamma_3(t)=\sqrt{t}$ for $t\geq 0$.

\begin{example}\label{ex4}
 $X_1,\ldots,X_4\stackrel{iid}{\sim}N(0,1)$ and $Y_1,\ldots,Y_4\stackrel{iid}{\sim}N(0,1/3)$.
\end{example}
\begin{example}\label{ex5}
 $X_1,\ldots,X_4\stackrel{iid}{\sim}C(0,1)$, while $Y_1,\ldots,Y_4\stackrel{iid}{\sim}C(2,1)$.
\end{example}
\begin{example}\label{ex6}
 $X_1,\ldots,X_4\stackrel{iid}{\sim}C(0,1)$ and $Y_1,\ldots,Y_4\stackrel{iid}{\sim}C(0,5)$.
\end{example}

\begin{example}\label{ex7}
$X_1,\ldots,X_4\stackrel{iid}{\sim}N(0,4)$ and $Y_1,\ldots,Y_4$ are iid from the mixture distribution $1/2 N(-\mu, 4-\mu^2) + 1/2 N(\mu, 4-\mu^2)$ with $\mu = 1.95$.
\end{example}

\begin{example}\label{ex8}
$X_1,\ldots,X_4\stackrel{iid}{\sim}N(0,1)$, while the pairs $\bY_{\{1,2\}}$ and $\bY_{\{3,4\}}$ are iid from the bivariate signed normal distribution \citep[see][]{DUTTA201482}.
\end{example}
\noindent
Example \ref{ex4} is a scale problem with Gaussian distributions. Examples \ref{ex5} and \ref{ex6} correspond to a location and a scale problem, respectively, but involve heavy-tailed distributions. The parameters of the two distributions in Example \ref{ex7} have been set in such a way that the means and variances are the same, but they have differences in their shapes. In Example \ref{ex8}, both $\bY_{\{1,2\}}$ and $\bY_{\{3,4\}}$ follow the bivariate signed normal distribution. The one-dimensional marginals are $N(0,1)$ here, which implies that $F_k=G_k$ for all $1 \leq k \leq d_n$ in this example. Therefore, only $\{1,2\}$ and $\{3,4\}$ constitute the paired signals.

Throughout this study, we have simulated data with $d_n=1000$. In Examples \ref{ex1}-\ref{ex7}, the training set is formed with $200$ observations ($100$ from each class), while in Example \ref{ex8} the training sample size is $400$ ($200$ observations from each class).
All numerical results are based on $100$ independent replications.
The {\tt R} package {\tt VariableScreening} was used for implementing MV-SIS. We used the {\tt R} package {\tt e1071} for implementing SVMLIN and SVMRBF (related to MCS). For SVMRBF with kernel $K_\theta({\bf x},{\bf y})=\mathrm{exp}\{-\theta\|{\bf x}-{\bf y}\|^2\}$, we considered the default value $1/d$ of the tuning parameter $\theta$.
The implementation of NBP algorithm was done using the {\tt R} package {\tt nbpMatching} \citep[see, e.g.,][]{lu2011optimal}. Due to the sparsity of features in the considered examples, we impose a restriction on the estimated signal set so that its cardinality is less than, or equal to $[d_n/2]$ (see Appendix \ref{Appendix_B} for more details). However, the proposed algorithms MarS and MixS can be readily used even when the number of signals is relatively large. Discarding half of the covariates might not be desired if the presence of a large number of signals is suspected. Under such circumstances, the proposed algorithms can be easily modified by discarding $[d_n/M]$ energy values for some large, positive value of $M$. One can also discard the lowest $d_{min}$ and the highest $d_{max}$ energies, where $d_{min}$ and $d_{max}$ are user-defined constants \citep[see, e.g., ][]{MR3514508}.

\begin{figure}[ht!]
		\centering
		\begin{tabular}{cc}
			\multicolumn{1}{c}{\hspace{2cm}Example \ref{ex1}} &\multicolumn{1}{c}{\hspace{2cm}Example \ref{ex2}}\\
                \hspace{2cm} (marginal signal) & \hspace{2cm}(paired signal)\\
                \multicolumn{2}{c}{\hspace{-0.65cm}\includegraphics[height = 0.35\linewidth, width = 0.9\linewidth]{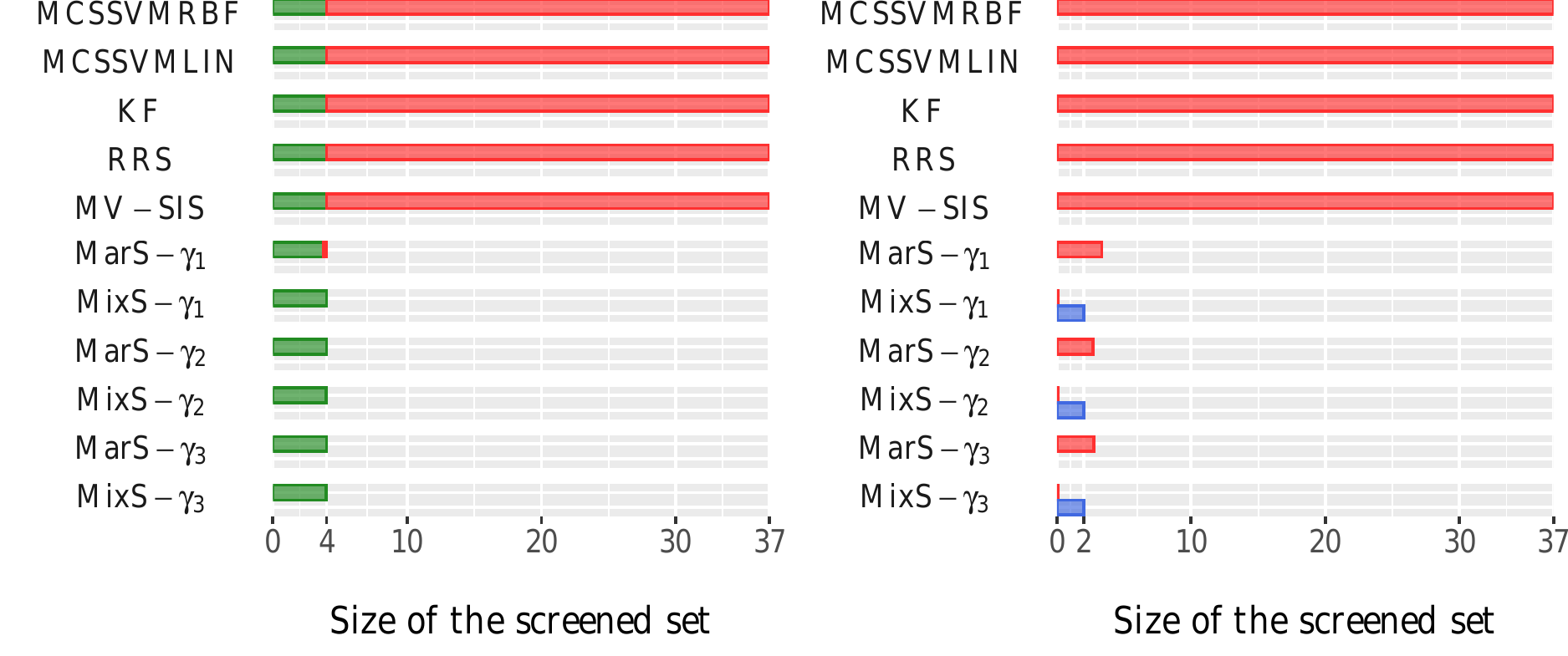}}\\
                \hspace{2cm} Example \ref{ex3} & \hspace{2cm} Example \ref{ex4} \\
                \hspace{2cm} (mixed signal) & \hspace{2cm} (marginal signal) \\
                \multicolumn{2}{c}{\hspace{-0.65cm}\includegraphics[height = 0.4\linewidth, width = 0.9\linewidth]{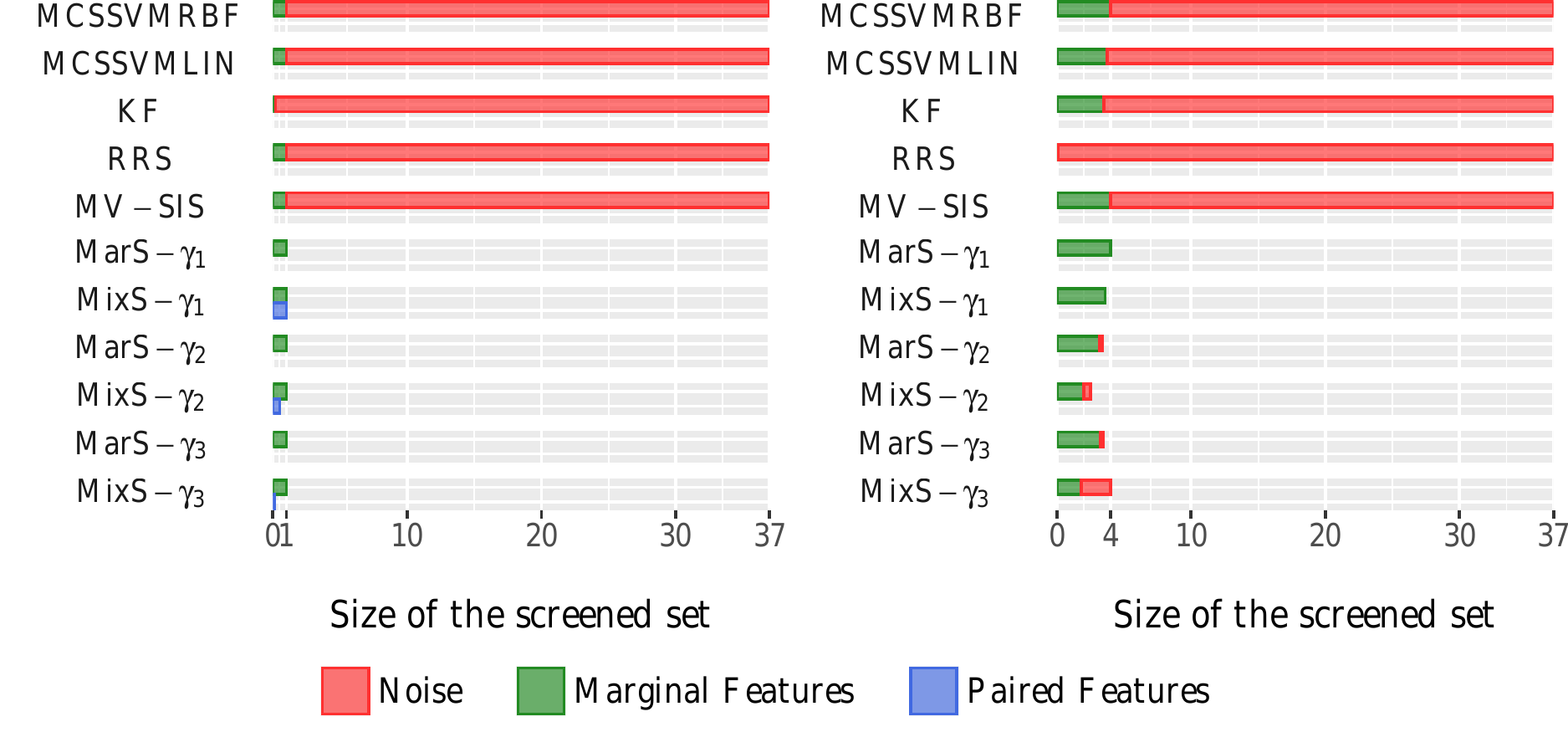}}
		\end{tabular}
		\caption{\it Bar plots indicating the average number of variables selected by competing methods as well as the MarS and MixS algorithms in Examples \ref{ex1}--\ref{ex4} over 100 simulation runs.}
	    \label{figsim1}
\end{figure}

The top left panel of Figure \ref{figsim1} shows that our proposed methods screen the signal components without selecting any additional noise in Example \ref{ex1}. We observe that the performance of MarS for $\gamma_1$ is a bit inferior when compared with $\gamma_2$ and $\gamma_3$.
Since the univariate marginals are equal in Example \ref{ex2}, competing methods as well as MarS were totally useless here. 
However, the bivariate joint distribution had scale differences in the pairs $\{1,2\}$ and $\{3,4\}$, and the advantage of the MixS algorithm can be clearly seen from this example (see the top right panel of Figure \ref{figsim1}). 

The competing methods (except KF) managed to successfully screen the marginal features in Example \ref{ex3}, but at the cost of including $36$ noise variables. In contrast, MarS selected the marginal features and discarded noise variables from $\widehat{S}_n$. The MixS algorithm with $\gamma_1$ outperformed all other choices of $\gamma$ by effectively screening both $\{1,2\}$ as well as $\{3\}$ (see the bottom left panel of Figure \ref{figsim1}). 
The bottom right panel of Figure \ref{figsim1} shows that the performance of both MarS and MixS is sensitive to the choice of $\gamma$ function in Example~\ref{ex4}.
In particular, $\gamma_1$ outperformed the other two choices here. Among the competing methods, RRS failed to retain any of the signals and detected $37$ noise components, while the others managed to retain the marginal features (along with $33$ noise variables).

\begin{figure}[ht!]
	\centering
	\begin{tabular}{cc}
		\multicolumn{1}{c}{\hspace{2cm}Example \ref{ex5}} &\multicolumn{1}{c}{\hspace{2cm}Example \ref{ex6}}\\
		\hspace{2cm} (Marginal signal) & \hspace{2cm} (Marginal signal)\\
	\multicolumn{2}{c}{\hspace{-0.65cm}\includegraphics[height = 0.35\linewidth, width = 0.9\linewidth]{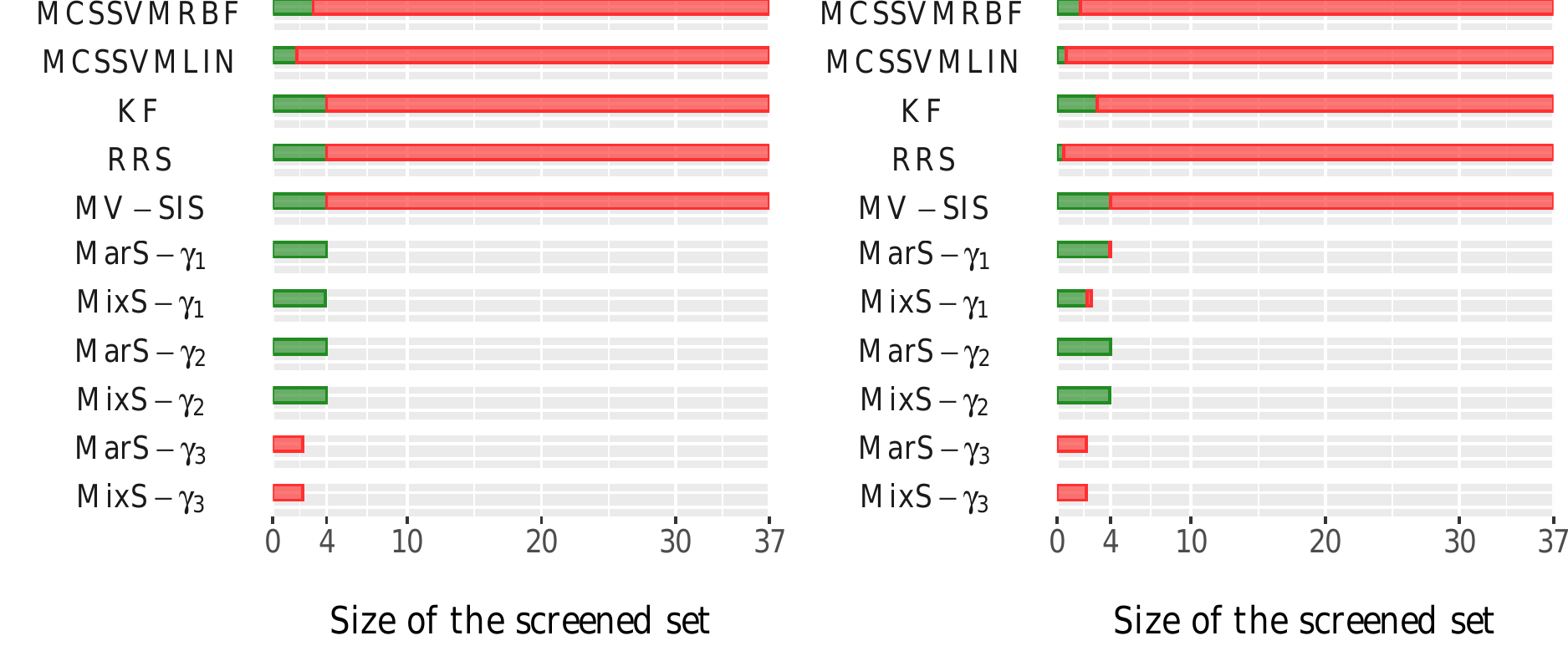}}	\\
		\hspace{2cm}Example \ref{ex7} & \hspace{2cm}Example \ref{ex8} \\
		\hspace{2cm} (Marginal signal) & \hspace{2cm} (Paired signal)\\
		\multicolumn{2}{c}{\hspace{-0.65cm}\includegraphics[height = 0.4\linewidth, width = 0.9\linewidth]{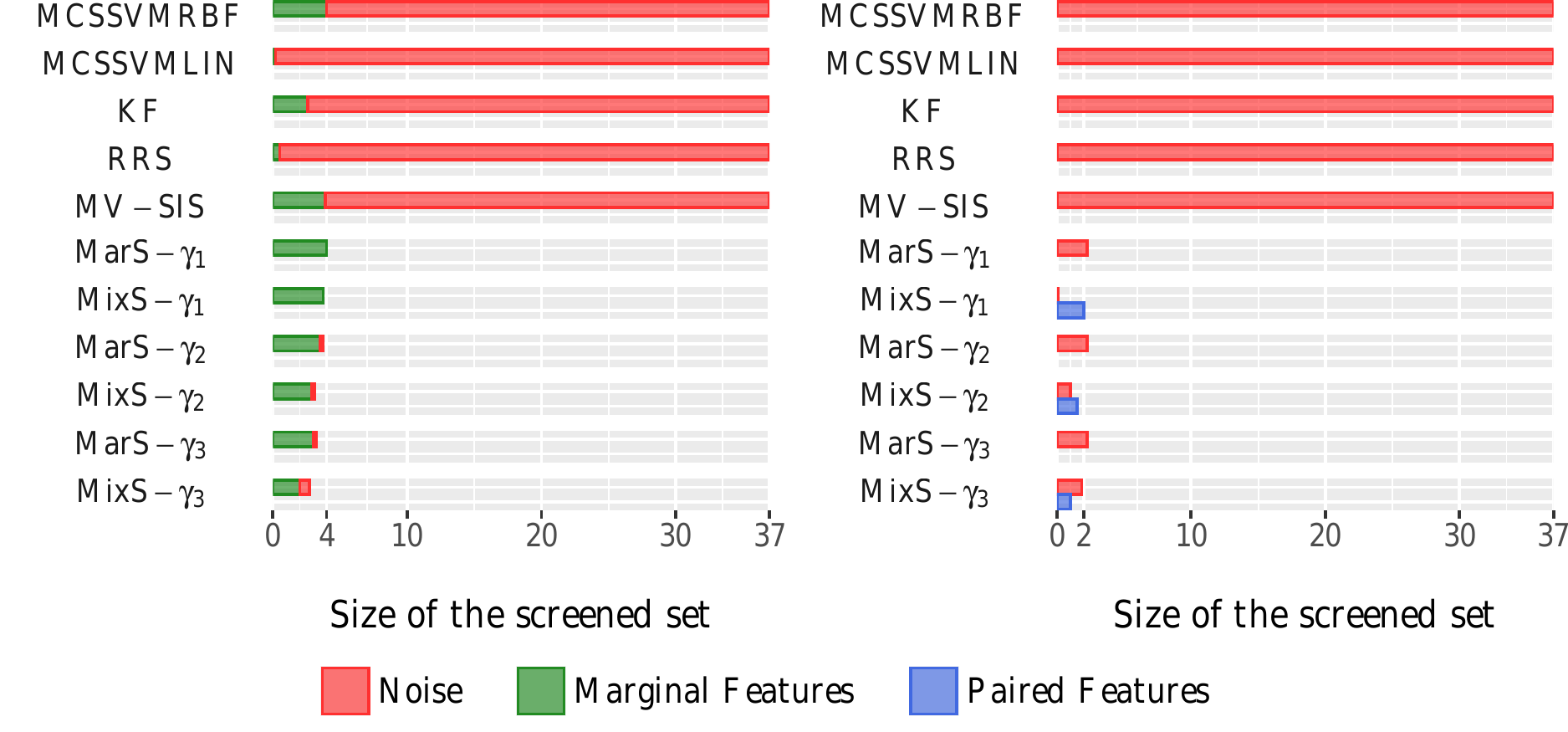}}
	\end{tabular}
	\caption{\it Bar plots indicating the average number of variables selected by competing methods as well as the MarS and MixS algorithms in Examples \ref{ex5}--\ref{ex8} over 100 simulation runs.}
	\label{figsim2}
\end{figure}

Examples \ref{ex5} and \ref{ex6} involve heavy-tailed distributions. 
Observe from the top panel of Figure \ref{figsim2} that MCS-SVM with linear as well as RBF kernels were unsuccessful in both these examples. RRS too had poor performance in Example~\ref{ex6}, which is a scale problem. Other competing methods screened all four signals but at the expense of gathering $33$ noise components. 
In fact, none of the competing methods (except MV-SIS) achieved sure screening in these two examples. 
It is evident from the top panel of Figure \ref{figsim2} that $\gamma_3$ (an unbounded function) is a poor choice when dealing with heavy-tailed distributions in Examples~\ref{ex5}~and~\ref{ex6}. In contrast, $\gamma_1$ (a bounded function) yielded satisfactory results. However, MixS with $\gamma_1$ could not retain all four signals in Example \ref{ex6}.  Recall that the screening accuracy of MixS relies on the solution of the NBP matching problem. 
Probably, the matching algorithm found a sub-optimal solution, which led to this deterioration in the overall performance. It is interesting to note that $\gamma_2$ outperformed $\gamma_1$ in both examples.

We dealt with a mixture distribution in Example \ref{ex7}. 
Among competing methods, only MCS-SVM (with the RBF kernel) and MV-SIS successfully retained all four relevant features (see the bottom left panel of Figure \ref{figsim2}). MCS-SVMLIN could not perform well because the two classes do not have any difference in their mean vectors. 
KF picked only three important features, whereas the performance of RRS was worse as it failed to procure even a single signal. 
In stark contrast, MarS with $\gamma_1$ selected only the useful features and managed to discard noise, while MixS with $\gamma_1$ yielded quite encouraging results. Our proposals with $\gamma_2$ and $\gamma_3$ are less promising, but still performed better than the competing methods.
Example~\ref{ex8} is more complex. 
Inescapably, all the competing methods as well as the MarS algorithm failed to identify the paired signals. But, MixS successfully retained only the paired signals 
(see the bottom right panel of Figure \ref{figsim2}). Among the three choices, $\gamma_1$ again led to the best result followed by $\gamma_2$ and $\gamma_3$. Remark \ref{gamma_C} in Appendix \ref{Appendix_C} contains a discussion on the relative performance of these three choices of $\gamma$.

Based on the above, we clearly see the usefulness of the proposed MarS and MixS algorithms. Among the competing methods, MV-SIS and MCS are generally quite promising in dealing with cases having marginal differences only. For the final task of classification, MV-SIS does not have a clear choice of classifier, which is left to the user. In contrast, MCS has a clear choice of the classifier (SVMLIN or SVMRBF). Hence, we decided to carry this method forward for a comparative performance in classification problems (see Section \ref{numstud_class} for more details).

\section{Classification Using Energy Distances} \label{class}

Our primary motivation behind using energy distances for screening was that they provide a measure of separation between the underlying distributions. As the next step, here we use energy distances to build a classifier based on the signal set $S_n$. Recall that ${S}_n$ constitutes of two subsets, namely, $S_{1n}$ and $S_{2n}$ comprising of the marginal signals and paired signals, respectively.
Given $S_n$, we can define the average of energy distances between $\bF$ and $\bG$ as: 
\[
\bar{\mathcal{E}}_n = \frac{1}{s_{1n}} \sum \limits_{k \in {S}_{1n}} \mathcal{E}_k + \frac{1}{s_{2n}} \sum \limits_{\{i,j\} \in {S}_{2n}} \mathcal{E}_{\{i,j\}},
\]
where $s_{1n}:=|S_{1n}|$ and $s_{2n}:=|S_{2n}|$. Fix $\bz\in\mathbb{R}^{d_n}$, and consider the following:
\begin{center}
 $\xi^\gamma(\bz) = E[h^{\gamma}_0(\bz,\bY_1)]-E[h^{\gamma}_0(\bz,\bX_1)]-\frac{1}{2}\left \{E[h^{\gamma}_0(\bY_1,\bY_2)]-E[h^{\gamma}_0(\bX_1,\bX_2)]\right \}$.
\end{center}

\noindent 
Here, $h_0^{\gamma}(\bu,\bv)$ is a measure of dissimilarity between $\bu,\bv \in \mathbb{R}^{d_n}$, which is defined as
\begin{equation} \label{h_0trans}
h_0^{\gamma}(\bu,\bv) = \frac{1}{s_{1n}}\sum_{k \in S_{1n}}\ \gamma\left (|u_l - v_l|^2\right ) + \frac{1}{s_{2n}}\sum_{\{i,j\}\in S_{2n}}\ \gamma\left (\frac{|u_{i} - v_{i}|^2+|u_{j} - v_{j}|^2}{2}\right ).
\end{equation}

\noindent 
If $\bZ$ is independent of $\bX_1,\bX_2$ and $\bY_1,\bY_2$, then it is easy to show that $E_{\bZ\sim\bF}[{\xi}^\gamma({\bf Z})]=\bar{\mathcal{E}}_n$, while $E_{\bZ\sim\bG}[{\xi}^\gamma({\bf Z})]=-\bar{\mathcal{E}}_n$. 

Since $\bar{\mathcal{E}}_n$ is always positive, we expect ${\xi}^\gamma({\bf Z})$ to take positive (respectively, negative) values if $\bZ\sim\bF$ (respectively, $\bZ\sim\bG$). This motivates the use of ${\xi}^\gamma$ for classifying a test observation. 
Using ${\xi}^\gamma$, we define a classifier $\delta_{0}$ as follows:
\[
\delta_0(\bz) = \begin{cases}
1,&\text{ if }\ \xi^{\gamma}(\bz)>0,\\
2, &\text{ otherwise.}
\end{cases}
\]

\noindent 
To use $\delta_0$ in practice, one needs to estimate the related quantities.
Using the screened set $\widehat S_n$ in place of $S_n$ (subsequently, $\widehat S_{1n}$ and $\widehat S_{2n}$ in places of $S_{1n}$ and $S_{2n}$, respectively) we obtain the sample counterpart of $h^\gamma_0(\bu,\bv)$ as 

\begin{equation} \label{h_trans}
h_n^{\gamma}(\bu,\bv) = \frac{1}{\widehat{s}_{1n}}\sum_{k\in\widehat{S}_{1n}}\ \gamma\left (|u_l - v_l|^2\right ) + \frac{1}{\widehat{s}_{2n}}\sum_{\{i,j\}\in\widehat{S}_{2n}}\ \gamma\left (\frac{|u_{i} - v_{i}|^2+|u_{j} - v_{j}|^2}{2}\right),
\end{equation}
{where $\widehat s_{1n} := |\widehat S_{1n}|$ and $\widehat s_{2n} := |\widehat S_{2n}|$}. Further, define $\widehat{\xi}^{\gamma}_{n}({\bf z})=\widehat{\xi}^\gamma_{2n}({\bf z})-\widehat{\xi}^\gamma_{1n}({\bf z})$ with
\begin{align}
&\ \widehat{\xi}^{\gamma}_{1n}(\bz) =\frac{1}{n_1} \sum_{i=1}^{n_1}h^{\gamma}_n(\bz,\bX_i)-\frac{1}{2n_1(n_1-1)} \sum_{1\leq i\neq j\leq n_1} h^{\gamma}_n(\bX_i,\bX_j) \mbox{ and }\\\nonumber
&\ \widehat{\xi}^{\gamma}_{2n}(\bz) =\frac{1}{n_2} \sum_{j=1}^{n_2}h^{\gamma}_n(\bz,\bY_j)-\frac{1}{2n_2(n_2-1)} \sum_{1\leq i\neq j\leq n_2} h^{\gamma}_n(\bY_i,\bY_j).
\end{align}
Using $\widehat{\xi}^{\gamma}_{n}$, we define the sample version of $\delta_0$ as
\begin{align}\label{gSAVG}
\delta_{\rm bgSAVG}(\bz)=\begin{cases}
1, \text{ if } {\widehat{\xi}^{\gamma}_{n}(\bz)}>0,\\
2, \text{ otherwise}.
\end{cases}
\end{align}
This sample version is in fact the block generalized scale-adjusted average distance classifier (abbreviated as the bgSAVG classifier) proposed in \cite{roy2020}.

\subsection{Consistency} \label{consistency}



Throughout this section, we assume that $S_n$ consists of marginal signals only, i.e., $S_n=S_{1n}$ and $S_{2n}=\emptyset$ is the empty set. Consequently, $\bar{\mathcal{E}}_n$ simplifies to $\sum_{k \in {S}_{1n}} \mathcal{E}_k/s_{1n}$. The definitions of $h^\gamma_0$ in \eqref{h_0trans} and $h^\gamma_n$ in \eqref{h_trans} are also reduced to
\begin{align*}
h_0^{\gamma}(\bu,\bv) = \frac{1}{s_{1n}}\sum_{k\in S_{1n}}\ \gamma\big(|u_l - v_l|^2\big) \text{ and }\
h_n^{\gamma}(\bu,\bv) = \frac{1}{\widehat{s}_{1n}}\sum_{k\in\widehat{S}_{1n}}\ \gamma\big(|u_l - v_l|^2\big),
\end{align*}
respectively, for $\bu,\bv\in\mathbb{R}^{d_n}$. The resulting classifier is referred to as the generalized scale-adjusted average distance (in short, gSAVG) classifier. 
We will study consistency properties of the gSAVG classifier combined with the marginal screening method MarS. Recall from Theorem \ref{mainthm_E} that if assumptions A1 and A2 are satisfied, then MarS possesses ESP. Firstly, we state a convergence result for the discriminant $\widehat{\xi}^\gamma_n$. 
\begin{lemma}\label{exbddisc}
If assumptions {\rm A1} and {\rm A2} are satisfied for $0<\alpha_1<(1-\beta)/2$ and $0<\alpha_2<(1-\beta)/2$, respectively, and $\gamma$ is a bounded function, then there exists $b_3>0$ such that
\[
{\rm P}\left [|\widehat{\xi}^\gamma_n(\bz) - \xi^\gamma(\bz)|>n^{-\alpha} \right ] \leq O \left (e^{-b_3\{ n^{1-2\alpha}-n^\beta\}}\right )
\]
for all $\bz\in\mathbb{R}^{d_n}$ and $0 < \max\{\alpha_1,\alpha_2\}\le \alpha<(1-\beta)/2$.
\end{lemma}
\noindent
From Lemma \ref{exbddisc}, it follows that $|\widehat{\xi}^\gamma_n(\bz) - \xi^\gamma(\bz)|$ converges in probability to 0 at an exponential rate as $n$ goes to infinity.

The classifier $\delta_0$ is basically the ``oracle" version of the classifier $\delta_{\rm gSAVG}$. 
Let $\Delta_0$ and $\Delta_{\rm gSAVG}$ denote the misclassification probabilities of $\delta_0$ and $\delta_{\rm gSAVG}$, respectively. We now derive an upper bound for $\Delta_{\rm gSAVG}-\Delta_0$, and analyze the terms in this bound as~$n\to\infty$. 

\vspace{0.15in}
\begin{theorem}\label{thmclass}
If assumptions {\rm A1} and {\rm A2} are satisfied for $0<\alpha_1<(1-\beta)/2$ and $0<\alpha_2<(1-\beta)/2$, respectively, and $\gamma$ is bounded, then there exists a constant $b_3>0$ such that
$$\Delta_{\rm gSAVG} - \Delta_0 \leq {\rm P}\left [|\xi^\gamma(\bZ)|< n^{-\alpha}\right ] + O \left ( e^{-b_3\{ n^{1-2\alpha}- n^\beta\}} \right )$$
for all $0 < \max\{\alpha_1,\alpha_2\}\le \alpha<(1-\beta)/2$.
\end{theorem}

\noindent
The second term in the upper bound goes to zero as $n \to \infty$. Under some additional conditions, we can also show that $P[|\xi^\gamma(\bZ)|< n^{-\alpha}] \to 0$ and $\Delta_0 \to 0$ as $n \to \infty$ as long as $s_n \to \infty$ (see Lemmas \ref{P0_conv} and \ref{Delta0_conv} in Appendix \ref{Appendix_A} for details). This now readily implies that $\Delta_{\rm gSAVG} \to 0$ (i.e., the gSAVG classifier achieves {\it perfect classification}) as $n \to \infty$. If $s_n \to \infty$, then this is expected because we accumulate more signal as $n \to \infty$, while ESP of MarS ensures that the noise variables get discarded. However, in our setup there is no restriction on the rate of growth of $s_n$, which can even be arbitrarily slow compared to $n$.

\begin{remark}
Alike Theorem \ref{mainthm}, if we assume $\bF$ and $\bG$ to be Gaussian and $\gamma$ to be $L$-Lipschitz continuous for some $L>0$ (e.g., $\gamma_2$), then Theorem \ref{thmclass} holds.
\end{remark}

\begin{remark}
In Theorem \ref{thmclass}, we have derived results for the case with marginal signals only. 
Similar derivations can also be made when we have paired, or mixed signals. While it is not difficult to make such an extension, the corresponding assumptions would be rather complicated. Hence, we chose not to pursue those to keep the exposition simple.
\end{remark}

\subsection{Comparison with Existing Classifiers} \label{numstud_class}

In this section, we conduct a comparative study of the performance of the proposed screening methods, namely, MarS and MixS in conjunction with the corresponding classifiers gSAVG and bgSAVG, respectively. We also include some state-of-the-art classifiers for comparison. 
Recall that the MCS screening method with SVMLIN and SVMRBF classifiers were denoted by MCS-SVMLIN and MCS-SVMRBF, respectively. We shall continue to use the same notation for the respective classifiers. 
Apart from MCS, we also include some popular {\it linear} classifiers suitable for sparse data, viz., sparse Support Vector Machines \citep[Sparse SVMLIN,][]{yi2017semismooth}, GLMNET \citep{hastie2009elements}, sparse group LASSO \citep[Sparse gLASSO,][]{simon2013sparse} and the ({\it non-linear}) nearest neighbor classifier based on sparse random projections~\citep[Sparse NN-RAND,][]{deegalla2006reducing}.

The {\tt R} packages {\tt sparseSVM}, {\tt glmnet}, {\tt SGL} and {\tt RandPro} were used for implementation of Sparse SVMLIN, GLMNET, Sparse gLASSO and Sparse NN-RAND, respectively. In this study, the parameters of the above methods were set to default (unless specified otherwise). For Sparse SVMLIN, the parameter associated with the penalty function was tuned using cross-validation. In GLMNET, the tuning parameter $\lambda$ was chosen by cross-validation (with the number of folds set to be the default value $10$). 
We provided the group memberships during our implementation of Sparse gLASSO, while the penalty term was chosen using cross-validation.
For Sparse NN-RAND, we used the projection method `li' that generates very sparse random matrices. {\tt R} codes for the proposed classifiers are available at \sloppy{\url{https://www.dropbox.com/sh/nd1v8b8nmb4cf86/AADkCaOStf2W6gHAKl8LD_h4a?dl=0}}.

We first implemented the gSAVG classifier without any screening (WoS). Then, the gSAVG classifier was combined with MarS, while the bgSAVG classifier was used with MixS.
Recall Examples \ref{ex1}--\ref{ex8} from Section \ref{numstud}. The training sample in each of Examples \ref{ex1}--\ref{ex7} was formed by generating $100$ observations from each class and a test set of size $500$ ($250$ from each class) was used. In Example \ref{ex8}, the training sample constituted of $200$ observations from each class, while the test size remained $500$. This procedure was repeated $100$ times to compute the average misclassification rates, which are reported in Table~\ref{tablesimulation}.

Superiority of the gSAVG classifier with MarS is clear from the examples with marginal differences, namely, Examples \ref{ex1}, \ref{ex4}, \ref{ex5}, \ref{ex6} and  \ref{ex7} (see Table \ref{tablesimulation}). On the other hand, the bgSAVG classifier with MixS yielded promising results in the non-marginal scenarios (namely, Examples \ref{ex2}, \ref{ex3} and  \ref{ex8}). Among the competing methods, linear classifiers showed satisfactory performance only in Examples \ref{ex1} and \ref{ex3}. 
They failed in Examples \ref{ex2}, \ref{ex4}, \ref{ex7} and \ref{ex8} where the difference lies either in scale or shape, and led to a misclassification rate of almost 50\%. The performance of Sparse NN-RAND was generally quite poor across all examples. Only MCS-SVMRBF showed some improvement in location and scale problems. In Examples \ref{ex5} and \ref{ex6} (with heavy-tailed distributions), none of the competing methods yielded satisfactory results due to lack of robustness. 
\newcommand{\se}[1]{\textit{#1}}

\begin{table}[t]

\caption{Misclassification rates and standard errors (in italics) of different classifiers in simulated data sets. The figure in {\bf bold} indicates the minimum misclassification rate.}

	\vspace*{0.1in}
	\footnotesize
	\setlength{\tabcolsep}{1pt}
	\centering
	\begin{tabular}{|c|c|c|c|c|c|c|c|ccc|ccc|ccc|}\hline
	& & Sparse & & Sparse & Sparse & MCS & MCS & \multicolumn{3}{c|}{gSAVG} & \multicolumn{3}{c|}{gSAVG} & \multicolumn{3}{c|}{bgSAVG} \\
	& & SVM- & GLM- & Group & NN- & SVM- & SVM- & \multicolumn{3}{c|}{WoS} & \multicolumn{3}{c|}{MarS} & \multicolumn{3}{c|}{MixS} \\
	Ex & Bayes & LIN & NET & LASSO & RAND & LIN & RBF & $\gamma_1$ & $\gamma_2$ & $\gamma_3$ & $\gamma_1$ & $\gamma_2$ & $\gamma_3$ & $\gamma_1$ & $\gamma_2$ & $\gamma_3$ \\\hline
	1 & 15.87 & 20.20 & 17.60 & 19.70 & 44.21 & 27.01 & 21.87 & 41.37 & 36.97 & 36.24 & 20.89 & 17.45 & {\bf 17.11} & 19.94 & 17.51 & {\bf 17.10} \\
	& \se{0.18} & \se{0.52} & \se{0.22} & \se{0.19} & \se{0.25} & \se{0.27} & \se{0.23} & \se{0.25} & \se{0.24} & \se{0.25} & \se{0.38} & \se{0.28} & \se{0.28} & \se{0.27} & \se{0.29} & \se{0.28} \\\hline
	
	2 & 5.07 & 49.20 & 49.98 & 49.92 & 49.45 & 50.30 & 49.57 & 44.03 & 46.94 & 46.51 & 49.96 & 49.66 & 49.54 & 13.01 & 13.88 & {\bf 12.82} \\
	& \se{0.10} & \se{0.17} & \se{0.20} & \se{0.20} & \se{0.20} & \se{0.22} & \se{0.26} & \se{0.24} & \se{0.23} & \se{0.23} & \se{0.21} & \se{0.21} & \se{0.21} & \se{0.25} & \se{0.19} & \se{0.43} \\\hline
	
	3 & 8.45 & 23.80 & 23.33 & 26.65 & 46.74 & 34.30 & 31.35 & 42.65 & 41.24 & 40.57 & 22.63 & 22.56 & 22.56 & {\bf 17.32} & 21.14 & 22.46 \\
	& \se{0.13} & \se{0.22} & \se{0.26} & \se{0.25} & \se{0.23} & \se{0.27} & \se{0.25} & \se{0.23} & \se{0.23} & \se{0.21} & \se{0.17} & \se{0.18} & \se{0.18} & \se{0.19} & \se{0.23} & \se{0.19} \\\hline
	
	4 & 7.50 & 47.60 & 50.11 & 49.92 & 50.19 & 49.99 & 38.35 & 40.99 & 45.17 & 45.23 & {\bf 10.14} & 18.80 & 19.09 & 12.45 & 24.23 & 33.09 \\
	& \se{0.12} & \se{0.20} & \se{0.19} & \se{0.19} & \se{0.20} & \se{0.22} & \se{0.27} & \se{0.26} & \se{0.23} & \se{0.24} & \se{0.23} & \se{0.75} & \se{0.73} & \se{0.57} & \se{0.92} & \se{0.51} \\\hline
	
	5 & 11.32 & 49.40 & 46.09 & 50.31 & 49.91 & 48.06 & 47.17 & 36.55 & 34.31 & 43.24 & 12.42 & {\bf 12.07} & 50.12 & 13.42 & 12.21 & 49.99 \\
	& \se{0.14} & \se{0.35} & \se{0.71} & \se{0.18} & \se{0.18} & \se{0.31} & \se{0.31} & \se{0.26} & \se{0.23} & \se{0.21} & \se{0.15} & \se{0.14} & \se{0.21} & \se{0.32} & \se{0.14} & \se{0.21} \\\hline
	
	6 & 13.07 & 50.60 & 49.87 & 49.87 & 49.81 & 50.23 & 48.91 & 41.90 & 33.80 & 39.95 & 17.02 & {\bf 14.15} & 49.30 & 22.43 & 14.39 & 49.28 \\
	& \se{0.15} & \se{0.19} & \se{0.21} & \se{0.18} & \se{0.18} & \se{0.22} & \se{0.25} & \se{0.24} & \se{0.22} & \se{0.21} & \se{0.34} & \se{0.16} & \se{0.45} & \se{0.49} & \se{0.23} & \se{0.45} \\\hline
	
	7 & 5.09 & 50.24 & 50.01 & 50.19 & 50.15 & 50.12 & 49.81 & 40.16 & 43.70 & 44.79 & {\bf 9.34} & 22.22 & 27.91 & 10.64 & 25.01 & 33.28 \\
	& \se{0.10} & \se{0.19} & \se{0.21} & \se{0.21} & \se{0.22} & \se{0.22} & \se{0.22} & \se{0.22} & \se{0.22} & \se{0.22} & \se{0.16} & \se{0.73} & \se{0.75} & \se{0.44} & \se{0.86} & \se{0.73} \\\hline
	
	8 & 12.83 & 50.11 & 50.17 & 49.92 & 50.04 & 50.08 & 50.16 & 48.55 & 49.36 & 49.32 & 50.13 & 49.68 & 49.54 & {\bf 36.64} & 42.25 & 45.17 \\
	& \se{0.14} & \se{0.21} & \se{0.23} & \se{0.22} & \se{0.22} & \se{0.25} & \se{0.23} & \se{0.22} & \se{0.21} & \se{0.21} & \se{0.25} & \se{0.22} & \se{0.24} & \se{0.27} & \se{0.44} & \se{0.40} \\\hline
	\end{tabular}
	\label{tablesimulation}
\end{table}

\subsection{Real Data Analysis}\label{real}

We now study the performance of our proposed methodology on two benchmark and four real data sets. The first data is {\tt Madelon}, which is an artificial data set that was part of the NIPS 2003 feature selection challenge \citep{NIPS2004_5e751896}. The data points are of 500 dimension 
with $5$ informative covariates. An additional $15$ linear combinations of these covariates were added to form a signal set having cardinality $20$. The remaining $480$ components are all noise with no predictive power. 
{\tt Madelon} has a total of $2600$ observations ($1300$ from each class).
The second data set {\tt CorrAL} \citep{john1994irrelevant} is also an artificial data of dimension $10000$. It has $128$ observations from two classes, $72$ from the first class and $56$ from the second class. Out of the $10000$ features, only the first $6$ components are highly correlated with the response variable, while the rest are all noise. We used a subset of this data set containing the first $1000$ features. Both {\tt Madelon} and {\tt CorrAL} are available from the {\tt R} package {\tt sbfc}.
The third and fourth data sets, {\tt Bittner} and {\tt Shipp} are from the \href{https://schlieplab.org/Static/Supplements/CompCancer/datasets.htm}{Compcancer} database. {\tt Bittner} is a microarray data set with $38$ observations of dimension $2201$. This data set was first studied by \cite{Bittner2000}, where the authors analyzed gene expression profiles for the $38$ samples.
Class labels of $19$ tightly clustered samples was denoted by ML1, while the remaining 19 samples were labelled as ML2. {\tt Shipp} was introduced by \cite{shipp2002diffuse}, where the authors aimed at distinguishing diffuse large B-cell lymphoma (DLBCL) from a related GC B-cell lymphoma, follicular (FL). It has $77$ data points ($58$ and $19$ samples corresponding to DLBCL and FL, respectively) of dimension $798$.
The final two data sets are related to \href{https://file.biolab.si/biolab/supp/bi-cancer/projections/}{Microarray} gene expressions. The {\tt GSE3726} data has $52$ gene expressions of dimension 22283 from breast and colon cancer patients with $31$ and $21$ samples, respectively. In the {\tt GSE967} data set, we have 23 observations of dimension 9945. These gene expression values correspond to two childhood tumors, namely, Ewing's sarcoma (EWS) and rhabdomyosarcoma (RMS) containing $11$ and $12$ data points.

For each of these six data sets, we randomly selected $50$\% of the observations (without replacement) corresponding to each class to form the training set. The remaining observations were considered as test cases. This procedure was repeated $100$ times over different splits of the data to obtain a stable estimate of the misclassification rate.
As already noted in our simulations, the number of features selected by MarS and MixS varies with the choice of $\gamma$. For each real data set, we have summarized this information in the respective paragraphs, while the complete numerical result is presented in Appendix \ref{Appendix_C} (see Table \ref{tabcardi}).

\begin{table}[t]
\caption{Misclassification rates and standard errors (in italics) of different classifiers in real and benchmark data sets. The figure in {\bf bold} indicates the minimum misclassification~rate.}

	\vspace*{0.1in}
	\footnotesize
	\setlength{\tabcolsep}{2pt}
	\centering
	\begin{tabular}{|c|c|c|c|c|c|ccc|ccc|ccc|}\hline
	& Sparse & & Sparse & MCS & MCS & \multicolumn{3}{c|}{gSAVG} & \multicolumn{3}{c|}{gSAVG} & \multicolumn{3}{c|}{bgSAVG} \\
	& SVM- & GLM- & NN- & SVM- & SVM- & \multicolumn{3}{c|}{WoS} & \multicolumn{3}{c|}{MarS} & \multicolumn{3}{c|}{MixS} \\
	Data set & LIN & NET & RAND & LIN & RBF & $\gamma_1$ & $\gamma_2$ & $\gamma_3$ & $\gamma_1$ & $\gamma_2$ & $\gamma_3$ & $\gamma_1$ & $\gamma_2$ & $\gamma_3$ \\\hline

    Madelon & 40.66 & {\bf 38.65} & 42.23 & 43.39 & 40.33 & 40.07 & 40.07 & 40.09 & 38.73 & 38.73 & 38.74 & 38.72 & 38.70 & {\bf 38.63} \\
    & \se{0.12} & \se{0.16} & \se{0.14} & \se{0.14} & \se{0.16} & \se{0.17} & \se{0.15} & \se{0.18} & \se{0.13} & \se{0.13} & \se{0.13} & \se{0.13} & \se{0.14} & \se{0.13} \\\hline

    CorrAL & 43.52 & 27.89 & 45.58 & 29.52 & {\bf 26.06} & 47.50 & 47.50 & 47.79 & 31.42 & 31.41 & 31.41 & 29.70 & 27.72 & 27.51 \\
    & \se{0.39} & \se{0.60} & \se{0.57} & \se{0.74} & \se{0.65} & \se{0.54} & \se{0.55} & \se{0.54} & \se{0.66} & \se{0.66} & \se{0.69} & \se{0.65} & \se{0.66} & \se{0.78} \\\hline
    
    Bittner & 28.39 & 26.67 & 26.94 & 25.50 & 27.33 & 32.89 & 26.11 & 27.78 & 29.44 & 25.06 & 23.89 & 27.23 & 24.05 & {\bf 23.37} \\
    & \se{1.36} & \se{0.92} & \se{0.11} & \se{1.06} & \se{1.07} & \se{1.14} & \se{1.08} & \se{1.20} & \se{1.12} & \se{1.02} & \se{0.89} & \se{0.87} & \se{0.93} & \se{0.87} \\\hline
    
    Shipp & 19.29 & 21.28 & {\bf 18.50} & 20.79 & 20.66 & 23.21 & 22.39 & 23.34 & 22.24 & 21.61 & 23.13 & 22.05 & 18.92 & 20.29 \\
    & \se{0.53} & \se{0.57} & \se{0.48} & \se{0.67} & \se{0.62} & \se{0.73} & \se{0.70} & \se{0.65} & \se{0.76} & \se{0.77} & \se{0.88} & \se{0.78} & \se{0.76} & \se{0.73} \\\hline
    
    GSE3726 & 22.23 & 16.54 & 39.55 & 10.30 & 11.81 & 17.20 & 13.38 & 21.31 & 17.19 & 9.69 & 11.65 & 16.31 & {\bf 8.73} & 10.62 \\
    & \se{1.36} & \se{0.97} & \se{1.19} & \se{0.64} & \se{0.60} & \se{0.73} & \se{0.70} & \se{0.65} & \se{0.89} & \se{0.61} & \se{0.44} & \se{0.89} & \se{0.42} & \se{0.37} \\\hline
    
    GSE967 & 27.27 & 28.00 & 39.82 & 29.27 & 36.73 & 34.64 & 33.27 & 30.72 & 30.91 & 27.27 & 26.36 & 28.55 & {\bf 20.82} & 35.64 \\
    & \se{1.30} & \se{1.45} & \se{1.29} & \se{1.09} & \se{1.05} & \se{1.32} & \se{1.50} & \se{1.43} & \se{1.01} & \se{1.01} & \se{1.01} & \se{1.36} & \se{1.28} & \se{1.29} \\\hline
    
	\end{tabular}
	\label{tablereal}
\end{table}

We start by analyzing the {\tt Madelon} data. The number of components screened by MCS was $181$. Even if all $5$ signals are retained by MCS, its screened set still contained about $97\%$ noise. In contrast, MarS selected $3.31$ components, while the number of features screened by MixS ranged between $7.78$ and $8.39$ on an average for the three choices of $\gamma$. 
Usefulness of feature screening is clear from this data set (see Table \ref{tablereal}). The gSAVG classifier without any screening (WoS) performs worse than gSAVG with MarS and bgSAVG with MixS. In particular, MixS for $\gamma_3$ when combined with the bgSAVG classifier yielded the minimum misclassification rate. 
Among the competing classifiers, only GLMNET led to a competitive~performance.

MCS screened $15$ components in the {\tt CorrAL} data. 
MarS screened $1.54$ components, while MixS selected $2.46$ components on an average for varying $\gamma$. 
Again, we observe a significant reduction in misclassification rate of the proposed screening methods w.r.t.\ the WoS version of the gSAVG classifier.
Among the competing methods, MCS-SVM with the RBF kernel produced the best result. While GLMNET successfully detected sparsity and its performance was at par with the proposed methods, the others did not fare so well.

In {\tt Bittner}, MCS screened about $6$ components out of $2201$. On an average, the number of components retained by MarS ranged from $3.51$ to $3.6$, whereas for MixS the range was from $4.9$ to $5.6$ for the three choices of $\gamma$. Interestingly, Table \ref{tablereal} shows that the performance of gSAVG and bgSAVG is subjective to the choice of the $\gamma$ function. Overall, bgSAVG with MixS for $\gamma_3$ led to the minimum misclassification rate, while gSAVG with MarS for $\gamma_2$ secured the second position. 
MCS-SVMLIN led to the lowest misclassification rate among the competing methods, closely followed by GLMNET and Sparse~NN-RAND.

For different choices of $\gamma$, the average number of variables screened by MarS in the {\tt Shipp} data ranged from $2.52$ to $4.75$, while it ranged from $3.66$ to $5.20$ for MixS. 
Superiority of MixS over MarS is transparent from this data set (see Table \ref{tablereal}). 
In particular, bgSAVG with MixS had a significantly improved performance when compared to gSAVG with MarS for the choices $\gamma_2$ and $\gamma_3$. 
The performances of MCS-SVM (selected $40$ signals) with linear and RBF kernels were comparable with Sparse NN-RAND, which yielded the minimum misclassification rate here. Meanwhile, Sparse SVMLIN led to a promising performance~too.

With $d = 22283$, {\tt GSE3726} has the highest number of covariates among all the data sets, but the sample size is only $32$. On an average, the number of variables selected by MarS and MixS ranged from $1.01$ to $2.65$ and $1.03$ to $2.85$, respectively. For $\gamma_2$ and $\gamma_3$, we observed considerable improvement in the misclassification rates for gSAVG with MarS when compared against gSAVG implemented without screening. 
As expected, MixS yielded further improvement in the performance of the bgSAVG classifier by accounting for the differences in joint distributions. In fact, bgSAVG when combined with MixS for $\gamma_2$ yielded the lowest misclassification rate. The MCS classifiers retained $30$ components, and only they could outperform the WoS version of gSAVG. 
All the other classifiers failed to capture sparsity in the covariates, and led to significantly higher misclassification rates.

In the {\tt GSE967} data set, MCS screened $7$ components, while the number of components retained by MixS ranged from $4.68$ to $28.7$. Surprisingly, MixS with $\gamma_1$ screened several signal components (see Table \ref{tabcardi} in Appendix \ref{Appendix_C}).
The choice $\gamma_2$ led to the best misclassification rate when combined with MixS and bgSAVG. 
Among the competitors, only the linear classifiers, namely, Sparse SVMLIN, GLMNET and MCS-SVMLIN showed improvement when compared with the gSAVG classifier without any screening. On the other hand, the non-linear classifiers, namely, Sparse NN-RAND and MCS-SVMRBF performed even worse than that classifier.

\section{Discussion} \label{Conc}

Using energy distances, we have developed some methods of variable screening for classification. The novelty with the proposed method MarS is that it not only retains all the signals but also disposes of the noise variables with probability tending to 1 as $n\to\infty$. We have also built {the MixS algorithm} which screens pairs of components having differences in their joint distributions. Furthermore, a discrimination criterion that is coherent with the screening method has been proposed. The classifier is shown to yield {\it perfect classification} under fairly general conditions. We have used a variety of simulated examples and some benchmarks as well as real data sets to amply demonstrate the {superiority} of the proposed classifiers when compared with several popular classification methods.

It is evident that computing the $d\times d$ matrix of pair-wise energy distances of component variables makes the complexity of MixS quadratic in $d$. This is the minimum price one needs to pay in order to look beyond marginal signals (see the recent work by \cite{MR3852003, MR4041813, MR4359625}).
We believe that the choice of whether to look for discriminatory information beyond one-dimensional marginals remains largely in the practitioner's hands, and one should make the choice depending on the available prior information related to the classification problem itself. Nevertheless, due to the increased power of computational systems and parallel processing techniques, screening of paired features can be executed within a moderate time. 
The bottleneck in the complexity of MixS is the step where permutation tests are conducted for identifying paired features. To avoid this problem, one may think of using a parametric test instead of the permutation test. 
Note that the asymptotic null distribution of the energy statistic is an infinite weighted sum of independent chi-square random variables, where the weights depend on the underlying distribution (see \cite{szekely2004testing}). This distribution has been approximated using a gamma distribution with appropriate choices of shape and scale parameters \citep[see, e.g.,][]{pfister2018kernel}. On a different note, a recent work by \cite{Panda2022} approximates the tail behavior (of an appropriately centered and scaled version) of the infinite sum by a single chi-square random variable. However, adapting these methods to our situation is not straight-forward, although it is worth investigating and is a topic of future research.


\section*{Acknowledgments}

The authors would like to thank the Editor, Associate Editor and two anonymous reviewers for their careful reading of the manuscript and thoughtful comments.

\bibliographystyle{chicago}      
{\small
\bibliography{citation}   
}


\newpage
\appendix

\begin{center}
  {\large\bf Supplementary to
  
  \vspace*{0.05in}
  \LARGE\bf ``On Exact Feature Screening in Ultrahigh-dimensional Binary Classification" \vspace*{0.1in}}
\end{center}

\section{Proofs and Mathematical Details} \label{Appendix_A}
\allowdisplaybreaks

Define
\begin{align}\label{Tdef}
&\ T_{11k} = \frac{1}{\binom{n_1}{2}} \mathop{\sum\sum}_{1\le i < j\le n_1} \gamma(|X_{ik}-X_{jk}|^2),\  T_{22k} = \frac{1}{\binom{n_2}{2}} \mathop{\sum\sum}_{1\le i < j\le n_2} \gamma(|Y_{ik}-Y_{jk}|^2)\text{ and }\nonumber\\
&\ T_{12k} = \frac{1}{n_1n_2} \sum\limits_{i=1}^{n_1}\sum\limits_{j=1}^{n_2} \gamma(|X_{ik}-Y_{jk}|^2)\text{ for  } 1\leq k\leq d_n.
\end{align}
Recall that $\widehat{\mathcal{E}}_{k}$ is defined as follows:
\begin{align}\label{samen.def}
\widehat{\mathcal{E}}_{k}= 2T_{12k}-\{T_{11k}+T_{22k}\} \mbox{ for } 1\le k \le d_n.
\end{align}
Fix $k\in\{1,\ldots, d_n\}$. The sample energy $\widehat{\mathcal{E}}_{k}$ consistently estimates  ${\mathcal{E}}_{k}$ as $n$ goes to infinity (see \cite{szekely2005new}). We now state the bounded differences inequality that we will use to derive the rate of convergence for this estimator.\newline

\noindent
For any two vectors $\bx,\bx^\prime\in\mathbb{R}^n$ and an index $1 \leq l \leq n$, define a new vector $\bx^{\backslash l}\in\mathbb{R}^n$ as
\begin{align*}
\bx^{\backslash l}=\begin{cases}
x_j,& \text{if } j\neq l,\\
x^{\prime}_l, &           \text{if } j= l.
\end{cases}
\end{align*}
Using this notation, we say that $f:\mathbb{R}^n\to\mathbb{R}$ satisfies the bounded difference inequality with parameters $M_1,\ldots, M_n$ if
\begin{align}\label{bdddiffprop}
|f(\bx)-f(\bx^{\backslash l})|\leq M_l\text{ for each } 1 \leq l \leq n.
\end{align}

\begin{lemma}\label{bdddiffineq}
	\citep[page 37]{wainwright2019high} Suppose that $f$ satisfies the bounded difference inequality stated in \eqref{bdddiffprop} with parameters $M_1,\ldots, M_n$ and that the random vector $\bU = (U_1,\ldots, U_n)^\top$ has independent components. Then,
	$${\rm P}\left[\left|f(\bU)-{\rm E}[f(\bU)]\right|>\epsilon\right ]\leq 2e^{-\frac{2\epsilon^2}{\sum_{l=1}^nM_l^2}}\text{ for all }\epsilon>0.$$
\end{lemma}

\begin{lemma}\label{expbound}
	Suppose that $\gamma$ is a bounded function. Then, there exists a constant $b>0$ such that
	$$P\big [|\widehat{\mathcal{E}}_{k} - \mathcal{E}_k|\leq n^{-\alpha}\big ]\geq 1 -O\big ( e^{-b n^{1-2\alpha}}\big )\text{ for all }0 <\alpha<1/2\text{ and } 1\leq k\leq d_n.$$
\end{lemma}

\noindent {\bf Proof :}
Fix $0 <\alpha<1/2$ and $k\in\{1,\ldots , d_n\}$.  Recall the definitions of $T_{11k},T_{22k}$ and $T_{12k}$ in \eqref{Tdef}. Observe that $T_{11k}$ and $T_{22k}$ are one sample U-statistics with a kernel of order 2, while $T_{12k}$ is a two-sample U-statistic with a kernel of order $(1,1) $. Using the union bound, it follows from \eqref{samen.def} that
\begin{align}\label{0X}
&\ P\big [|\widehat{\mathcal{E}}_{k} - \mathcal{E}_k| > n^{-\alpha}\big ]= P[|\{2T_{12k} - T_{11k} - T_{22k}\}-\{2E[T_{12k}] - E[T_{11k}]-E[T_{22k}]\}|>n^{-\alpha}],\nonumber \\
&\ \hspace{1.25in} \leq P[|T_{11k}-E[T_{11k}]|>n^{-\alpha}/3]+ P [|T_{12k}- E[T_{12k}]|>n^{-\alpha}/6 ]\nonumber \\
&\hspace{3.35in} + P [|T_{22k} - E[T_{22k}]|>n^{-\alpha}/3].
\end{align}
\noindent
Consider the first term in the right hand side (RHS) of \eqref{0X}. Note that the random variables $X_{1k},\ldots,X_{n_1k}$ are independently distributed. Let us denote the vector $(X_{1k},\ldots,X_{n_1k})^\top$ by $\mathcal{X}_k$. Since $\gamma$ is bounded, Lemma \ref{bdddiffineq} can be used to obtain an upper bound on $P[|T_{11k}-E[T_{11k}]|>n^{-\alpha}/3]$. We can view $T_{11k}$ as the function $f(X_{1k},\ldots,X_{n_1k}) = f(\bf{\mathcal{X}}_k)$. For any given co-ordinate $l \in \{1,\ldots,n_1\}$, we have
\begin{align*}
\left |f(\mathcal{X}_k)-f(\mathcal{X}_k^{\backslash l})\right |&\leq \frac{1}{n_1(n_1-1)}\sum\limits_{j\neq l}\bigl|h(X_{jk},X_{lk})-h(X_{jk},X^\prime_{lk})\bigr|\leq (n_1-1)\frac{2}{n_1(n_1-1)}=\frac{2}{n_1}.
\end{align*}
So, the bounded difference property holds for $T_{11k}$ with parameter $M_l=2/n_1$ for all $1 \leq l \leq n_1$. Using Lemma \ref{bdddiffineq}, we have
\begin{align*}
{\rm P}\big [|{T}_{11k}-{\rm E}[{T}_{11k}]|>n^{-\alpha}/3 \big]\leq 2e^{- \frac{n_1 n^{-2\alpha}}{18}}.
\end{align*}
Since $\lim_{n\to\infty}n_1/n=\pi_1$ with $0<\pi_1<1$, there exist constants $b_{11}>0$ and $N_{11}\in\mathbb{N}$ such that
\begin{align}\label{1X}
P[|{T}_{11k} - {\rm E}[{T}_{11k}]|\geq n^{-\alpha}/3]\leq 2e^{-{b_{11} n^{1-2\alpha}}}\text{ for all }n\geq N_{11}.
\end{align}
Similar arguments lead us to the following bounds:
\begin{align*}
{\rm P}\big [|{T}_{12k}-{\rm E}[{T}_{12k}]|>n^{-\alpha}/6 \big]\leq 2e^{- \frac{\min{\{n_1,n_2\}} n^{-2\alpha}}{72}} \text{ and } {\rm P}\big [|{T}_{22k}-{\rm E}[{T}_{22k}]|>n^{-\alpha}/3 \big]\leq 2e^{- \frac{n_2 n^{-2\alpha}}{18}}.
\end{align*}

\noindent
Again, using $\lim_{n\to\infty}n_1/n=\pi_1$ with $0<\pi_1<1$, there exist constants $b_{12}>0, b_{22}>0$ and positive integers $ N_{12},N_{22}\in\mathbb{N}$ such that
\begin{align}\label{ref12}
&{\rm P}\big [|{T}_{12k}-{\rm E}[{T}_{12k}]|>n^{-\alpha}/6 \big]\leq 2e^{- b_{12} n^{1-2\alpha}} \text{ for all }n\geq N_{12} \text{ and }\nonumber \\
&{\rm P}\big [|{T}_{22k}-{\rm E}[{T}_{22k}]|>n^{-\alpha}/3 \big]\leq 2e^{- b_{22} n^{1-2\alpha}} \text{ for all }n\geq N_{22}.
\end{align}
Define $N_0 = \max\{N_{11},N_{12},N_{22}\}$. Combining \eqref{0X}, \eqref{1X} and \eqref{ref12}, we obtain
\begin{align*}
&\ P[|\widehat{\mathcal{E}}_{k,n} - \mathcal{E}_k|> n^{-\alpha}]\leq 2e^{- b_{11} n^{1-2\alpha}} + 2e^{- b_{12} n^{1-2\alpha}} + 2e^{- b_{22} n^{1-2\alpha}}\leq 6e^{- b n^{1-2\alpha}}\text{ for all } n\geq N,
\end{align*}
where $b = \min\{b_{11},b_{12},b_{22}\}$. Observe that the above inequality holds for any $0 <\alpha<1/2$ and $1\le k\le d_n.$ Hence, the proof.\hfill \QEDB
\begin{remark}
 Lemma \ref{expbound} indeed holds for all $\alpha>0$. However, for $\alpha\ge 1/2$, the sequence $e^{-bn^{1-2\alpha}}$ diverges as $n\to\infty$ and the result states a trivial upper bound.
\end{remark}

\noindent
In ultra high-dimensional settings, the dimension $d_n$ is allowed to grow with the sample size $n$. In particular, we assume that there exists $0\leq \beta <1$ such that $\log\ d_n = O(n^\beta)$. The following results are obtained under this assumption. The conventional setting (i.e., the dimension is fixed and the sample size increases) is a special case of this setting when~$\beta=0$.

\noindent
Since $\mathcal{E}_k>0$ for all $k\in S_n$ and $\mathcal{E}_k=0$ for all $k\in S^c_n$, Lemma \ref{expbound} suggests that $\widehat{\mathcal{E}}_{k}$ for $k\in S_n$ converges in probability to a positive quantity, whereas $\widehat{\mathcal{E}}_{k}$ for $k\in S^c_n$ converges to 0 as $n\to\infty$.
Next, we derive this bound uniformly over all indices.

\begin{lemma}\label{maxen_expbound}
There exists a constant $b_1>0$ such that
$$P \left [\max\limits_{1\leq k\leq d_n}|\widehat{\mathcal{E}}_{k} - \mathcal{E}_k|> n^{-\alpha} \right ] \leq O \Big ( e^{-b_1 \{n^{1-2\alpha}-n^\beta\}} \Big )\text{ for all }0 <\alpha<(1-\beta)/2.$$
\end{lemma}

\noindent {\bf Proof :}
Fix $0 <\alpha<(1-\beta)/2$. It is easy to see that following Lemma \ref{expbound}, we have
\begin{align*}
P\left[\max\limits_{1\leq k\leq d_n}|\widehat{\mathcal{E}}_{k} - \mathcal{E}_k|>n^{-\alpha}\right]
\leq \sum\limits_{k=1}^{ d_n}P\left [|\widehat{\mathcal{E}}_{k} - \mathcal{E}_k|>n^{-\alpha}\right ] \leq O \Big (d_n e^{-b n^{1-2\alpha}} \Big ).
\end{align*}
Since $\log\ d_n = O(n^\beta)$ for $0\leq \beta <1,$ there exist $M>0$ and $N\in\mathbb{N}$ such that
\begin{align*}
d_n & \leq e^{Mn^\beta}\text{ for all }n\geq N\\
\text{i.e., } d_n e^{-b n^{1-2\alpha}} & \leq e^{-(b n^{1-2\alpha} - M n^\beta)}\text{ for all }n\geq N.
\end{align*}
Observe that there exist constants $0<b_1 <b$ and $N^\prime\in\mathbb{N}$ such that
\begin{align*}
&\ e^{-\{b n^{1-2\alpha} - M n^\beta\}}\leq e^{-b_1\{ n^{1-2\alpha} - n^\beta\}}\text{ for all }n\geq N^\prime\\
\text{i.e.,}&\ d_n e^{-b n^{1-2\alpha}}\leq e^{-b_1\{ n^{1-2\alpha} - n^\beta\}}\text{ for all }n\geq N_1 = \max\{N,N^\prime\}.
\end{align*}
Hence, the proof.
\hspace*{\fill} \QEDB\newline

\vspace*{-0.1in}
\noindent 
The upper bound in the above result is free of $1\le k\le d_n$. We now prove that a similar inequality (with the same uniform bound) continues to hold for the ordered energy~distances.

\begin{corollary}\label{order_e_conv}
There exists a constant $b_1>0$ such that
$$P\left [\max\limits_{1\leq k\leq d_n}|\widehat{\mathcal{E}}_{(k)} - \mathcal{E}_{(k)}|> n^{-\alpha}\right ]\leq O \Big ( e^{-b_1 \{n^{1-2\alpha}-n^\beta\}} \Big )\text{ for all }0 <\alpha<(1-\beta)/2.$$
\end{corollary}

\noindent {\bf Proof :}
Let $(u_1,\ldots, u_{d_n})^\top$ and $(v_1,\ldots, v_{d_n})^\top$ denote two vectors in $\mathbb{R}^{d_n}$. Then, for any $1\leq k\leq d_n,$ we have $|u_{(k)}-v_{(k)}|\leq |u_i - v_j|$ for $1\leq i,j\leq d_n,$ where $i$ and $j$ are such that $u_i\geq u_{(k)}$ and $v_j\leq v_{(k)}.$ There are $(d_n-k+1)$ and $k$  such choices for $i$ and $j$, respectively. It follows from the {\it pigeon-hole principle} that for each $1\leq k \leq d_n$ there exists at least one $l$ satisfying $1\leq l \leq d_n$ such that $|u_{(k)}-v_{(k)}|\leq |u_l - v_l|$ (see \cite{wainwright2019high}). Therefore,
\begin{align}\label{php}
&\ |u_{(k)}-v_{(k)}|\leq \max\limits_{1\leq l\leq d_n}|u_l - v_l| \text{ for all }1\le k\le d_n\nonumber \\
\Rightarrow &\ \max\limits_{1\leq k\leq d_n}|u_{(k)}-v_{(k)}|\leq \max\limits_{1\leq l\leq d_n}|u_l - v_l|.
\end{align}
Using this result for the vectors $(\widehat{\mathcal{E}}_{1},\ldots, \widehat{\mathcal{E}}_{d_n})^\top$ and $(\mathcal{E}_{1},\ldots, \mathcal{E}_{d_n})^\top ,$ we obtain the following:
\begin{align*}
&\ \max\limits_{1\leq k\leq d_n}|\widehat{\mathcal{E}}_{(k)} - {\mathcal{E}}_{(k)}|> n^{-\alpha}\Rightarrow \max\limits_{1\leq k\leq d_n}|\widehat{\mathcal{E}}_{k} - {\mathcal{E}}_{k}|> n^{-\alpha}\\
\Rightarrow &\ P\left [\max\limits_{1\leq k\leq d_n}|\widehat{\mathcal{E}}_{(k)} - {\mathcal{E}}_{(k)}|> n^{-\alpha}\right ]
\leq P\left [ \max\limits_{1\leq k\leq d_n}|\widehat{\mathcal{E}}_{k} - {\mathcal{E}}_{k}|> n^{-\alpha}\right ]\text{ for all }\alpha>0.
\end{align*}
Now, it follows from Corollary \ref{maxen_expbound} that there exists a constant $b_1>0$ such that $$P\left [\max\limits_{1\leq k\leq d_n}|\widehat{\mathcal{E}}_{(k)} - {\mathcal{E}}_{(k)}|> n^{-\alpha}\right ]\leq O \Big (    e^{-b_1\{ n^{1-2\alpha}-n^\beta\}} \Big )\text{ for all }0 <\alpha<(1-\beta)/2.$$ Hence, the proof.
\QEDB\newline

Recall that $\widehat{R}=\widehat{\mathcal{E}}_{(k+1)}/\widehat{\mathcal{E}}_{(k)}$ for $1 \leq k \leq (d_n-1)$. The next result shows that $\widehat{R}_{t_n}$ takes values smaller than $\max\limits_{1\leq j\leq (s_n -1)}\widehat{R}_{t_n+j}$ with a high probability.

\begin{lemma}\label{teststatresult_s}
If assumption A1 is satisfied for $0 <\alpha_1<(1-\beta)/2,$ then there exists a constant $b_1>0$ such that
\begin{align*}
P\bigg[\widehat{R}_{t_n}\leq \ \max\limits_{1\leq j\leq (s_n -1)}\widehat{R}_{t_n+j} \bigg]\leq O\left (    e^{-b_1\{ n^{1-2\alpha_1}-n^\beta\}}\right )\text{ for all }0 <\alpha_1\le \alpha<(1-\beta)/2.
\end{align*}
\end{lemma}

\noindent 
{\bf Proof :} Consider $0 < \alpha_1 < (1-\beta)/2$ that satisfies the assumption A1. Recall that $\mathcal{E}_{(k)}=0$ for $1 \leq k \leq t_n$. Therefore, for $\alpha>0$, we have the following:
\begin{align}\label{X0}
&\ P \Big [\max_{1\leq k\leq d_n}|\widehat{\mathcal{E}}_{(k)} - {\mathcal{E}}_{(k)}|\leq n^{-\alpha} \Big]\nonumber\\
=&\ P\big[\widehat{\mathcal{E}}_{(t_n)}\leq n^{-\alpha},\ \mathcal{E}_{(t_n + j)} - n^{-\alpha}\leq \widehat{\mathcal{E}}_{(t_n + j )} \leq \mathcal{E}_{(t_n + j)} + n^{-\alpha}\text{ for all }1\leq j\leq (s_n -1) \big]\nonumber\\
\leq &\ P\bigg[\frac{\widehat{\mathcal{E}}_{(t_n + 1)}}{\widehat{\mathcal{E}}_{(t_n)}}\geq \frac{\mathcal{E}_{(t_n + 1)} - n^{-\alpha}}{n^{-\alpha}},\ \frac{\widehat{\mathcal{E}}_{(t_n + j + 1)}}{\widehat{\mathcal{E}}_{(t_n+j)}} \leq \frac{\mathcal{E}_{(t_n + j + 1)} + n^{-\alpha}}{\mathcal{E}_{(t_n + j)} - n^{-\alpha}}\text{ for all }1\leq j\leq (s_n -1) \bigg]\nonumber\\
= &\ P\bigg[\widehat{R}_{t_n}\geq \frac{\mathcal{E}_{(t_n + 1)} - n^{-\alpha}}{n^{-\alpha}},\ \widehat{R}_{t_n+j} \leq \frac{\mathcal{E}_{(t_n + j + 1)} + n^{-\alpha}}{\mathcal{E}_{(t_n + j)} - n^{-\alpha}}\text{ for all }1\leq j\leq (s_n -1) \bigg]\nonumber\\
\leq &\ P\bigg[\widehat{R}_{t_n}\geq \frac{\mathcal{E}_{(t_n + 1)} - n^{-\alpha}}{n^{-\alpha}},\ \max\limits_{1\leq j\leq (s_n -1)}\widehat{R}_{t_n+j} \leq \max\limits_{1\leq j\leq (s_n -1)}\frac{\mathcal{E}_{(t_n + j + 1)} + n^{-\alpha}}{\mathcal{E}_{(t_n + j)} - n^{-\alpha1}} \bigg].
\end{align}
Let us consider $0<\epsilon <1$. Since A1.2 is satisfied for $0<\alpha_1<(1-\beta)/2$, we have $N_2\in\mathbb{N}$ such that
\begin{align}\label{X1}
\max\limits_{1\leq j\leq (s_n -1)}{R}_{t_n+j}\frac{1}{n^{\alpha}\mathcal{E}_{(t_n + 1)}} < \epsilon < 1 \text{ for all } \alpha_1\le \alpha<(1-\beta)/2\text{ and }n\geq N_2.
\end{align}
Again, since A1.1 is satisfied, we have $N_3\in\mathbb{N}$ such that
\begin{align}\label{X2}
0<\epsilon <1 - \frac{2}{n^{\alpha}\mathcal{E}_{(t_n + 1)}}<1\text{ for all }\alpha_1\le \alpha<(1-\beta)/2\text{ and }n\geq N_3.
\end{align}
Combining \eqref{X1} and \eqref{X2}, for all $\alpha_1\le \alpha<(1-\beta)/2$ and $n\geq \max\{N_2,N_3\}$  we obtain the following:
\begin{align}\label{nref0}
&\max\limits_{1\leq j\leq (s_n -1)}{R}_{t_n+j}\frac{1}{n^{\alpha}\mathcal{E}_{(t_n + 1)}}< 1 - \frac{2}{n^{\alpha}\mathcal{E}_{(t_n + 1)}} \nonumber \\
\Rightarrow &\max\limits_{1\leq j\leq (s_n -1)}\frac{n^{\alpha}\mathcal{E}_{(t_n + j + 1)}}{n^{\alpha}\mathcal{E}_{(t_n + j)}}< {n^{\alpha}\mathcal{E}_{(t_n + 1)}} - 2 \nonumber \\
\Rightarrow  &\ \frac{n^{\alpha}\mathcal{E}_{(t_n + j + 1)}}{n^{\alpha}\mathcal{E}_{(t_n + j)}}< {n^{\alpha}\mathcal{E}_{(t_n + 1)}} - 2 \text{ for all } 1\leq j\leq (s_n -1) \nonumber \\
\Rightarrow  &\ n^{\alpha}\mathcal{E}_{(t_n + j + 1)}< n^{\alpha}\mathcal{E}_{(t_n + j)}\ {n^{\alpha}\mathcal{E}_{(t_n + 1)}} - 2n^{\alpha}\mathcal{E}_{(t_n + j)}  \text{ for all } 1\leq j\leq (s_n -1) \nonumber \\
\Rightarrow &\ n^{\alpha}\mathcal{E}_{(t_n + j + 1)} < n^{\alpha}\mathcal{E}_{(t_n + j)}\ {n^{\alpha}\mathcal{E}_{(t_n + 1)}} - n^{\alpha}\mathcal{E}_{(t_n + j)} - n^{\alpha}\mathcal{E}_{(t_n + 1)}  \text{ for all } 1\leq j\leq (s_n -1) \nonumber \\
\Rightarrow  &\ n^{\alpha}\mathcal{E}_{(t_n + j + 1)} + 1< (n^{\alpha}\mathcal{E}_{(t_n + j)} -1) (n^{\alpha}\mathcal{E}_{(t_n + 1)} - 1)  \text{ for all } 1\leq j\leq (s_n -1) \nonumber \\
\Rightarrow  &\ \frac{\mathcal{E}_{(t_n + j + 1)} + n^{-\alpha}}{\mathcal{E}_{(t_n + j)} - n^{-\alpha}}<  \frac{\mathcal{E}_{(t_n + 1)} - n^{-\alpha}}{n^{-\alpha}}  \text{ for all } 1\leq j\leq (s_n -1) \nonumber \\
\Rightarrow  &\ \max\limits_{1\leq j\leq (s_n -1)}\frac{\mathcal{E}_{(t_n + j + 1)} + n^{-\alpha}}{\mathcal{E}_{(t_n + j)} - n^{-\alpha}}<  \frac{\mathcal{E}_{(t_n + 1)} - n^{-\alpha}}{n^{-\alpha}}.
\end{align}
Therefore, it follows from \eqref{X0} and \eqref{nref0} that
\begin{align*}
&\ P \Big [\max_{1\leq k\leq d_n}|\widehat{\mathcal{E}}_{(k)} - {\mathcal{E}}_{(k)}|\leq n^{-\alpha} \Big ]\\
\leq &\ P\bigg[\widehat{R}_{t_n}\geq \frac{\mathcal{E}_{(t_n + 1)} - n^{-\alpha}}{n^{-\alpha}},\ \max\limits_{1\leq j\leq (s_n -1)}\widehat{R}_{t_n+j} < \frac{]\mathcal{E}_{(t_n + 1)} - n^{-\alpha}}{n^{-\alpha}} \bigg]\\
= &\ P\bigg[\widehat{R}_{t_n} \geq \frac{\mathcal{E}_{(t_n + 1)} - n^{-\alpha}}{n^{-\alpha}} > \max\limits_{1\leq j\leq (s_n -1)}\widehat{R}_{t_n+j}\bigg]\\
\leq & \ P\bigg[\widehat{R}_{t_n} > \max\limits_{1\leq j\leq (s_n -1)}\widehat{R}_{t_n+j}\bigg]\text{ for all }\alpha_1\le \alpha<(1-\beta)/2\text{ and }n\geq \max\{N_2,N_3\}.
\end{align*}

\noindent Corollary \ref{order_e_conv} implies that
\begin{align*}
&1 - O\left (e^{-b_1\{ n^{1-2\alpha} - n^\beta\}}\right )
\leq  P\left[\max_{1\leq k\leq d_n}|\widehat{\mathcal{E}}_{(k)} - {\mathcal{E}}_{(k)}|\leq n^{-\alpha}\right]
\leq P\bigg[\widehat{R}_{t_n}> \ \max\limits_{1\leq j\leq (s_n -1)}\widehat{R}_{t_n+j} \bigg]\\
\Rightarrow &\ P\bigg[\widehat{R}_{t_n}\leq \ \max\limits_{1\leq j\leq (s_n -1)}\widehat{R}_{t_n+j} \bigg]\leq O \left (e^{-b_1\{ n^{1-2\alpha} - n^\beta\}}\right )\text{ for all }\alpha_1<\alpha<(1-\beta)/2.
\end{align*}

\noindent Hence, the proof.
\QEDB
\newline

\vspace*{-0.2in}
\noindent
We now prove the {\it sure screening property } (SSP) of the proposed screening method MarS.
\newline

\vspace*{-0.1in}
\noindent
{\bf Proof of Theorem \ref{mainthm} :}
If possible, assume that $S_n$ is not a subset of $\widehat{S}_n$. Now, $S_n\not\subseteq \widehat{S}_n$ means that the set $\{\widehat{\mathcal{E}}_{l}\leq \widehat{\mathcal{E}}_{(\widehat{t}_n)}\text{ for some }l\in S_n\}$ is non-empty, where $\widehat{t}_n (= d_n - \widehat{s}_n)$ is the estimated number of noise components (see \eqref{prop_est_sn}). Suppose that $S_n=\{k_1,\ldots, k_{s_n}\}$, where $1 \leq k_i \leq d_n$ and $k_i\neq k_j$ for $1\leq i\neq j\leq d_n.$ Consider an $\alpha_1$ that satisfies assumption A1. Then,
\begin{align*}
&\ P\left [S_n\not\subseteq \widehat{S}_n\right ] \\
=&\ P\left[\widehat{\mathcal{E}}_{l}\leq \widehat{\mathcal{E}}_{(\widehat{t}_n)}\text{ for some }l\in S_n\right ] \\
\leq &\ P\left[\widehat{\mathcal{E}}_{l}\leq \widehat{\mathcal{E}}_{(\widehat{t}_n)}\text{ for some }l\in S_n,\max\limits_{1\leq k\leq d_n}|\widehat{\mathcal{E}}_{k} - {\mathcal{E}}_{k}|\leq n^{-\alpha} \right]+P\left [\max\limits_{1\leq k\leq d_n}|\widehat{\mathcal{E}}_{k} - {\mathcal{E}}_{k}|> n^{-\alpha} \right ]\\
\leq &\ \sum\limits_{i=1}^{s_n}P\left [\widehat{\mathcal{E}}_{k_i}\leq \widehat{\mathcal{E}}_{(\widehat{t}_n)},\max\limits_{1\leq k\leq d_n}|\widehat{\mathcal{E}}_{k} - {\mathcal{E}}_{k}|\leq n^{-\alpha} \right ]+P\left [\max\limits_{1\leq k\leq d_n}|\widehat{\mathcal{E}}_{k} - {\mathcal{E}}_{k}|> n^{-\alpha} \right ]\ \text{ for all }\alpha>0.
\end{align*}

\noindent Fix $\alpha_1\le \alpha<(1-\beta)/2.$ Using Corollary \ref{order_e_conv}, we have $P\left [\max\limits_{1\leq k\leq d_n}|\widehat{\mathcal{E}}_{k} - {\mathcal{E}}_{k}|> n^{-\alpha} \right ]=O\left (e^{-b_1 \{n^{1-2\alpha}-n^\beta\}}\right )$. Consequently,
\begin{align}\label{e2}
P\left [S_n\not\subseteq \widehat{S}_n\right ]\le \sum\limits_{i=1}^{s_n}P\left [\widehat{\mathcal{E}}_{k_i}\leq \widehat{\mathcal{E}}_{(\widehat{t}_n)},\max\limits_{1\leq k\leq d_n}|\widehat{\mathcal{E}}_{k} - {\mathcal{E}}_{k}|\leq n^{-\alpha} \right ]+O\left (e^{-b_1 \{n^{1-2\alpha}-n^\beta\}}\right ).
\end{align}
Now, in the proof of Corollary \ref{order_e_conv} we have shown that
\begin{align*}
\max\limits_{1\leq k\leq d_n}|\widehat{\mathcal{E}}_{k} - {\mathcal{E}}_{k}|\leq n^{-\alpha}\Rightarrow\max\limits_{1\leq k\leq d_n}|\widehat{\mathcal{E}}_{(k)} - {\mathcal{E}}_{(k)}|\leq n^{-\alpha}\ (\text{follows from }\eqref{php}).
\end{align*}
Further, in the proof of Lemma \ref{teststatresult_s} we have shown that if A1 is satisfied, then there exists $N\in\mathbb{N}$ such that
\begin{align}\label{nref17}
&\max_{1\leq k\leq d_n}|\widehat{\mathcal{E}}_{(k)} - {\mathcal{E}}_{(k)}|\leq n^{-\alpha}\Rightarrow \widehat{R}_{t_n} > \max\limits_{1\leq j\leq (s_n -1)}\widehat{R}_{t_n+j} \nonumber \\
\text{i.e., }&\max_{1\leq k\leq d_n}|\widehat{\mathcal{E}}_{k} - {\mathcal{E}}_{k}|\leq n^{-\alpha}\Rightarrow \argmax\limits_{1\leq j\leq (d_n -1)}\widehat{R}_{j}\leq t_n
\end{align}
for all $n\geq N$.
Observe that $\argmax_{1\leq j\leq (d_n -1)}\frac{\widehat{\mathcal{E}}_{(j+1)}}{\widehat{\mathcal{E}}_{(j)}}= \widehat{t}_n$ (follows from the definition of $\widehat{s}_n$ in \eqref{prop_est_sn}). Hence, it follows from \eqref{nref17} that $\max_{1\leq k\leq d_n}|\widehat{\mathcal{E}}_{k} - {\mathcal{E}}_{k}|\leq n^{-\alpha}\Rightarrow \widehat{t}_n = d_n - \widehat{s}_n \leq d_n - s_n = t_n$ for all $n\geq N$. Using \eqref{e2}, we now have
\begin{align}\label{sspproof}
&\ P[S_n\not\subseteq \widehat{S}_n]\nonumber \\
\leq &\sum\limits_{i=1}^{s_n} P \Big [\widehat{\mathcal{E}}_{k_i}\leq \widehat{\mathcal{E}}_{(d_n-\widehat{s}_n)},\max\limits_{1\leq k\leq d_n}|\widehat{\mathcal{E}}_{k} - {\mathcal{E}}_{k}|\leq n^{-\alpha}, d_n-\widehat{s}_n\leq d_n-s_n \Big ] + O \Big ( e^{-b_1 \{n^{1-2\alpha}-n^\beta\}} \Big )\nonumber \\
\leq &\sum\limits_{i=1}^{s_n} P \Big [\widehat{\mathcal{E}}_{k_i}\leq \widehat{\mathcal{E}}_{(d_n-s_n)},\max\limits_{1\leq k\leq d_n}|\widehat{\mathcal{E}}_{k} - {\mathcal{E}}_{k}|\leq n^{-\alpha} \Big ] + O \Big (e^{-b_1 \{n^{1-2\alpha}-n^\beta\}} \Big )\ \text{ for all }n\geq N.
\end{align}
Further, observe that
\begin{align*}
&\ P \Big [\widehat{\mathcal{E}}_{k_i}\leq \widehat{\mathcal{E}}_{(t_n)},\max\limits_{1\leq k\leq d_n}|\widehat{\mathcal{E}}_{k} - {\mathcal{E}}_{k}|\leq n^{-\alpha} \Big ]\nonumber \\
=&\ P \Big [\widehat{\mathcal{E}}_{k_i}\leq \widehat{\mathcal{E}}_{(t_n)},\max\limits_{1\leq k\leq d_n}|\widehat{\mathcal{E}}_{k} - {\mathcal{E}}_{k}|\leq n^{-\alpha},\max\limits_{1\leq k\leq d_n}|\widehat{\mathcal{E}}_{(k)} - {\mathcal{E}}_{(k)}|\leq n^{-\alpha} \Big ]\ (\text{using }\eqref{php})\nonumber \\
\leq&\ P\big [{ \mathcal{E}}_{k_i} - n^{-\alpha}\leq \widehat{\mathcal{E}}_{k_i}\leq \widehat{\mathcal{E}}_{(t_n)}\leq n^{-\alpha}\big ]\nonumber \\
\leq &\ \mathbb{I}[\mathcal{E}_{k_i}-n^{-\alpha}\leq n^{-\alpha}]= \mathbb{I}[n^{\alpha}\mathcal{E}_{k_i}\leq 2]\text{ for all } 1 \leq i \leq s_n.
\end{align*}
Since $\alpha$ satisfies assumption A1, there exists $N^\prime\in\mathbb{N}$ such that $\mathbb{I}[n^{\alpha}\mathcal{E}_{(t_n+1)}\leq 2]=0$ for all $n\geq N^\prime$. Consequently, $\mathbb{I}[n^{\alpha}\mathcal{E}_{k_i}\leq 2]=0$ for all $1 \leq i \leq s_n$ and $n\geq N^\prime$. As a result, we have $P \Big [\widehat{\mathcal{E}}_{l}\leq \widehat{\mathcal{E}}_{(t_n)},\max\limits_{1\leq k\leq d_n}|\widehat{\mathcal{E}}_{k} - {\mathcal{E}}_{k}|\leq n^{-\alpha} \Big ]=0$ for all $n\geq N^\prime$ and $l\in S_n$. Therefore, $\sum\limits_{l=1}^{s_n} P \Big [\widehat{\mathcal{E}}_{l}\leq \widehat{\mathcal{E}}_{(t_n)},\max\limits_{1\leq k\leq d_n}|\widehat{\mathcal{E}}_{k} - {\mathcal{E}}_{k}|\leq n^{-\alpha} \Big ]=0$ for all $n\geq N^\prime$. Now, it follows from \eqref{sspproof} that
$$P[S_n\not\subseteq \widehat{S}_n] \leq O \Big (e^{-b_1 \{n^{1-2\alpha}-n^\beta\}} \Big ).$$
This completes the proof. \hfill\QEDB\newline

\noindent
We have proved that if $\gamma$ is bounded, then the probability that the proposed screened set contains the true signal set converges to one at an exponential rate (with respect to the sample size $n$). Now, we show that exponential rate of convergence can be obtained under additional conditions like sub-Gaussianity (say, SG) of the random variables $\{\gamma(|U_k-V_k|^2):k\ge 1\}$ (see \cite{wainwright2019high, BLM_2013}).

\begin{lemma}\label{U_expbound_suff_Gaussian}
	Let $X_i \sim N(\mu_i, \sigma_i^2)$ for $i=1,2$ and $f$ is an $L$-Lipschitz continuous function. Define $\sigma^2=\sigma_1^2+\sigma_2^2.$ Then, $f(X_i-X_j)- E[f(X_i-X_j)] \in SG(L^2\sigma^2)$  for $1\le i\neq j \le 2$.
\end{lemma}

\noindent
{\bf Proof :}
Clearly, $X_i-X_j \sim N(\mu_i-\mu_j, \sigma_i^2+\sigma_j^2)$. Using Theorem 5.6 of \cite{BLM_2013}, we have $f(X_i-X_j-\mu_i+\mu_j) \in SG(L^2\sigma^2) \mbox{ with } \sigma^2=\sigma_i^2+\sigma_j^2$. We now argue that $f(X_i-X_j) \in SG(L^2\sigma^2)$. Define $Y=X_i-X_j-(\mu_i-\mu_j) \sim N(0,\sigma^2)$ and $g(x)=f(x+\mu_i-\mu_j)$.~So,
$$|g(x)-g(y)| = |f(x+\mu_i-\mu_j)-f(y+\mu_i-\mu_j)| \leq L |(x+\mu_i-\mu_j)-(y+\mu_i-\mu_j)| = L|x-y|,$$
i.e., $g$ is also $L$-Lipschitz continuous.
Using Theorem 5.6 of \cite{BLM_2013}, we have $g(Y) \in SG(L^2\sigma^2)$.

Fix $t>0$, and consider the following
\begin{align*}
&\ P[|f(X_i-X_j) - E(f(X_i-X_j))| \geq t] \\
=&\ P[|f(X_i-X_j-(\mu_i-\mu_j)+\mu_i-\mu_j) - E(f(X_i-X_j-(\mu_i-\mu_j)+\mu_i-\mu_j))| \geq t]\\
=&\  P[|f(Y+\mu_i-\mu_j) - E(f(Y+\mu_i-\mu_j))| \geq t]\\
= &\ P[|g(Y) - E(g(Y))| \geq t]\leq 2e^{-t^2/2\sigma^2L^2}.
\end{align*}
Hence, the proof.
\hspace*{\fill} \QEDB 
\newline

\vspace*{-0.2in}
\begin{lemma}\label{U_expbound_lip}
	Fix $k\in\{1,\ldots, d_n\}$.
	Let $X_{ik} \stackrel{iid}{\sim} N(\mu_{1k}, \sigma_{1k}^2)$ for all $1\leq i\leq n_1$ and $Y_{jk} \stackrel{iid}{\sim} N(\mu_{2k}, \sigma_{2k}^2)$ for all $1\leq j\leq n_2$. Define $\sigma^2_0=\max\{\sigma_{11}^2,\ldots, \sigma_{1d_n}^2,\sigma_{21}^2\ldots, \sigma_{2d_n}^2\}<\infty$. If $h(x,y) = f(x-y)$ with $f$ being an $L$-Lipschitz continuous function, then for any $t>0$, there exist positive constants $B_1$ and $B_2$ such that
	\begin{enumerate}[(a)]
		\item $P\left[\left|T_{11k}-E[T_{11k}]\right| \geq t\right] \leq 2 e^{-\frac{n_1 t^{2}}{B_1}}$ and $P\left[\left|T_{22k}-E[T_{22k}]\right| \geq t\right] \leq 2 e^{-\frac{n_2 t^{2}}{B_1}}$,
		\item $P\left[\left|T_{12k}-E[T_{12k}]\right| \geq t\right] \leq 2 e^{-\frac{\min\{n_1,n_2\}t^{2}}{B_2}}$.
	\end{enumerate}
\end{lemma}

\noindent {\bf Proof :}
\begin{enumerate}[(a)]
	\item Fix $k\in\{1, \ldots, d_n\}$. Let $l$ be the smallest integer larger than $\frac{n_1}{2}$, i.e., $l=\left[\frac{n_1}{2}\right]+1$. Define	$$W_k\left(\bx_{1}, \ldots, \bx_{n_1}\right)=\frac{h\left(x_{1k}, x_{2k}\right)+h\left(x_{3k}, x_{4k}\right)+\cdots+h\left(x_{2l-1\ k},x_{2 l\ k}\right)}{l},$$
	i.e., we break our sample into $l$ non-overlapping blocks of size $2$. Let $\mathcal{P}^{n_1}$ be the collection of all possible permutations of $\{1, \ldots, n_1\}$ and ${\bf p}=(p_1,\ldots,p_{n_1})^\top\in \mathcal{P}^{n_1}$. Then,
	$$l \sum_{{\bf p} \in \mathcal{P}^{n_1}} W_k\left(\bX_{p_{1}}, \ldots, \bX_{p_{n_1}}\right) = l\ 2! (n_1-2)! \sum_{i<j} h\left(X_{ik}, X_{jk}\right),$$
	We now have
	$$
	\sum_{{\bf p} \in \mathcal{P}^{n_1}} W_k\left(\bX_{p_{1}}, \ldots, \bX_{p_{n_1}}\right)=\underbrace{2 !(n_1-2) !{n_1\choose 2}}_{n_1 !} T_{11k}
	$$
	which gives
	$$T_{11k}=\frac{1}{n_1!} \sum_{{\bf p} \in \mathcal{P}^{n_1}} W_k\left(\bX_{p_{1}},\ldots, \bX_{p_{n_1}}\right).$$

	Define $\theta_{1k} = E[h(X_{1k},X_{2k})]$. Fix ${\bf p}\in \mathcal{P}^{n_1}$. Clearly, $\ W_k\left(\bX_{p_{1}}, \ldots, \bX_{p_{n_1}}\right)$ is an average of $l$ iid random variables $h(X_{ik},X_{jk})$ with $i,j\in {\bf p}$ and $i\neq j.$  Therefore, $E\left [W_k\left(\bX_{p_{1}}, \ldots, \bX_{p_{n_1}}\right)\right ]=\theta_{1k}$ and consequently, we have $E[T_{11k}]=\theta_{1k}$.

	\noindent
	Here, $h(x,y)=f(x-y)$ with $f$ an $L$-Lipschitz function. It now follows from Lemma \ref{U_expbound_suff_Gaussian} that $h(X_{ik},X_{jk})-\theta_{1k}\in SG(2L^2\sigma_{1k}^2)$ for all $i,j\in {\bf p}$ with $i\neq j$. Hence,
	\begin{equation} \label{Sub_G}
	W^*_{{\bf p}k}=W_k\left(\bX_{p_1}, \ldots, \bX_{p_{n_1}}\right)-\theta_{1k} \in SG\left(\frac{2L^{2}\sigma^2_{1k}}{l}\right).
	\end{equation}
	Now, if we sum over all possible values of ${\bf p}$, then we obtain:
	$$T_{11k}-\theta_{1k} = \frac{1}{n_1!} \sum_{{\bf p} \in \mathcal{P}^{n_1}} W^*_{{\bf p}k}.$$
	For $t>0$ and $\lambda>0$, we have
	$$
	\begin{aligned}
	P\left[T_{11k}-\theta_{1k} \geq t\right] & \leq e^{-\lambda t} \ \mathbb{E}\left[e^{\lambda\left(T_{11k}-\theta_{1k}\right)}\right] [\text{using Markov's inequality}] \\
	& = e^{-\lambda t}\ \mathbb{E}\left[ e^{\lambda \frac{1}{n_1!} \sum\limits_{{\bf p} \in \mathcal{P}^{n_1}} W^*_{pk}} \right]\\
	& \leq e^{-\lambda t} \frac{1}{n_1!} \sum_{{\bf p} \in \mathcal{P}^{n_1}} \mathbb{E}\left[ e^{\lambda  W^*_{pk}} \right]  \text { [using Jensen's inequality] } \\
	& \leq e^{-\lambda t} e^{\frac{\lambda^2 L^2\sigma^2_{1k}}{l}}\ \left [\text{since } W^*_{pk} \in S G\left(\frac{2L^{2}\sigma^2_{1k}}{l}\right)\right ].
	\end{aligned}
	$$
	Minimizing the upper bound w.r.t. $\lambda$, we obtain
	$$P\left[T_{11k}-\theta_{1k} \geq t\right] \leq e^{-\frac{t^2}{L^2\sigma^2_{1k}}l} \leq e^{-\frac{t^2}{L^2\sigma^2_{1k}} \frac {n_1}{ 2}}\leq e^{-\frac{n_1t^2}{B_1}},$$
	where $B_1 = 2L^2\sigma^2_0$.

	Repeating for the other side yields the following
	$$P\left[|T_{11k}-\theta_{1k}| \geq t\right] \leq 2e^{-\frac{n_1t^2}{B_1}}.$$

	Define $\theta_{2k} = E[h(Y_{1k},Y_{2k})]$ for $1\leq k\leq d_n.$ Following similar arguments, one can prove that $$P\left[|T_{22k}-\theta_{2k}| \geq t \right]\leq 2e^{-\frac{n_2t^2}{B_1}}\text{ for all }1\leq k\leq d_n.$$

	\item Fix $k\in\{1,\ldots, d_n\}$. Without loss of generality, let us assume that $n_1\leq n_2$ and define
	\begin{align*}
	\eta_k(i_1,\ldots,i_{n_1}) = \frac{1}{n_1}\sum\limits_{j=1}^{n_1}h(X_{jk},Y_{i_j k})
	\end{align*}
	for some permutation of $(i_1,\ldots,i_{n_1})$ of $n_1$ elements chosen from $\{1,\ldots,n_2\}.$ Note that $\eta_k$ is an average of iid random variables. Using $\eta_k$, we can express $T_{12k}$ as follows:
	\begin{align*}
	T_{12k} = \frac{(n_2-n_1)!}{n_2!}\sum\limits_{(i_1,\ldots,i_{n_1})\in \mathcal{P}^n}\eta_k(i_1,\ldots,i_{n_1}),
	\end{align*}
	where $\mathcal{P}^n$ denotes the set of all possible permutations $\{i_1,\ldots,i_{n_1}\}$ of the elements of the set $\{1,\ldots, n_2\}$.

	Define $\theta_k = E[h(X_{1k},Y_{1k})]$. Therefore, $E[\eta_k(i_1,\ldots, i_{n_1})]=\theta_{k}$ for any permutation $(i_1,\ldots,i_{n_1})$ and consequently, we have $E[T_{12k}]=\theta_k$. We are interested in an upper bound of the probability $P[|T_{12k}-\theta_k|>t]$ for $t>0$. It follows from Lemma \ref{U_expbound_suff_Gaussian} that $h(X_{jk},Y_{i_j k})-\theta_k \in SG(L^2(\sigma^2_{1k}+ \sigma^2_{2k}))$. Using sub-Gaussianity of $h(X_{jk},Y_{i_j k})$, we~have
	\begin{align}\label{2U_bdref1}
	E\left [e^{s\{ h(X_{jk},Y_{i_j k})-\theta_k\}}\right ]\leq e^{\frac{s^2L^2(\sigma^2_{1k}+ \sigma^2_{2k})}{2}}\leq e^{s^2L^2\sigma^2_0}\text{ for all }s\in\mathbb{R}.
	\end{align}
	Using Jensen's inequality on the convex function $e^{sx}$, we obtain
	\begin{align}\label{2Uref1}
	&\ e^{sT_{12k}}\leq \frac{(n_2-n_1)!}{n_2!}\sum\limits_{(i_1,\ldots,i_{n_1})\in \mathcal{P}^n}e^{s \eta_k(i_1,\ldots,i_{n_1})}\text{ for every }s>0 \nonumber \\
	\Rightarrow &\ E\left [e^{sT_{12k}}\right ]\leq E\left [e^{s \eta_k(i_1,\ldots,i_{n_1})}\right ]\leq \left\{E\left [e^{sh(X_{1k},Y_{1k})/{n_1}}\right ]\right\}^{n_1} \nonumber \\
	\Rightarrow &\ E\left [e^{s[T_{12k}- \theta_k]}\right ]\leq \left\{E\left [e^{s[h(X_{1k},Y_{1k})-\theta_k]/{n_1}}\right ]\right\}^{n_1}.
	\end{align}
	For any $s>0$ and $t>0$, it now follows from Markov's inequality and \eqref{2Uref1} that
	\begin{align*}
	&\ P[T_{12k}-\theta_k>t]\leq \frac{E\left [e^{s[T_{12k}-\theta_k]}\right ]}{e^{st}}\leq e^{-st}\left\{E\left [e^{s[h(X_{1k},Y_{1k})-\theta_k]/{n_1}}\right ]\right\}^{n_1}\\
	\text{i.e., }&\ P[T_{12k}-\theta_k>t]\leq e^{-st +\frac{s^2L^2\sigma^2_0}{n_1}}\hspace{1cm}[\text{follows from \eqref{2U_bdref1}}].
	\end{align*}
	Minimizing the upper bound with respect to $s$, we obtain $P[T_{12k}-\theta_k>t]\leq e^{-{n_1t^2}/{4L^2\sigma^2_0}}$ for all $t>0.$ Following similar arguments, it can be shown that $P[T_{12k}-\theta_k<-t]\leq e^{-{n_1t^2}/{4L^2\sigma^2_0}}$ for any $t>0$. Define $B_2 = L^2\sigma^2_0.$ Therefore,
	\begin{align*}
	P[|T_{12k}-\theta_k|>t]\leq 2e^{-{n_1t^2}/{B_2}}\ \text{ for all }t>0.
	\end{align*}
\end{enumerate}
Hence, the proof.\hspace*{\fill}\QEDB

\begin{lemma}\label{mainlemma_E}
	If assumptions {\rm A1} and {\rm A2} are satisfied for $0 <\alpha_1<(1-\beta)/2$ and $0 <\alpha_2<(1-\beta)/2,$ respectively, then there exists a constant $b_1>0$ such that
	$$P[\widehat{s}_n\neq s_n]\leq O\left (e^{-b_1\{ n^{1-2\alpha}- n^\beta\}}\right )\text{ for all }\max\{\alpha_1,\alpha_2\}\le \alpha<(1-\beta)/2.$$
\end{lemma}
\noindent {\bf Proof :}
Fix $0<\epsilon <1.$ If A2 is satisfied, then we have $N_4\in\mathbb{N}$ such that
$$P\bigg [\max\limits_{1\leq j\leq (t_n -1 ) } \frac{N_{(j+1)}}{N_{(j)}}\frac{1}{n^{\alpha_2} \mathcal{E}_{(t_n + 1)}}>\epsilon \bigg ]<\epsilon\text{ for all }n\geq N_4.$$
Define $\alpha_0 = \max\{\alpha_1,\alpha_2\}.$ Therefore, for all $\alpha\geq \alpha_0,$ we have
\begin{align}\label{nref}
P\bigg[\max\limits_{1\leq j\leq (t_n -1 ) } \frac{N_{(j+1)}}{N_{(j)}}\frac{1}{n^{\alpha} \mathcal{E}_{(t_n + 1)}}>\epsilon\bigg]\leq P\bigg[\max\limits_{1\leq j\leq (t_n -1 ) } \frac{N_{(j+1)}}{N_{(j)}}\frac{1}{n^{\alpha_2} \mathcal{E}_{(t_n + 1)}}>\epsilon\bigg]<\epsilon\text{ for all }n\geq N_4.
\end{align}
Fix $\alpha$ satisfying $\alpha_0\le \alpha<(1-\beta)/2$. Observe that
\begin{align}\label{X4}
&\ P\bigg[\max_{1\leq k\leq d_n}|\widehat{\mathcal{E}}_{(k)} - {\mathcal{E}}_{(k)}|\leq n^{-\alpha}\bigg]\nonumber\\
=&\ P\bigg[\max_{1\leq k\leq d_n}|\widehat{\mathcal{E}}_{(k)} - {\mathcal{E}}_{(k)}|\leq n^{-\alpha},\max\limits_{1\leq j\leq (t_n -1 ) } \frac{N_{(j+1)}}{N_{(j)}}\frac{1}{n^{\alpha} \mathcal{E}_{(t_n + 1)}}\leq \epsilon \bigg]\nonumber \\
&\ \ + P\bigg[\max_{1\leq k\leq d_n}|\widehat{\mathcal{E}}_{(k)} - {\mathcal{E}}_{(k)}|\leq n^{-\alpha},\max\limits_{1\leq j\leq (t_n -1 ) } \frac{N_{(j+1)}}{N_{(j)}}\frac{1}{n^{\alpha} \mathcal{E}_{(t_n + 1)}}>\epsilon \bigg]\nonumber\\
\leq &\ P\bigg[\max_{1\leq k\leq d_n}|\widehat{\mathcal{E}}_{(k)} - {\mathcal{E}}_{(k)}|\leq n^{-\alpha},\max\limits_{1\leq j\leq (t_n -1 ) } \frac{N_{(j+1)}}{N_{(j)}}\frac{1}{n^{\alpha} \mathcal{E}_{(t_n + 1)}}\leq \epsilon \bigg]\nonumber \\
&\ \ + P\bigg[\max\limits_{1\leq i\leq (t_n -1 ) } \frac{N_{(i+1)}}{N_{(i)}}\frac{1}{n^{\alpha} \mathcal{E}_{(t_n + 1)}}>\epsilon \bigg]\nonumber\\
\leq &\ P\bigg[\max_{1\leq k\leq d_n}|\widehat{\mathcal{E}}_{(k)} - {\mathcal{E}}_{(k)}|\leq n^{-\alpha},\max\limits_{1\leq j\leq (t_n -1 ) } \frac{N_{(j+1)}}{N_{(j)}}\frac{1}{n^{\alpha} \mathcal{E}_{(t_n + 1)}}\leq \epsilon \bigg] + \epsilon\text{ for all }n\geq N_4.
\end{align}
The last inequality follows from \eqref{nref}. Since assumption A1.1 is satisfied, we have $N_5\in\mathbb{N}$ such that $n^{\alpha} \mathcal{E}_{(t_n +1 )} - 1 > 1$ for all $n\geq N_5.$ Consequently,
\begin{align}\label{nref2}
&\ \max_{1\leq k\leq d_n}|\widehat{\mathcal{E}}_{(k)} - {\mathcal{E}}_{(k)}|\leq n^{-\alpha}\nonumber \\
\Rightarrow &\  \widehat{\mathcal{E}}_{(t_n)}\leq n^{-\alpha}
\Rightarrow  n^{\alpha}\widehat{\mathcal{E}}_{(t_n)}\leq 1
\Rightarrow  n^{\alpha}\widehat{\mathcal{E}}_{(t_n)}< n^{\alpha}{\mathcal{E}}_{(t_n + 1)} - 1\text{ for all } n\geq N_5 \nonumber\\
\Rightarrow &\  \widehat{\mathcal{E}}_{(t_n)}\leq n^{-\alpha}< {\mathcal{E}}_{(t_n + 1)} - n^{-\alpha}\text{ for all } n\geq N_5.
\end{align}
Recall the definition of $N_{(i)}$ for $1\leq i\leq t_n$ stated in Section \ref{Screen_Mars}. Equation \eqref{nref2} implies that $N_{(i)}$ and $\widehat{\mathcal{E}}_{(i)}$ are equal for all $1 \leq i \leq t_n$. In other words, the smallest $t_n$ sample energy distances correspond to the $t_n$ noise components. Therefore, it follows from \eqref{X4} that
\begin{align}\label{X5}
&\ P\bigg[\max_{1\leq k\leq d_n}|\widehat{\mathcal{E}}_{(k)} - {\mathcal{E}}_{(k)}|\leq n^{-\alpha}\bigg]\nonumber\\
\leq &\ P\bigg[\max\limits_{1\leq j\leq (t_n - 1 ) } \frac{\widehat{\mathcal{E}}_{(j+1)}}{\widehat{\mathcal{E}}_{(j)}}\frac{1}{n^{\alpha} \mathcal{E}_{(t_n + 1)}}\leq \epsilon , \max_{1\leq k\leq d_n}|\widehat{\mathcal{E}}_{(k)} - {\mathcal{E}}_{(k)}|\leq n^{-\alpha}\bigg] + \epsilon\nonumber\\
\leq &\ P\bigg [\   \max\limits_{1\leq j\leq (t_n - 1) } \frac{\widehat{\mathcal{E}}_{(j+1)}}{\widehat{\mathcal{E}}_{(j)}}\frac{1}{n^{\alpha} \mathcal{E}_{(t_n + 1)}}\leq \epsilon ,  \ \widehat{\mathcal{E}}_{(t_n)}\leq n^{-\alpha},\nonumber\\
&\hspace{1in}  \mathcal{E}_{(t_n + j)} -n^{-\alpha}\leq \widehat{\mathcal{E}}_{(t_n + j)} \leq  \mathcal{E}_{(t_n + j)} + n^{-\alpha} \text{ for all }1\leq j\leq s_n\bigg] + \epsilon\nonumber\\
\leq &\ P\bigg [\max\limits_{1\leq j\leq (t_n - 1)  } \frac{\widehat{\mathcal{E}}_{(j+1)}}{\widehat{\mathcal{E}}_{(j)}}\frac{1}{n^{\alpha} \mathcal{E}_{(t_n + 1)}}\leq \epsilon ,\ \frac{\widehat{\mathcal{E}}_{(t_n+1)}}{\widehat{\mathcal{E}}_{(t_n)}}\geq {\frac{\mathcal{E}_{(t_n + 1)} -n^{-\alpha}  }{n^{-\alpha}}},\nonumber\\
&\hspace{1.5in} \frac{\widehat{\mathcal{E}}_{(t_n + j + 1)}}{\widehat{\mathcal{E}}_{(t_n+j)}}\leq \frac{\mathcal{E}_{(t_n + j + 1)} + n^{-\alpha} }{\mathcal{E}_{(t_n + j)} - n^{-\alpha}} \text{ for all }1\leq j\leq (s_n -1) \bigg] + \epsilon\nonumber\\
\leq &\ P\bigg [\max\limits_{1\leq j\leq (t_n - 1)  } \widehat{R}_{j}\frac{1}{n^{\alpha} \mathcal{E}_{(t_n + 1)}}\leq \epsilon ,\ \widehat{R}_{t_n}\geq {\frac{\mathcal{E}_{(t_n + 1)} -n^{-\alpha}  }{n^{-\alpha}}},\nonumber\\
&\hspace{1.5in} \widehat{R}_{t_n + j}\leq \frac{\mathcal{E}_{(t_n + j + 1)} + n^{-\alpha} }{\mathcal{E}_{(t_n + j)} - n^{-\alpha}} \text{ for all }1\leq j\leq (s_n -1) \bigg] + \epsilon\nonumber\\
\leq &\ P\bigg [\max\limits_{1\leq j\leq (t_n - 1)  } \widehat{R}_{j}\frac{1}{n^{\alpha} \mathcal{E}_{(t_n + 1)}}\leq \epsilon ,\ \widehat{R}_{t_n}\geq {\frac{\mathcal{E}_{(t_n + 1)} -n^{-\alpha}  }{n^{-\alpha}}},\nonumber\\
&\hspace{2in}\max\limits_{t_n + 1\leq j\leq (d_n -1)} \widehat{R}_{j}\leq \max\limits_{t_n + 1\leq j\leq (d_n -1)}\frac{\mathcal{E}_{( j + 1)} + n^{-\alpha} }{\mathcal{E}_{(j)} - n^{-\alpha}}\bigg ] + \epsilon
\end{align}
for all $n\geq \max \{N_4, N_5\}$.
We have already shown in the proof of Lemma \ref{teststatresult_s} (see \eqref{nref0}) that there exists $N_6\in\mathbb{N}$ such that
\begin{align}\label{X7}
\ \max\limits_{1\leq j\leq (s_n -1)}\frac{\mathcal{E}_{(t_n + j + 1)} + n^{-\alpha}}{\mathcal{E}_{(t_n + j)} - n^{-\alpha}}<  \frac{\mathcal{E}_{(t_n + 1)} - n^{-\alpha}}{n^{-\alpha}}\ \  \text{for all }n\geq N_6.
\end{align}
Combining \eqref{X5} and \eqref{X7}, we observe that for all $n\geq \{N_4,N_5,N_6\}$
\begin{align}\label{nref3}
&\ P\bigg[\max_{1\leq k\leq d_n}|\widehat{\mathcal{E}}_{(k)} - {\mathcal{E}}_{(k)}|\leq n^{-\alpha}\bigg]\nonumber\\
\leq &\ P\bigg [\max\limits_{1\leq j\leq (t_n - 1)  } \widehat{R}_{j}\frac{1}{n^{\alpha} \mathcal{E}_{(t_n + 1)}}\leq \epsilon ,\ \widehat{R}_{t_n}\geq {\frac{\mathcal{E}_{(t_n + 1)} -n^{-\alpha}  }{n^{-\alpha}}},\max\limits_{t_n + 1\leq j\leq ((d_n -1))} \widehat{R}_{ j }\leq {\frac{\mathcal{E}_{(t_n + 1)} -n^{-\alpha}  }{n^{-\alpha}}}\bigg ] + \epsilon .
\end{align}
Recall that $0<\epsilon<1$. It follows from assumption A1.1 that there exists some $N_7\in\mathbb{N}$ such that for all $n\geq N_7$, we have
\begin{align}\label{X6}
&\ \epsilon\ <1 - \frac{1}{n^{\alpha}\mathcal{E}_{(t_n + 1)} }<1 \Rightarrow \epsilon\  n^{\alpha}\mathcal{E}_{(t_n + 1)}<n^{\alpha}\mathcal{E}_{(t_n + 1)}-1 \Rightarrow \epsilon\  n^{\alpha}\mathcal{E}_{(t_n + 1)}<\frac{\mathcal{E}_{(t_n + 1)}-n^{-\alpha}}{n^{-\alpha}}.
\end{align}
\noindent Combining \eqref{X5}, \eqref{nref3} and \eqref{X6}, we conclude that for all $n\geq \max \{N_4, N_5, N_6, N_7\}$
\begin{align*}
&\ P\left [\max_{1\leq k\leq d_n}|\widehat{\mathcal{E}}_{(k)} - {\mathcal{E}}_{(k)}|\leq n^{-\alpha}\right ]\nonumber\\
\leq &\ \epsilon + P\bigg [\max\limits_{1\leq j\leq (t_n - 1) } \widehat{R}_{j}\leq \epsilon\ n^{\alpha} \mathcal{E}_{(t_n+1)},\ \widehat{R}_{t_n}\geq {\frac{\mathcal{E}_{(t_n + 1)} -n^{-\alpha}  }{n^{-\alpha}}},\max\limits_{(t_n + 1)\leq j\leq (d_n -1)} \widehat{R}_{ j }\leq \frac{\mathcal{E}_{(t_n + 1)}-n^{-\alpha}}{n^{-\alpha}}\bigg ]\\
\leq &\ \epsilon + P\left  [\max\limits_{1\leq j\leq (t_n - 1) } \widehat{R}_{i}\leq \frac{\mathcal{E}_{(t_n + 1)}-n^{-\alpha}}{n^{-\alpha}},\ \widehat{R}_{t_n}\geq {\frac{\mathcal{E}_{(t_n + 1)} -n^{-\alpha}  }{n^{-\alpha}}},\max\limits_{(t_n + 1)\leq j\leq (d_n -1)} \widehat{R}_{j}\leq \frac{\mathcal{E}_{(t_n + 1)}-n^{-\alpha}}{n^{-\alpha}}\right ]\\
=&\ \epsilon +  P\bigg [ \widehat{R}_{t_n} =  \max\limits_{1\leq j\leq (d_n -1)} \widehat{R}_{j}\bigg ] =\  \epsilon +  P\bigg [ \argmax\limits_{1\leq j\leq (d_n -1)} \widehat{R}_{j} = t_n\bigg ] =\ \epsilon +  P [ \widehat{s}_n  =  s_n ].
\end{align*}

\noindent
Since $\epsilon >0$ is arbitrary, we have
$$P\bigg[\max_{1\leq k\leq d_n}|\widehat{\mathcal{E}}_{(k)} - {\mathcal{E}}_{(k)}|\leq n^{-\alpha}\bigg]\leq P [ \widehat{s}_n  =  s_n ] \text{ for all }n\geq \max\{N_4, N_5, N_6, N_7\}.$$

\noindent
It follows from Corollary \ref{order_e_conv} that
$P [ \widehat{s}_n  =  s_n ] > 1 - O\bigl(e^{-b_1 \{n^{1-2\alpha}-n^\beta\}}\bigr)$. Hence, the proof.
\hfill  \QEDB\newline

\noindent
We now prove the main result of this article, {\it exact screening property} (ESP) of the estimator $\widehat{S}_n.$\newline

\noindent
{\bf Proof of Theorem \ref{mainthm_E} :}
In the proof of Theorem \ref{mainthm}, we have already established that if assumption A1 is satisfied for $0<\alpha_1<(1-\beta)/2$, then we have $b_1>0$ such that $$P[S_n\not\subseteq \widehat{S}_n]\leq  O\left (e^{-b_1 \{n^{1-2\alpha}-n^\beta\}}\right )\text{ for all }\alpha_1\le\alpha <(1-\beta)/2.$$

\noindent
Further, we assume that assumption A2 is also satisfied for $0 <\alpha_2<(1-\beta)/2.$ We now show that $$P[\widehat{S}_n\neq S_n]\leq  O\left (e^{-b_1\{ n^{1-2\alpha}- n^{\beta}\}}\right )\text{ for all }\max\{\alpha_1,\alpha_2\}\le\alpha <(1-\beta)/2.$$
Note that
\begin{align*}
P[\widehat{S}_n\neq S_n] =  P[\widehat{S}_n\neq S_n ,\ \widehat{s}_n=s_n] + P[\widehat{S}_n\neq S_n, \widehat{s}_n\neq s_n] \leq P[\widehat{S}_n\neq S_n ,\ \widehat{s}_n=s_n] + P[\widehat{s}_n\neq s_n].
\end{align*}
Define $\alpha_0=\max\{\alpha_1,\alpha_2\}$. Fix $\alpha_0\le \alpha<(1-\beta)/2$. It follows from Lemma \ref{mainlemma_E} that
\begin{align*}
P[\widehat{S}_n\neq S_n]&\leq P[\widehat{S}_n\neq S_n ,\ \widehat{s}_n=s_n] + O\left (e^{-b_1\{ n^{1-2\alpha}-n^\beta\}}\right ).
\end{align*}
If $\widehat{S}_n$ has the same number of elements as $S_n$, then $\widehat{S}_n\neq S_n $ implies that there exist at least one $i$ (with $i\in\{1,\ldots,s_n\}$) such that $k_i\in S_n,\ l_i\in S^c_n$ and
$\widehat{\mathcal{E}}_{k_i} < \widehat{\mathcal{E}}_{(t_n+1)}\leq \widehat{\mathcal{E}}_{l_i}.$ Thus, we have
\begin{align}\label{Sn_ref}
P[\widehat{S}_n\neq S_n] \leq  \sum\limits_{i=1}^{s_n} P[\widehat{\mathcal{E}}_{k_i} < \widehat{\mathcal{E}}_{(t_n+1)}\leq \widehat{\mathcal{E}}_{l_i},\ k_i\in S_n,\ l_i\in S^c_n] + O\left (e^{-b_1\{ n^{1-2\alpha}-n^\beta\}}\right ).
\end{align}
Let us now consider the probability $P[\widehat{\mathcal{E}}_{k_i} < \widehat{\mathcal{E}}_{(t_n+1)}\leq \widehat{\mathcal{E}}_{l_i},\ k_i\in S_n,\ l_i\in S^c_n]$ for any fixed $i$ (with $1 \leq i \leq s_n$) as follows
\begin{align}\label{arbit0}
&\ P[\widehat{\mathcal{E}}_{k_i} < \widehat{\mathcal{E}}_{(t_n+1)}\leq \widehat{\mathcal{E}}_{l_i},\ k_i\in S_n,\ l_i\in S^c_n]\nonumber\\
\leq &\ P[\widehat{\mathcal{E}}_{k_i} < \widehat{\mathcal{E}}_{l_i},\ k_i\in S_n,\ l_i\in S^c_n]\nonumber\\
\leq &\ P[ \widehat{\mathcal{E}}_{k_i} < \widehat{\mathcal{E}}_{l_i}, ~\widehat{\mathcal{E}}_{l_i} ~{\leq}~ n^{-\alpha}, ~|\widehat{\mathcal{E}}_{k_i}-\mathcal{E}_{k_i}|~{\leq}~ n^{-\alpha},\ k_i\in S_n,\ l_i\in S^c_n]\nonumber\\
&\ + P[ \widehat{\mathcal{E}}_{k_i} < \widehat{\mathcal{E}}_{l_i}, ~\widehat{\mathcal{E}}_{l_i} ~{>}~ n^{-\alpha},~|\widehat{\mathcal{E}}_{k_i}-\mathcal{E}_{k_i}| ~{\leq}~ n^{-\alpha},\ k_i\in S_n,\ l_i\in S^c_n]\nonumber\\
&\ + P[ \widehat{\mathcal{E}}_{k_i} < \widehat{\mathcal{E}}_{l_i},~ \widehat{\mathcal{E}}_{l_i} ~{\leq}~ n^{-\alpha},~|\widehat{\mathcal{E}}_{k_i}-\mathcal{E}_{k_i}| ~{>}~ n^{-\alpha},\ k_i\in S_n,\ l_i\in S^c_n]\nonumber\\
&\ + P[ \widehat{\mathcal{E}}_{k_i} < \widehat{\mathcal{E}}_{l_i},~ \widehat{\mathcal{E}}_{l_i} ~{>}~ n^{-\alpha},~|\widehat{\mathcal{E}}_{k_i}-\mathcal{E}_{k_i}| ~{>}~ n^{-\alpha},\ k_i\in S_n,\ l_i\in S^c_n]\nonumber \\
= &\ P_{1i} + P_{2i} + P_{3i} + P_{4i}\text{ (say)}.
\end{align}
Let us take a look at the term $P_{1i}$ first.
\begin{align*}
P_{1i} = &  P[ \widehat{\mathcal{E}}_{k_i} < \widehat{\mathcal{E}}_{l_i},\widehat{\mathcal{E}}_{l_i}\leq n^{-\alpha},|\widehat{\mathcal{E}}_{k_i}-\mathcal{E}_{k_i}|\leq n^{-\alpha},\ k_i\in S_n,\ l_i\in S^c_n ]\nonumber \\
\leq & \ P[\mathcal{E}_{k_i} - n^{-\alpha} \leq \widehat{\mathcal{E}}_{k_i} < \widehat{\mathcal{E}}_{l_i}\leq n^{-\alpha},\ k_i\in S_n,\ l_i\in S^c_n]\nonumber\\
\leq & \ \mathbb{I}[\mathcal{E}_{k_i} - n^{-\alpha}\leq n^{-\alpha},\ k_i\in S_n]\ \ (\text{since }P[a_1\leq X\leq a_2]\leq \mathbb{I}[a_1\leq a_2] )\nonumber\\
= & \ \mathbb{I}[\mathcal{E}_{k_i} \leq 2n^{-\alpha},\ k_i\in S_n].
\end{align*}
Since assumption A.1 is satisfied for $\alpha_1$ and $\alpha\ge \alpha_0\geq \alpha_1$, there exists $N\in \mathbb{N}$ such that $\mathbb{I}[\mathcal{E}_{k_i} \leq 2n^{-\alpha} \mbox{ for } k_i\in S_n]=0$ for all $n\geq N$. This holds true for every $i$ for $1\leq i\leq s_n$. Thus, we get
\begin{align}\label{arbit1}
\sum_{i=1}^{s_n}P_{1i}\leq \sum_{i=1}^{s_n} \mathbb{I}[\mathcal{E}_{k_i} \leq 2n^{-\alpha} \mbox{ for } k_i\in S_n]=0\text{ for all }n\ge N.
\end{align}
\noindent Next, we consider the second term $P_{2i}$.
\begin{align*}
P_{2i}\leq &\ P[ \widehat{\mathcal{E}}_{k_i} < \widehat{\mathcal{E}}_{l_i}, \widehat{\mathcal{E}}_{l_i}> n^{-\alpha},|\widehat{\mathcal{E}}_{k_i}-\mathcal{E}_{k_i}|\leq n^{-\alpha},\ k_i\in S_n,\ l_i\in S^c_n]
\leq P[ \widehat{\mathcal{E}}_{l_i}> n^{-\alpha_0},\ l_i\in S^c_n].
\end{align*}
\noindent Following Lemma \ref{expbound}, we have
\begin{align}\label{arbit2}
&\ P[ \widehat{\mathcal{E}}_{l_i}> n^{-\alpha} \mbox{ for } l_i\in S^c_n]= O\left (e^{-b n^{1-2\alpha}}\right )\Rightarrow P_{2i}\leq O\left (e^{-b n^{1-2\alpha}}\right )\Rightarrow \sum\limits_{i=1}^{s_n} P_{2i} = \ O\left (s_n e^{-b n^{1-2\alpha}}\right ).
\end{align}
Observe that
\begin{align*}
s_n e^{-b n^{1-2\alpha}}
\leq d_n e^{-b n^{1-2\alpha}}
= &\ O\left (e^{-b n^{1-2\alpha}+ M n^\beta}\right )\ (\text{since }\log\ d_n \leq Mn^\beta\text{ for all }n\geq N)\\
=&\ O\left (e^{-b_1\{ n^{1-2\alpha}- n^\beta\}}\right ), \text{ where } 0<b_1 <b.
\end{align*}
Therefore, $\sum\limits_{i=1}^{s_n} P_{2i} = O\left (e^{-b_1\{ n^{1-2\alpha}- n^\beta\}}\right )$. Following a similar line of arguments and using Lemma \ref{expbound}, one can now obtain the following:
\begin{align}\label{arbit3}
\sum\limits_{i=1}^{s_n} P_{3i}=  O\left (e^{-b_1\{ n^{1-2\alpha}- n^\beta\}}\right ) \text{ and }\sum\limits_{i=1}^{s_n} P_{4i}=  O\left (e^{-b_1\{ n^{1-2\alpha}- n^\beta\}}\right ).
\end{align}
Finally, combining \eqref{Sn_ref}-\eqref{arbit3}, we get
\begin{align*}
P[S_n\neq \widehat{S}_n] &\leq O\left (e^{-b_1\{ n^{1-2\alpha}- n^\beta\}}\right ) = O\left (e^{-b_1\{ n^{1-2\alpha}- n^\beta\}}\right ).
\end{align*}
Hence, the proof.\hfill \QEDB
\newline

\noindent
\begin{remark} \label{relaxed_conditions}
We have proved {\it exact screening property} (ESP) of the marginal screening method (MarS) under assumptions A1 and A2. We would like to point out that Theorem \ref{mainthm} can be proved under a set of weaker conditions. Let us first introduce assumption $A1.2^{\prime}$ and $A2^\prime$, which are weaker versions of A1.2 and A2, respectively. Assumption A1.1 remains unaltered.

\begin{enumerate}
	\item[$A1.2^\prime$.] There exist constants $0 <\alpha_1 <(1-\beta)/2,$ $0<M_1<1$ and $N_1\in\mathbb{N}$ such that
	$$\max\limits_{1\leq j\leq (s_n -1)}R_{t_n + j} \leq M_1 n^{\alpha_1}\mathcal{E}_{(t_n+1)}\text{ for all } n\geq N_1.$$
	\item[$A2^\prime$.] For every $\epsilon >0$ there exist constants $ 0 <\alpha_2 <(1-\beta)/2,\  0<M_2<1$ and $N_2\in\mathbb{N}$ such that
	$$P\bigg [
	\max\limits_{1\leq j\leq (t_n-1) } \frac{N_{(j+1)}}{N_{(j)}}>M_2 n^{\alpha_2} \mathcal{E}_{(t_n + 1)}\bigg ]<\epsilon\text{ for  all }n\geq N_2.$$
\end{enumerate}
\noindent
Assumptions $A1.2^\prime$ and $A2^\prime$ are satisfied if the components of the signal set are all iid Similar to the earlier case, A1.1 and $A1.2^\prime$ are needed to prove SSP of $\widehat{S}_n$. $A2^\prime$ is required to establish ESP. 
\vspace*{0.1in}
	
\noindent
To ensure SSP of the proposed estimator $\widehat{S}_n$, we had used
$$\max\limits_{1\leq j\leq (s_n -1)}\widehat{R}_{t_n+j} < {n^{\alpha_1}\mathcal{E}_{t_n + 1}} \text{ for all } n\geq N_2$$
as obtained in \eqref{X1} (see the proof of Lemma \ref{teststatresult_s}) under assumption A1.1, where it was assumed that $\max_{1\leq j\leq (s_n -1)}R_{t_n + j} = o(n^{\alpha_1}\mathcal{E}_{(t_n + 1)})$. However, $\max_{1\leq j\leq (s_n -1)}R_{t_n + j}$ can be of order $O(n^{\alpha_1}\mathcal{E}_{(t_n + 1)})$ and \eqref{X1} still holds if assumption $A1.2^\prime$ is satisfied. 
\vspace*{0.1in}
	
\noindent
Similarly, one can prove inequality \eqref{X6} and subsequently, Lemma \ref{mainlemma_E} under $A2^\prime$ which allows the sequence of random variables $\max\limits_{1\leq j\leq (t_n-1) } {N_{(j+1)}}/{N_{(j)}}$ to be of the order $O_{\rm P}(n^{\alpha_2} \mathcal{E}_{(t_n + 1)})$ for an  appropriate choice of  $\alpha_2$.
\end{remark}

\vspace*{0.1in}
\noindent {\bf Proof of Lemma \ref{joint_indep} :}
Fix $P_n \in \mathcal{P}_n$. The pairs of indices that belong to $P_n$ can be divided into the following four categories:
\begin{enumerate}
	\item $\{i,j^\prime\}$ with $\{i,j\}\in S_n$ and $\{i^\prime, j^\prime\}\in S_n$,
	\item $\{i,l\}$ with $\{i,j\}\in S_n$ and $l\in S^c_n$,
	\item $\{l, l^\prime\}$ with $l$ and $l^\prime\in S^c_n$,
	\item $\{i,j\}$ with $\{i,j\}\in S_n$.
\end{enumerate}
Let us consider case 1, i.e., the pair $\{i,j^\prime\}$, where $\{i,j\}\in S_n$ and $\{i^\prime, j^\prime\}\in S_n$. Since $\{i,j\}$ and $\{i^\prime,j^\prime\}$ are paired signals, we have $F_{l}= G_l$ for all $l\in\{i,j,i^\prime,j^\prime\}$. Further, since $\bX_{\{i,j\}}$ and $\bX_{\{i^\prime,j^\prime\}}$ are mutually independent, the following holds true:
\begin{align*}
\bF_{\{i,j^\prime\}} = F_iF_{j^\prime} = G_iG_{j^\prime} = \bG_{\{i,j^\prime\}},\text{ i.e., } \mathcal{E}_{\{i,j^\prime\}}=0.
\end{align*}
For case 2, it can be similarly shown that $\mathcal{E}_{\{i,l\}}=0$ when $\{i,j\}\in S_n$ and $l\in S^c_n$.
If $l$ and $l^\prime$ both are elements of $S^c_n$ (case 3), then it follows from the definition of noise variables that $\bF_{\{l,l^\prime\}}=\bG_{\{l,l^\prime\}}$. Consequently, we have $\mathcal{E}_{\{l,l^\prime\}}=0$.

\noindent 
Finally, $\mathcal{E}_{\{i,j\}}>0$ since $\{i,j\}\in S_n$ for case 4. It is easy to see that $\mathcal{E}(P_n)$ takes its maximum value iff all the paired signals are present in $P_n$, i.e., $S_n \subseteq P_n$ and the maximum value is~$\sum\limits_{\{i,j\}\in S_n} \mathcal{E}_{\{i,j\}}$. \hfill \QEDB

\vspace*{0.2in}

\vspace*{0.1in}
\noindent 
In the next result, we prove that the discriminant of gSAVG classifier converges in probability to its  population counterpart, and the rate of convergence is exponential in $n$.

\vspace*{0.1in}
\noindent {\bf Proof of Lemma \ref{exbddisc} :}
Suppose that assumptions A1 and A2 are satisfied for $0<\alpha_1<(1-\beta)/2$ and $0<\alpha_2<(1-\beta)/2,$ respectively. Recall the definitions of $\widehat{\xi}^\gamma_{n}(\bz)$ and $\xi^\gamma(\bz)$ for $\bz\in\mathbb{R}^{d_n}$. Fix $\max\{\alpha_1,\alpha_2\}\le \alpha<(1-\beta)/2$. Now, we have the following:
\begin{align}\label{nref4}
& P[|\widehat{\xi}_n^\gamma(\bz)-\xi^\gamma(\bz)|>n^{-\alpha}]\nonumber \\
=&\ P[|\widehat{\xi}_n^\gamma(\bz)-\xi^\gamma(\bz)|>n^{-\alpha},\widehat{S}_n=S_n]+P[|\widehat{\xi}_n^\gamma(\bz)-\xi^\gamma(\bz)|>n^{-\alpha},\widehat{S}_n\neq S_n]\nonumber \\
\leq &\ P[|\widehat{\xi}_n^\gamma(\bz)-\xi^\gamma(\bz)|>n^{-\alpha}, \widehat{S}_n=S_n]+P[\widehat{S}_n\neq S_n]\nonumber \\
= &\ P[|\widehat{\xi}_n^\gamma(\bz)-\xi^\gamma(\bz)|>n^{-\alpha}, \widehat{S}_n=S_n]+O\left (e^{-b_1\{ n^{1-2\alpha}- n^\beta\}}\right ) \ (\text{using Theorem }\ref{mainthm_E}).
\end{align}
Since $S_n=S_{1n}$, it follows from the definition of $h^\gamma_0$ in \eqref{h_0trans} that
 \begin{align*}
  E\left [h^{\gamma}_n(\bz,\bX_1)\big |\widehat{S}_n=S_n\right ] = E\left [\frac{1}{\widehat{s}_n}\sum\limits_{k\in \widehat{S}_n}\gamma(|z_k-X_{1k}|^2)\bigg |\widehat{S}_n=S_n\right ]=&\ \frac{1}{{s}_n}\sum\limits_{k\in {S}_n}E\left [\gamma(|z_k-X_{1k}|^2)\right ] \\
  =&\ h^\gamma_0(\bz,\bX_1).
\end{align*}
Similarly, $E\left [h^{\gamma}_n(\bz,\bY_1)\big |\widehat{S}_n=S_n\right ]=h^\gamma_0(\bz,\bY_1)$, $E\left [h^{\gamma}_n(\bX_1,\bX_2)\big |\widehat{S}_n=S_n\right ]=h^\gamma_0(\bX_1,\bX_2)$ and \\
$E\left [h^{\gamma}_n(\bY_1,\bY_2)\big |\widehat{S}_n=S_n\right ]=h^\gamma_0(\bY_1,\bY_2).$ Therefore, we can write the following:
\begin{align*}
 &\ \widehat{\xi}^\gamma(\bz)-\xi^\gamma(\bz)\\
 =&\ \{\widehat{\xi}_2^\gamma(\bz)-\widehat{\xi}_1^\gamma(\bz)\}-\{\xi_2^\gamma(\bz)-\xi_1^\gamma(\bz)\}\\
  =&\ \left\{\frac{1}{n_1} \sum_{i=1}^{n_1}h^{\gamma}_n(\bz,\bX_i)-\frac{1}{{n_1\choose 2}} \sum_{i< j} h^{\gamma}_n(\bX_i,\bX_j)\right\} -\left \{\frac{1}{n_2} \sum_{i=1}^{n_2}h^{\gamma}_n(\bz,\bY_i)-\frac{1}{{n_2\choose 2}} \sum_{i< j} h^{\gamma}_n(\bY_i,\bY_j)\right\}\\
 -&\ \left\{ E\left [h^{\gamma}_n(\bz,\bX_1)|\widehat{S}_n=S_n\right ] -  E\left [h^{\gamma}_n(\bX_1,\bX_2)|\widehat{S}_n=S_n\right ]\right\} \\
 -&\ + \left \{E\left [h^{\gamma}_n(\bz,\bY_1)|\widehat{S}_n=S_n\right ] E\left [h^{\gamma}_n(\bY_1,\bY_2)|\widehat{S}_n=S_n\right ]\right \}\\
 =&\ \left\{\frac{1}{n_1} \sum_{i=1}^{n_1}h^{\gamma}_n(\bz,\bX_i)-E[h^{\gamma}_n(\bz,\bX_1)|\widehat{S}_n=S_n]\right\} -\left \{\frac{1}{n_2} \sum_{i=1}^{n_2}h^{\gamma}_n(\bz,\bY_i)-E[h^{\gamma}_n(\bz,\bY_1)|\widehat{S}_n=S_n]\right\}\\
 -&\ \left\{ \frac{1}{n_1(n_1-1)} \sum_{i\neq j} h^{\gamma}_n(\bX_i,\bX_j) -  E[h^{\gamma}_n(\bX_1,\bX_2)|\widehat{S}_n=S_n]\right\}\\
 +&\ \left \{\frac{1}{n_2(n_2-1)} \sum_{i\neq j} h^{\gamma}_n(\bY_i,\bY_j) - E[h^{\gamma}_n(\bY_1,\bY_2)|\widehat{S}_n=S_n]\right \}.
\end{align*}
Consequently, we have
\begin{align}\label{nref5.0}
&\ P[|\widehat{\xi}^\gamma(\bz)-\xi^\gamma(\bz)|>n^{-\alpha}, \widehat{S}_n=S_n]\nonumber \\
\leq &\ P \left [\left |\frac{1}{n_1} \sum_{i=1}^{n_1}h^{\gamma}_n(\bz,\bX_i) - h^{\gamma}_0(\bz,\bX_1)\right |> \frac{n^{-\alpha}}{4}, \widehat{S}_n=S_n\right ]\nonumber\\
+&\  P\left [\left |\frac{1}{n_2} \sum_{i=1}^{n_2}h^{\gamma}_n(\bz,\bY_i) - h^{\gamma}_0(\bz,\bY_1)\right |> \frac{n^{-\alpha}}{4}, \widehat{S}_n=S_n\right ]\nonumber \\
+&\ P\left [\left |\frac{1}{n_1(n_1-1)} \sum_{i\neq j} h^{\gamma}_n(\bX_i,\bX_j) - h^{\gamma}_0(\bX_1,\bX_2)\right |> \frac{n^{-\alpha}}{2}, \widehat{S}_n=S_n\right ]\nonumber \\
+&\   P\left [\left |\frac{1}{n_2(n_2-1)} \sum_{i\neq j} h^{\gamma}_n(\bY_i,\bY_j) - h^{\gamma}_0(\bY_1,\bY_2)]\right |> \frac{n^{-\alpha}}{2}, \widehat{S}_n=S_n\right ].
\end{align}
Let us look at the first and second term in the RHS of \eqref{nref5.0}. Note that $\sum_{i=1}^{n_1}h^{\gamma}_n(\bz,\bX_i)/n_1$ is an average of independently distributed random variables for each fixed $\bz\in\mathbb{R}^{d_n}$. Also, $\sum_{i=1}^{n_1}E[h^{\gamma}_n(\bz,\bX_i)|\widehat{S}_n=S_n]/n_1 = E[h^{\gamma}_n(\bz,\bX_1)|\widehat{S}_n=S_n] = h^{\gamma}_0(\bz,\bX_1)$. Therefore, for every fixed $\bz$, the following holds using Hoeffding's inequality \citep[see][]{wainwright2019high}:
\begin{align}\label{nref6}
&\ P\left [\left |\frac{1}{n_1} \sum_{i=1}^{n_1}\left \{h^{\gamma}_n(\bz,\bX_i) - E[h^{\gamma}_n(\bz,\bX_i)]\right \}\right |> \frac{n^{-\alpha}}{4}, \widehat{S}_n=S_n\right ]\leq \ 2e^{-\frac{n_1n^{-2\alpha}}{16}}\nonumber \\
\Rightarrow &\ P\left [\left |\frac{1}{n_1} \sum_{i=1}^{n_1}h^{\gamma}_n(\bz,\bX_i) - h^{\gamma}_0(\bz,\bX_1)\right |> \frac{n^{-\alpha}}{4}, \widehat{S}_n=S_n\right ]\leq O(e^{-C_3n^{1-2\alpha}})\text{ for some }C_3>0.
\end{align}
Similarly,
\begin{align}\label{nref7}
&\ P\left [\left |\frac{1}{n_2} \sum_{i=1}^{n_2}\left \{h^{\gamma}_n(\bz,\bY_i) - E[h^{\gamma}_n(\bz,\bY_i)]\right \}\right |> \frac{n^{-\alpha}}{4}, \widehat{S}_n=S_n\right ]\leq \ 2e^{-\frac{n_2n^{-2\alpha}}{16}}\nonumber \\
\Rightarrow &\ P\left [\left |\frac{1}{n_2} \sum_{i=1}^{n_2}h^{\gamma}_n(\bz,\bY_i) - h^{\gamma}_0(\bz,\bY_1)\right |> \frac{n^{-\alpha}}{4}, \widehat{S}_n=S_n\right ]\leq O(e^{-C_4n^{1-2\alpha}})\text{ for some }C_4>0.
\end{align}
Let us now look at the third term in the RHS of \eqref{nref5.0}. Recall the definition of $T_{11k}$ and $T_{22k}$ for $1 \leq k \leq d_n$ given in Lemma \ref{expbound}. Note that
\begin{align*}
&\ P\left [\left |\frac{1}{n_1(n_1-1)} \sum_{i\neq j} h^{\gamma}_n(\bX_i,\bX_j) - h^{\gamma}_0(\bX_1,\bX_2)\right |> \frac{n^{-\alpha}}{2}, \widehat{S}_n=S_n\right ]\\
=&\ P\left [\left |\frac{1}{n_1(n_1-1)} \sum_{i\neq j} \frac{1}{s_n}\sum_{k\in S_n}\gamma(|X_{ik}-X_{jk}|^2) -  \frac{1}{s_n}\sum_{k\in S_n} E[\gamma(|X_{1k}-X_{2k}|^2)]\right |> \frac{n^{-\alpha}}{2}\right ]\\
\leq &\ P\left [\frac{1}{s_n}\sum_{k\in S_n}\left |\frac{1}{n_1(n_1-1)} \sum_{i\neq j} \gamma(|X_{ik}-X_{jk}|^2) - E[\gamma(|X_{1k}-X_{2k}|^2)]\right |> \frac{n^{-\alpha}}{2}\right ]\\
\leq &\ \sum_{k\in S_n}P\left [\left |\frac{1}{n_1(n_1-1)} \sum_{i\neq j} \gamma(|X_{ik}-X_{jk}|^2) - E[\gamma(|X_{1k}-X_{2k}|^2)]\right |> \frac{n^{-\alpha}}{2}\right ]\\
\leq &\ \sum_{k\in S_n}P\left [\left |T_{11k} - E[T_{11k}]\right |> \frac{n^{-\alpha}}{2}\right ].
\end{align*}
We have already shown in the proof of Lemma \ref{expbound} that $P\left [\left |T_{11k} - E[T_{11k}]\right |> \frac{n^{-\alpha}}{2}\right ]\leq O\left (e^{-b^{*}n^{1-2\alpha}}\right )$ for some $b^*>0.$ Therefore,
\begin{align}\label{nref8}
&\ P\left [\left |\frac{1}{n_1(n_1-1)} \sum_{i\neq j} h^{\gamma}_n(\bX_i,\bX_j) - E[h^{\gamma}_n(\bX_1,\bX_2)]\right |> \frac{n^{-\alpha}}{2}, \widehat{S}_n=S_n\right ]\leq s_n O\left (e^{-b^{*}n^{1-2\alpha}}\right )\nonumber \\
\Rightarrow &\ P\left [\left |\frac{1}{n_1(n_1-1)} \sum_{i\neq j} h^{\gamma}_n(\bX_i,\bX_j) - E[h^{\gamma}_n(\bX_1,\bX_2)]\right |> \frac{n^{-\alpha}}{2}, \widehat{S}_n=S_n\right ]\leq d_n O\left (e^{-b^{*}n^{1-2\alpha}}\right )\nonumber \\
\Rightarrow &\ P\left [\left |\frac{1}{n_1(n_1-1)} \sum_{i\neq j} h^{\gamma}_n(\bX_i,\bX_j) - E[h^{\gamma}_n(\bX_1,\bX_2)]\right |> \frac{n^{-\alpha}}{2}, \widehat{S}_n=S_n\right ]\leq O\left (e^{-C_1\{n^{1-2\alpha}-n^\beta\}}\right )
\end{align}
for $0<C_1<b^{*}.$ Similarly, we have $C_2>0$ such that
\begin{align}\label{nref9}
&\ P\left [\left |\frac{1}{n_2(n_2-1)} \sum_{i\neq j} h^{\gamma}_n(\bY_i,\bY_j) - E[h^{\gamma}_n(\bY_1,\bY_2)]\right |> \frac{n^{-\alpha}}{2}, \widehat{S}_n=S_n\right ]\nonumber \\
\leq &\ \sum_{k\in S_n}P\left [\left |T_{22k} - E[T_{22k}]\right |> \frac{n^{-\alpha}}{2}\right ]= O\left (e^{-C_2\{n^{1-2\alpha}-n^\beta\}}\right ).
\end{align}
Combining \eqref{nref5.0} - \eqref{nref9}, we obtain
\begin{align}\label{nref5}
&\  P\left [|\widehat{\xi}^\gamma(\bz)-\xi^\gamma(\bz)|>n^{-\alpha}, \widehat{S}_n=S_n\right ]\nonumber \\
\leq &\ O\left (e^{-C_1n^{1-2\alpha}}\right )+O\left (e^{-C_2n^{1-2\alpha}}\right )+O\left (e^{-C_3\{n^{1-2\alpha}-n^{\beta}\}}\right )+O\left (e^{-C_4\{n^{1-2\alpha}-n^{\beta}\}}\right )\nonumber \\
\leq &\ O\left (e^{-C_5\{n^{1-2\alpha}-n^{\beta}\}}\right )\text{ for some }C_5>0.
\end{align}
Consequently, it follows from \eqref{nref4} and \eqref{nref5} that
\begin{align*}
&P\left [|\widehat{\xi}^\gamma(\bz)-\xi^\gamma(\bz)|>n^{-\alpha}\right ]\leq O\left (e^{-b_1n^{1-2\alpha}}\right )+ O\left (e^{-C_5\{n^{1-2\alpha}-n^{\beta}\}}\right )\nonumber \\
\text{i.e., }&P\left [|\widehat{\xi}^\gamma(\bz)-\xi^\gamma(\bz)|>n^{-\alpha}\right ]\leq O(e^{-b_3\{n^{1-2\alpha}-n^{\beta}\}})\text{ for some }b_3>0.
\end{align*}
Hence, the proof.\hfill\QEDB\newline

\vspace*{-0.1in}
\noindent{\bf Proof of Theorem \ref{thmclass} :} Let ${\rm C}$ denote the true class label of a test observation $\bZ$. Clearly, ${\rm C}$ is a dichotomous random variable with $P[{\rm C}=i]=\pi_i$ for $i=1,2$ and $\pi_1+\pi_2=1$. We have $\bZ\mid {\rm C}=1\sim\bF$ and $\bZ\mid {\rm C}=2\sim\bG$. The distribution of $\bZ$ is given by $\bH(\bz) = \pi_1\bF(\bz) + \pi_2\bG(\bz)$ for $\bz\in\mathbb{R}^{d_n}$. The misclassification probabilities of $\delta_{0}$ and $\delta_{\rm gSAVG}$ are defined as $\Delta_0 = P[\delta_{0}(\bZ)\neq {\rm C}]$ and   $\Delta_{\rm gSAVG} = P[\delta_{\rm gSAVG}(\bZ)\neq {\rm C}]$, respectively. Now, observe that 
\vspace*{-0.1in}
\begin{align}\label{nref11}
&\ \Delta_{\rm gSAVG} - \Delta_{0}\nonumber \\
=&\ P[\delta_{\rm gSAVG}(\bZ)\neq {\rm C}]- P[\delta_{0}(\bZ)\neq {\rm C}]\nonumber\\
=&\ \int \left \{P[\delta_{\rm gSAVG}(\bz)\neq {\rm C}]- P[\delta_{0}(\bz)\neq {\rm C}]\right \}\ d\bH(\bz)\nonumber\\
=&\ \int \left \{P[\delta_{0}(\bz)= {\rm C}]- P[\delta_{\rm gSAVG}(\bz)= {\rm C}]\right\}\ d\bH(\bz)\nonumber\\
=&\ \int \big \{\big (I[\delta_{0}(\bz)=1]P[ {\rm C}=1] + I[\delta_{0}(\bz)=2]P[ {\rm C}=2]\big )-\nonumber\\
&\ \ \big (P[\delta_{\rm gSAVG}(\bz)=1]P[ {\rm C}=1] + P[\delta_{\rm gSAVG}(\bz)=2]P[ {\rm C}=2]\big)\big \}\ d\bH(\bz)\nonumber\\
=&\ \int \big \{(I[\delta_{0}(\bz)=1] - P[\delta_{\rm gSAVG}(\bz)=1])P[ {\rm C}=1] +\nonumber\\
&\ \ (I[\delta_{0}(\bz)=2] - P[\delta_{\rm gSAVG}(\bz)=2])P[ {\rm C}=2]\big \}\ d\bH(\bz)\nonumber\\
=&\ \int (I[\delta_{0}(\bz)=1] - E\big [I[\delta_{\rm gSAVG}(\bz)=1]\big ])(2P[ {\rm C}=1]-1)\ d\bH(\bz)\nonumber\\
\leq &\ \int \big |E\big [I[\delta_{0}(\bz)=1] - I[\delta_{\rm gSAVG}(\bz)=1]\big ]\big |\ |2P[ {\rm C}=1]-1|\ d\bH(\bz)\nonumber\\
= &\ \int E\big [|I[\delta_{0}(\bz)=1] - I[\delta_{\rm gSAVG}(\bz)=1]|\big ]\ d\bH(\bz)\nonumber\\
= &\ \int E\big [I[\delta_{0}(\bz)\neq \delta_{\rm gSAVG}(\bz)]\big ]\ d\bH(\bz)\nonumber\\
= &\ \int P[\delta_{0}(\bz)\neq \delta_{\rm gSAVG}(\bz)]\ d\bH(\bz)\nonumber\\
= &\ \int \left \{P[\delta_{0}(\bz)\neq \delta_{\rm gSAVG}(\bz), \widehat{S}_n=S_n] + P[\delta_{0}(\bz)\neq \delta_{\rm gSAVG}(\bz), \widehat{S}_n\neq S_n]\right \}\ d\bH(\bz)\nonumber\\
\leq  &\ \int \left \{ P[\delta_{0}(\bz)\neq \delta_{\rm gSAVG}(\bz), \widehat{S}_n=S_n] +P[\widehat{S}_n\neq S_n]\right \}\ d\bH(\bz)\nonumber\\
= &\ \int P[\delta_{0}(\bz)\neq \delta_{\rm gSAVG}(\bz), \widehat{S}_n=S_n]\ d\bH(\bz) +P[\widehat{S}_n\neq S_n]\nonumber\\
= &\ \int P[\xi^\gamma_1(\bz)-\xi^\gamma_2(\bz)>0,\widehat{\xi}^\gamma_{1n}(\bz)-\widehat{\xi}^\gamma_{2n}(\bz)\leq 0, \widehat{S}_n=S_n]\ d\bH(\bz)\nonumber\\
&\ +\ \int P[\xi^\gamma_1(\bz)-\xi^\gamma_2(\bz)\leq 0,\widehat{\xi}^\gamma_{1n}(\bz)-\widehat{\xi}^\gamma_{2n}(\bz)> 0, \widehat{S}_n=S_n]\ d\bH(\bz)+P[\widehat{S}_n\neq S_n]\nonumber\\
=&\ P_1 + P_2 + P[\widehat{S}_n\neq S_n].
\end{align}
Fix $\max\{\alpha_1,\alpha_2\}\le \alpha<(1-\beta)/2$. For $P_1$, we obtain the following:
\begin{align}\label{nref12}
P_1 &= \int P[\xi^\gamma(\bz)>0,\widehat{\xi}^\gamma_{n}(\bz)\leq 0, \widehat{S}_n=S_n]\ d\bH(\bz)\nonumber\\
&= \int P[\xi^\gamma(\bz)>0,\widehat{\xi}^\gamma_{n}(\bz)\leq 0, |\xi^\gamma(\bz)-\widehat{\xi}^\gamma_{n}(\bz)|\leq n^{-\alpha}, \widehat{S}_n=S_n]\ d\bH(\bz)\nonumber\\
&\ +\ \int P[\xi^\gamma(\bz)>0,\widehat{\xi}^\gamma_{n}(\bz)\leq 0, |\xi^\gamma(\bz)-\widehat{\xi}^\gamma_{n}(\bz)|> n^{-\alpha}, \widehat{S}_n=S_n]\ d\bH(\bz)\nonumber\\
&\leq \int P[\xi^\gamma(\bz)>0,\widehat{\xi}^\gamma_{n}(\bz)\leq 0, \xi^\gamma(\bz)-\widehat{\xi}^\gamma_{n}(\bz)\leq n^{-\alpha}, \widehat{S}_n=S_n]\ d\bH(\bz)\nonumber\\
&\ +\ \int P[|\xi^\gamma(\bz)-\widehat{\xi}^\gamma_{n}(\bz)|> n^{-\alpha}, \widehat{S}_n=S_n]\ d\bH(\bz)\nonumber\\
&= P_{11}(\alpha) + P_{12}(\alpha).
\end{align}
Note that
\begin{align}\label{nref13}
P_{11}(\alpha) =& \int P[\xi^\gamma(\bz)-\widehat{\xi}^\gamma_{n}(\bz)\leq n^{-\alpha}, \xi^\gamma(\bz)>0,\widehat{\xi}^\gamma_{n}(\bz)\leq 0, \widehat{S}_n=S_n]\nonumber\\
\leq & \int P[\xi^\gamma(\bz)\leq n^{-\alpha}, \xi^\gamma(\bz)>0,\widehat{\xi}^\gamma_{n}(\bz)\leq 0, \widehat{S}_n=S_n]\ d\bH(\bz)\nonumber\\
\leq & \int P[\xi^\gamma(\bz)\leq n^{-\alpha}, \xi^\gamma(\bz)>0]\ d\bH(\bz)\nonumber\\
= &\ P[0<\xi^\gamma(\bZ)\leq n^{-\alpha}].
\end{align}
Combining \eqref{nref12} and \eqref{nref13}, we obtain
\begin{align}\label{nrefP1}
P_1 \leq P[0<\xi^\gamma(\bZ)\leq n^{-\alpha}] + P_{12}(\alpha).
\end{align}

\noindent Following similar arguments, we can write $P_2$ as follows:
\begin{align}\label{nrefP2}
P_2 &= \int P[\xi^\gamma(\bz)\leq 0,\widehat{\xi}^\gamma_{n}(\bz)> 0, \widehat{S}_n=S_n]\ d\bH(\bz)\nonumber\\
&\leq \int P[\xi^\gamma(\bz)\leq 0,\widehat{\xi}^\gamma_{n}(\bz)> 0, |\xi^\gamma(\bz)-\widehat{\xi}^\gamma_{n}(\bz)|\leq n^{-\alpha}, \widehat{S}_n=S_n]\ d\bH(\bz)\nonumber\\
+&\ \int P[|\xi^\gamma(\bz)-\widehat{\xi}^\gamma_{n}(\bz)|> n^{-\alpha}, \widehat{S}_n=S_n]\ d\bH(\bz)\nonumber\\
&= \int P[\xi^\gamma(\bz)\leq 0,\widehat{\xi}^\gamma_{n}(\bz)> 0, |\xi^\gamma(\bz)-\widehat{\xi}^\gamma_{n}(\bz)|\leq n^{-\alpha}, \widehat{S}_n=S_n]\ d\bH(\bz) +P_{12}(\alpha)\nonumber\\
&\leq \int P[\xi^\gamma(\bz)\leq 0,\widehat{\xi}^\gamma_{n}(\bz)> 0, -\xi^\gamma(\bz)+\widehat{\xi}^\gamma_{n}(\bz)\leq n^{-\alpha}, \widehat{S}_n=S_n]\ d\bH(\bz) +P_{12}(\alpha)\nonumber\\
&\leq \int P[-n^{\alpha}<\xi^\gamma(\bz)\leq 0, \widehat{S}_n=S_n]\ d\bH(\bz) +P_{12}(\alpha)\nonumber\\
&= P[-n^{-\alpha}<\xi^\gamma(\bZ)\leq 0] + P_{12}(\alpha).
\end{align}

\noindent 
Combining \eqref{nref11}, \eqref{nrefP1} and \eqref{nrefP2}, we obtain
\vspace{-0.1in}
\begin{align}\label{nref10}
\Delta_{\rm gSAVG} - \Delta_0\leq P[|\xi^\gamma(\bZ)|< n^{-\alpha}] + 2P_{12}(\alpha) + P[\widehat{S}_n\neq S_n].
\end{align}
It follows from Lemma \ref{exbddisc} that $P_{12}(\alpha)\le O\left (e^{-b_3\{ n^{1-2\alpha}- n^\beta\}}\right )$ for all $\max\{\alpha_1,\alpha_2\}\le \alpha_0<(1-\beta)/2$. Therefore, Theorem \ref{mainthm_E} and \eqref{nref10} suggest that
\vspace{-0.1in}
\begin{align*}
\Delta_{\rm gSAVG} - \Delta_0 &\leq P[|\xi^\gamma(\bZ)|< n^{-\alpha}] + O\left (e^{-b_3\{ n^{1-2\alpha}- n^\beta\}}\right ) + O\left (e^{-b_1\{ n^{1-2\alpha}- n^\beta\}}\right )\\
&\leq P[|\xi^\gamma(\bZ)|< n^{-\alpha}] + O\left (e^{-b_3\{ n^{1-2\alpha}- n^\beta\}}\right ).
\end{align*}
Hence, the proof.\hspace*{\fill}\QEDB\newline

\noindent
{\bf Remark 1 :} Recall that there exist a constant $N\in\mathbb{N}$ such that $d_n \leq e^{Mn^\beta}$ for all $n\geq N$, where $M>0$ and $0\leq \beta<1.$ If $\beta =0$, then $d_n(=d)$ is free of $n$. Therefore, if $\bF$ and $\bG$ are absolutely continuous distribution functions, then $P[|\xi^\gamma(\bZ)|< n^{-\alpha}]\to 0$ for all $\alpha>0$ as $n\to\infty$. Consequently, $\Delta_{\rm gSAVG} - \Delta_0\to 0$ as $n\to\infty$. 

\vspace*{0.2in}
\noindent
{\bf Remark 2 :} If $\beta >0,$ then $d_n$ grows with $n$. Under this setting, if $P[|\xi^\gamma(\bZ)|< n^{-\alpha}]\to 0$ and $\Delta_0\to 0$ as $n\to\infty$, then Theorem \ref{thmclass} suggests that $\Delta_{\rm gSAVG}\to 0$ as $n\to\infty$.\newline

\noindent 
We now present some sufficient conditions for $P[|\xi^\gamma(\bZ)|< n^{-\alpha}]$ and $\Delta_0$ to go to 0 as~$n\to\infty$.\newline
Under appropriate moment conditions and weak dependence among the component variables, we show that $P[|\xi^\gamma(\bZ)|< n^{-\alpha}]\to 0$ (for some $0< \alpha<1$), as $n\to\infty.$ 

\noindent
Since there exist $M>0,\ 0<\beta<1$ and $N\in\mathbb{N}$ such that for all $n\geq N$, we have
\begin{align*}
&\ \log\ d_n\leq Mn^\beta\\
\Rightarrow &\ \log\ s_n \leq \log\ d_n \leq Mn^\beta \ [\text{since } s_n\leq d_n]\\
\Rightarrow &\ n^{-\alpha}\leq (M/\log\ s_n)^\frac{\alpha}{\beta}\text{ for all }\alpha>0.
\end{align*}
We now assume the following:
\begin{itemize}
	\item[A1$^\prime$.]  There exist constants $\max\{\alpha_1,\alpha_2\}\le \alpha_3\le (1-\beta)/2$ and  $N\in\mathbb{N}$ such that $\mathcal{E}_{(t_n+1)}>2M_1/(\log \ s_n)^{\frac{\alpha_3}{\beta}}$ for all $n\geq N.$
\end{itemize}
If A1$^\prime$ is satisfied, then A1 also holds. Assumption A1$^\prime$ allows the minimum signal to decay to 0, but at a rate slower than $1/(\log \ s_n)^{\frac{\alpha_3}{\beta}}$.

Suppose that $\bU=(U_1,\ldots, U_{d_n})^\top\sim\bF_j$ and $\bU^\prime=(U^\prime_1,\ldots, U^\prime_{d_n})^\top\sim\bF_{j^\prime}$ for $j,j^\prime\in\{1,2\}$ and $\bU,\bU^\prime$ are independently distributed random vectors. We further assume the following:
\begin{enumerate}
	\item[A3.]  $E[\gamma^2(|U_{k}-U^\prime_{k}|^2)]<C<\infty$ for all $k\in S_n.$
	\item[A4.] There exists $\max\{\alpha_1,\alpha_2\}\le \alpha_4\le (1-\beta)/2$ such that $$\mathop{\sum\sum}\limits_{\substack{k,k^\prime \in S_n\\ k\neq k^\prime}}{\rm Corr}{\left(E[\gamma(|U_{k}-U^\prime_{k}|^2)\mid V_{k}],E[\gamma(|U_{k^\prime}-U^\prime_{k^\prime}|^2)\mid V_{k^\prime}]\right )}=o\left (\frac{s_n^2}{\left (\log\ s_n\right )^\frac{2\alpha_4}{\beta}}\right )$$
	where $\bV\in\{\bU,\bU^\prime\}$.
\end{enumerate}

\noindent
Assumption A3 is trivially satisfied if $\gamma$ is a bounded function. If the underlying components are Gaussian and $\gamma$ is an $L$-Lipschitz continuous function, then assumption A3 is satisfied as well (see Lemma \ref{U_expbound_lip} for details. 

\noindent
Let us take a look at assumption A4 now. Clearly, A4 is satisfied if the component variables are independently distributed. It continues to hold if an additional structure on the dependence of the components is assumed. For instance, if the components are $m$-dependent \citep{billingsley2008probability} for some fixed integer $m$, then the LHS of assumption A4 is bounded above by $m(2s_n- m - 1)$ and it holds as $n\to\infty$.
\begin{lemma}\label{P0_conv}
	If $A1^\prime$, A3 and A4 are satisfied with $\alpha_3<\alpha_4$, then $P[|\xi^\gamma(\bZ)|<n^{-\alpha}]\to 0$ as $n\to\infty$ for all $\alpha_3\le \alpha\le  \alpha_4.$
\end{lemma}
\noindent{\bf Proof :}
Fix $\alpha_3\le \alpha\le \alpha_4.$ Recall that there exist $M_1>0,0<\beta<1$ and $N\in\mathbb{N}$ such that $\log\ s_n\leq M_1n^\beta,$ i.e., $n^{-\alpha}\leq (M_1/\log\ s_n)^\frac{\alpha}{\beta}$ for all $n\geq N.$ Therefore,
\begin{align*}
& \ P[|\xi^\gamma(\bZ)|<n^{-\alpha}]\leq  P\left [|\xi^\gamma(\bZ)|<\left (\frac{M_1}{\log\ s_n}\right )^\frac{\alpha}{\beta}\right ]\\
=&\ \pi_1 P\left [|\xi^\gamma(\bX_1)|<\left (\frac{M_1}{\log\ s_n}\right )^\frac{\alpha}{\beta}\right ] + \pi_2 P\left [|\xi^\gamma(\bY_1)|<\left (\frac{M_1}{\log\ s_n}\right )^\frac{\alpha}{\beta}\right ]\text{ for all }n\geq N.
\end{align*}
Further, we have
\begin{align}\label{P0_1}
&\ P\left [|\xi^\gamma(\bX_1)-E[\xi^\gamma(\bX_1)]|\geq \left (\frac{M_1}{\log\ s_n}\right )^\frac{\alpha}{\beta}\right ]\nonumber \\
\leq &\ P\left [|\xi_1^\gamma(\bX_1)-E[\xi_1^\gamma(\bX_1)]|\geq \frac{1}{2}\left (\frac{M_1}{\log\ s_n}\right )^\frac{\alpha}{\beta}\right ]+P\left [|\xi_2^\gamma(\bX_1)-E[\xi_2^\gamma(\bX_1)]|\geq \frac{1}{2}\left (\frac{M_1}{\log\ s_n}\right )^\frac{\alpha}{\beta}\right ]\nonumber \\
\leq &\ \frac{1}{4}\left (\frac{\log\ s_n}{M_1}\right )^\frac{2\alpha}{\beta} \big [Var[\xi_1^\gamma(\bX_1)]+Var[\xi_2^\gamma(\bX_1)] \big ]\ (\text{using Markov's inequality}).
\end{align}
Recall the definition of $\xi_1^\gamma$ and $\xi_2^\gamma$ stated in Section \ref{consistency}. Define $W(X_{1k})=E[\gamma(|X_{1k}-X_{2k}|^2)\mid X_{1k}]$ for $k\in S_n$. Then, we have
\begin{align}
&\ \left (\log\ s_n\right )^\frac{2\alpha}{\beta}Var[\xi_1^\gamma(\bX_1)]\nonumber \\
=&\ \left (\log\ s_n\right )^\frac{2\alpha}{\beta}Var\left [\frac{1}{s_n}\sum\limits_{k\in S_n} W(X_{1k})\right ]\nonumber \\
\leq &\ \frac{\left (\log\ s_n\right )^\frac{2\alpha}{\beta}}{s^2_n}\left \{\sum\limits_{k\in S_n}E\left [\{W(X_{1k})\}^2\right ] +\mathop{\sum\sum}\limits_{\substack {k,k^\prime\in S_n\\ k\neq k^\prime}}\rm{Cov}(W(X_{1k}),W(X_{1k^\prime}))\right\}\nonumber \\
\leq &\ \frac{\left (\log\ s_n\right )^\frac{2\alpha}{\beta}}{s^2_n}\sum\limits_{k\in S_n}E\left [\{W(X_{1k})\}^2\right ]\nonumber \\ &\hspace{0.1cm}+\frac{\left (\log\ s_n\right )^\frac{2\alpha}{\beta}}{s^2_n}\left\{\mathop{\sum\sum}\limits_{\substack{k,k^\prime\in S_n\\ k\neq k^\prime}}{\rm Corr}(W(X_{1k}),W(X_{1k^\prime}))\sqrt{E\left [\{W(X_{1k})\}^2\right ]E\left [\{W(X_{1k^\prime})\}^2\right ]}\right \}\nonumber \\
\leq &\ \frac{\left (\log\ s_n\right )^\frac{2\alpha}{\beta}}{s_n}C+\frac{\left (\log\ s_n\right )^\frac{2\alpha}{\beta}}{s^2_n}C\left\{\mathop{\sum\sum}\limits_{\substack{k,k^\prime\in S_n\\ k\neq k^\prime}}{\rm Corr}{(W(X_{1k}),W(X_{1k^\prime}))}\right \}\ [\text{using A3}].
\end{align}
The first term in the RHS of the above inequality is clearly $o(1)$ for all $\alpha>0$ and $0<\beta <1$. The second term is also $o(1)$ due to assumption A4 since $\alpha\le \alpha_4$. Hence, $\left (\log\ s_n\right )^\frac{2\alpha}{\beta}Var[\xi_1^\gamma(\bX_1)]\to 0$ as $n\to\infty$. 

\noindent 
Define $W^\prime(X_{1k})=E[\gamma(|Y_{1k}-X_{1k}|^2)\mid X_{1k}]$ for $k\in S_n$. Following similar arguments, we can show that $$\left (\log\ s_n\right )^\frac{2\alpha}{\beta}Var\left [\sum_{k\in S_n}W^\prime_{k}/s_n\right ] = \left (\log\ s_n\right )^\frac{2\alpha}{\beta}Var[\xi_2^\gamma(\bX_1)]\to 0\text{ as }n\to\infty.$$ As a result, \eqref{P0_1} implies that $\displaystyle P\left [|\xi^\gamma(\bX_1)-E[\xi^\gamma(\bX_1)]|\geq \left (\frac{M_1}{\log\ s_n}\right )^\frac{\alpha}{\beta}\right ]\to 0$ as $n\to\infty.$ Now,
\begin{align*}
P\left [|\xi^\gamma(\bX_1)-E[\xi^\gamma(\bX_1)]|\geq \left (\frac{M_1}{\log\ s_n}\right )^\frac{\alpha}{\beta}\right ]\geq  P\left [\xi^\gamma(\bX_1)\leq E[\xi^\gamma(\bX_1)]-\left (\frac{M_1}{\log\ s_n}\right )^\frac{\alpha}{\beta}\right ]
\end{align*}
If assumption A1$^\prime$ is satisfied, then $\displaystyle E[\xi^\gamma(\bX_1)]=\sum_{k\in S_n}\mathcal{E}_k/s_n\geq 2\left (\frac{M_1}{\log\ s_n}\right )^\frac{\alpha}{\beta},$ i.e., $\displaystyle E[\xi^\gamma(\bX_1)]-\left (\frac{M_1}{\log\ s_n}\right )^\frac{\alpha}{\beta}\geq \left (\frac{M_1}{\log\ s_n}\right )^\frac{\alpha}{\beta}$ for all $n\geq N.$ Therefore,
\begin{align*}
\ P\left [|\xi^\gamma(\bX_1)-E[\xi^\gamma(\bX_1)]|\geq \left (\frac{M_1}{\log\ s_n}\right )^\frac{\alpha}{\beta}\right ]
\geq &\  P\left [\xi^\gamma(\bX_1)\leq E[\xi^\gamma(\bX_1)]-\left (\frac{M_1}{\log\ s_n}\right )^\frac{\alpha}{\beta}\right ]\\
\geq &\  P\left [\xi^\gamma(\bX_1)\leq \left (\frac{M_1}{\log\ s_n}\right )^\frac{\alpha}{\beta}\right ]\\
\geq &\  P\left [|\xi^\gamma(\bX_1)|\leq \left (\frac{M_1}{\log\ s_n}\right )^\frac{\alpha}{\beta}\right ]\text{ for all }n\geq N.
\end{align*}
If assumptions A1$^\prime$, A3 and A4 are satisfied, then $\displaystyle P\left [|\xi^\gamma(\bX_1)|\leq \left (\frac{M_1}{\log\ s_n}\right )^\frac{\alpha}{\beta}\right ]\to 0 $ as $n\to\infty$.

\noindent Arguments for proving $\displaystyle P\left [|\xi^\gamma(\bY_1)|\leq \left (\frac{M_1}{\log\ s_n}\right )^\frac{\alpha}{\beta}\right ]\to 0 $ as $n\to\infty$ are similar to those presented above. First, one needs to show that $\displaystyle P\left [|\xi^\gamma(\bY_1)-E[\xi^\gamma(\bY_1)]|\geq \left (\frac{M_1}{\log\ s_n}\right )^\frac{\alpha}{\beta}\right ]\to 0$ as $n\to\infty$. If assumption A1$^\prime$ is satisfied, then $\displaystyle E[\xi^\gamma(\bY_1)]=-\sum_{k\in S_n}\mathcal{E}_k/s_n\leq -2\left (\frac{M_1}{\log\ s_n}\right )^\frac{\alpha}{\beta},$ i.e., $\displaystyle E[\xi^\gamma(\bY_1)]-\left (\frac{M_1}{\log\ s_n}\right )^\frac{\alpha}{\beta}\geq \left (\frac{M_1}{\log\ s_n}\right )^\frac{\alpha}{\beta}$ for all $n\geq N.$ Therefore,
\begin{align*}
\ P\left [|\xi^\gamma(\bY_1)-E[\xi^\gamma(\bY_1)]|\geq \left (\frac{M_1}{\log\ s_n}\right )^\frac{\alpha}{\beta}\right ]
\geq &\  P\left [\xi^\gamma(\bY_1)\leq E[\xi^\gamma(\bY_1)]-\left (\frac{M_1}{\log\ s_n}\right )^\frac{\alpha}{\beta}\right ]\\
\geq &\  P\left [\xi^\gamma(\bY_1)\leq \left (\frac{M_1}{\log\ s_n}\right )^\frac{\alpha}{\beta}\right ]\\
\geq &\  P\left [|\xi^\gamma(\bY_1)|\leq \left (\frac{M_1}{\log\ s_n}\right )^\frac{\alpha}{\beta}\right ]\text{ for all }n\geq N.
\end{align*}
If assumptions A1$^\prime$, A3 and A4 are satisfied, then $\displaystyle P\left [|\xi^\gamma(\bX_1)|\leq \left (\frac{M_1}{\log\ s_n}\right )^\frac{\alpha}{\beta}\right ]\to 0 $ as $n\to\infty.$ Hence, we have $P[|\xi^\gamma(\bZ)|<n^{-\alpha}]\to 0$ as $n\to \infty$. \hfill \QEDB
\newline

\begin{lemma}\label{Delta0_conv}
	If assumptions $A1^\prime$, A3 and A4 are satisfied, then $\Delta_0\to 0$ as $n\to\infty.$
\end{lemma}

\noindent {\bf Proof :} Recall that $\Delta_0$ is defined as
\begin{align*}
\Delta_0 =  \pi_1 P[\xi^{\gamma}_{1}(\bX_1)\geq \xi^{\gamma}_{2}(\bX_1)] + \pi_2 P[\xi^{\gamma}_{1}(\bY_1)<\xi^{\gamma}_{2}(\bY_1)].
\end{align*}
Observe that $E[\xi^{\gamma}_{2}(\bX_1)-\xi^{\gamma}_{1}(\bX_1)] = \sum_{k\in S_n}\mathcal{E}_k/s_n.$ Since assumption A1$^\prime$ holds, we have\\
$E\left [\xi^{\gamma}_{2}(\bX_1)- \xi^{\gamma}_{1}(\bX_1)\right ]=\sum_{k\in S_n}\mathcal{E}_k/s_n>2M_1(\log \ s_n)^{\frac{\alpha}{\beta}}$ for all $\alpha_3\le \alpha<(1-\beta)/2$ and  $n\geq N.$ Therefore,
\begin{align}\label{mref14}
P[\xi^{\gamma}_{1}(\bX_1)- \xi^{\gamma}_{2}(\bX_1)\geq 0]
\leq &\ P\left [\xi^{\gamma}_{1}(\bX_1)- \xi^{\gamma}_{2}(\bX_1)> -\frac{1}{s_n}{\sum_{k\in S_n}\mathcal{E}_k}+2M_1(\log \ s_n)^{-\frac{\alpha}{\beta}}\right ]\nonumber \\
\leq &\ P\left [\xi^{\gamma}_{1}(\bX_1)- \xi^{\gamma}_{2}(\bX_1)+\frac{1}{s_n}{\sum_{k\in S_n}\mathcal{E}_k}> 2M_1(\log \ s_n)^{-\frac{\alpha}{\beta}}\right ]\nonumber \\
\leq &\ P\left [\left |\xi^{\gamma}_{1}(\bX_1)- \xi^{\gamma}_{2}(\bX_1)+\frac{1}{s_n}{\sum_{k\in S_n}\mathcal{E}_k}\right |> 2M_1(\log \ s_n)^{-\frac{\alpha}{\beta}}\right ]\nonumber \\
=&\ P\left [\left |\xi^{\gamma}(\bX_1)-E[\xi^{\gamma}(\bX_1)]\right |> 2M_1(\log \ s_n)^{-\frac{\alpha}{\beta}}\right ]\text{ for all }n\geq N.
\end{align}

\noindent
We have already shown in the proof of Lemma \ref{P0_conv}
that if assumptions A3 and A4 are satisfied, then $P\left [\left |\xi^{\gamma}(\bX_1)-E[\xi^{\gamma}(\bX_1)]\right |> 2M_1(\log \ s_n)^{-\frac{\alpha}{\beta}}\right ]\to 0$ for all $\alpha_3\le \alpha\le \alpha_4$ as $n\to\infty$. Therefore, $P[\xi^{\gamma}_{1}(\bX_1)- \xi^{\gamma}_{2}(\bX_1)\ge 0]\to 0$ as $n\to\infty.$ The argument for proving $P[\xi^{\gamma}_{1}(\bY_1)- \xi^{\gamma}_{2}(\bY_1)< 0]\to 0$ as $n\to\infty$ is similar, and we skip it.

\noindent 
Now, from the definition of $\Delta_0$, it can be concluded that $\Delta_0\to 0$ as $n\to\infty$.
\hfill\QEDB
\vspace*{1in}
	
\section{Algorithms for MarS and MixS} \label{Appendix_B}

\begin{algorithm}
	\caption{MarS}
	\begin{algorithmic}
		\FOR{$k=1:d_n$}
		\STATE compute the sample energy distance $\widehat{\mathcal{E}}_{k}$;
		\ENDFOR
		\STATE sort in an increasing order $\widehat{\mathcal{E}}_{(1)} < \cdots < \widehat{\mathcal{E}}_{(d_n)}$;
		\FOR{$k=[d_n/2]:(d_n-1)$}
		\STATE compute $\widehat{R}_k = \widehat{\mathcal{E}}_{(k+1)}/\widehat{\mathcal{E}}_{(k)}$;
		\ENDFOR
		\STATE $\widehat{t}_n=\argmax_{[d_n/2] \leq k \leq (d_n-1)} \widehat{R}_{k}$;
		\STATE $\widehat{S}_n=\{k:\widehat{\mathcal{E}}_{k}\geq \widehat{\mathcal{E}}_{(\widehat{t}_n+1)} \mbox{ for } 1\leq k\leq d_n\}$;
		\STATE return $\widehat{S}_{n}$.
	\end{algorithmic}
\end{algorithm}

\begin{algorithm}[H]
	\caption{MixS}
	\begin{algorithmic}
		\STATE $\bZ_i\in\rchi_n$ (the training sample) for $1 \leq i \leq n$;
		\IF {$d_n$ is odd,}
		\STATE generate $Z_i\stackrel{iid}{\sim}N(0,1)$ for $1 \leq i \leq n$;
		\STATE define ${\bf Z}^\prime_i=({\bf Z}_i^\top,Z_i)^\top$
		\STATE $\tilde{d}_n = (d_n +1)/2$;
		\ELSE
		\STATE $\bZ^\prime_i=\bZ_i$ for $1 \leq i \leq n$;
		\STATE $\tilde{d}_n = d_n/2$;
		\ENDIF
		\STATE define $\rchi_n^\prime = \{{\bZ}^\prime_1,\ldots, {\bZ}^\prime_n\}$;
		\FOR {$i=1:\tilde{d}_n$}
		\FOR {$j=(i+1):\tilde{d}_n$}
		\STATE compute the sample energy distance $\widehat{\mathcal{E}}_{\{i,j\}}$ based on $\rchi_n^\prime$;
		\ENDFOR
		\ENDFOR
		\STATE define the constant $K=\max_{i,j} \widehat{\mathcal{E}}_{\{i,j\}}+1$;
		\STATE define the matrix $W=K-\widehat{\mathcal{E}}$;
		\STATE {$\widehat{P}_n = \argmin_{P\in\mathcal{P}}\mathcal{W}(P)$ (via the non-bipartite matching algorithm);} 
		\STATE $\widehat{\mathcal{E}}_{k}\equiv\widehat{\mathcal{E}}_{\{i_k,j_k\}}$ are estimates of $\mathcal{E}_{\{i_k,j_k\}}$ for $\{i_k,j_k\}\in \widehat{P}_n$ and $1 \leq k \leq \tilde{d}_n$;
		\STATE sort in an increasing order $\widehat{\mathcal{E}}_{(1)} < \cdots < \widehat{\mathcal{E}}_{(\tilde{d}_n)}$;
		\FOR{$k=[\tilde{d}_n/2]:(\tilde{d}_n-1)$}
		\STATE compute $\widehat{R}_k = \widehat{\mathcal{E}}_{(k+1)}/\widehat{\mathcal{E}}_{(k)}$;
		\ENDFOR
		\STATE $\widehat{t}_n = \argmax_{[\tilde{d}_n/2] \leq k \leq (\tilde{d}_n-1)} \widehat{R}_{k}$;
		\STATE $\widehat{S}^*_n=\{\{i_k,j_k\}:\widehat{\mathcal{E}}_{k}\geq \widehat{\mathcal{E}}_{(\widehat{t}_n+1)} \mbox{ for } 1\leq k\leq \tilde{d}_n\}$;
		\STATE define $\widehat{s}_n^* = |\widehat{S}^*_n|$ and $\widehat{S}^*_n=\{\{i_1,j_1\},\ldots,\{i_{\widehat{s}_n^* },j_{\widehat{s}_n^* }\}\}$;
		\STATE initialize $\widehat{S}_{1n}=\widehat{S}_{2n}=\phi$ (the null set);
		\algstore{myalg}
\end{algorithmic}
\end{algorithm}

\begin{algorithm}
    \begin{algorithmic}
        \algrestore{myalg}
        		\FOR{$k=1:\widehat{s}_n^*$}
				\FOR{$m=1:M$}
		\STATE let $\rchi^\prime_{m,n}$ denote the bootstrapped random sample of size $n$ from $\rchi^\prime_n$;
		\STATE $\widehat{\mathcal{E}}_{\{i_k\},m},\widehat{\mathcal{E}}_{\{j_k\},m}$ and $\widehat{\mathcal{E}}_{\{i_k,j_k\},m}$ are the estimates of $\mathcal{E}_{\{i_k\}},\mathcal{E}_{\{j_k\}}$ and $\mathcal{E}_{\{i_k,j_k\}}$ based on $\rchi^\prime_{m,n}$, respectively;
		\ENDFOR
		\STATE{$\begin{aligned}
		  &\widehat{p}_{1k}=\frac{1}{M}\sum\limits_{m=1}^M \mathbb{I}[\widehat{\mathcal{E}}_{\{i_k\},m}\geq \widehat{\mathcal{E}}_{\{i_k\}}],\ \widehat{p}_{2k}=\frac{1}{M}\sum\limits_{m=1}^M \mathbb{I}[\widehat{\mathcal{E}}_{\{j_k\},m}\geq \widehat{\mathcal{E}}_{\{j_k\}}],\\ &\widehat{p}_{3k}=\frac{1}{M}\sum\limits_{m=1}^M \mathbb{I}[\widehat{\mathcal{E}}_{\{i_k\},m}+\widehat{\mathcal{E}}_{\{j_k\},m}\geq \widehat{\mathcal{E}}_{\{i_k\}}+\widehat{\mathcal{E}}_{\{j_k\}}] \text{ and } \ \widehat{p}_{4k}=\frac{1}{M}\sum\limits_{m=1}^M \mathbb{I}[\widehat{\mathcal{E}}_{\{i_k,j_k\},m}\geq \widehat{\mathcal{E}}_{\{i_k,j_k\}}];
		\end{aligned}$}
		\STATE define $\widehat{p}_{0k} = \min\{\widehat{p}_{1k},\widehat{p}_{2k},\widehat{p}_{3k},\widehat{p}_{4k}\}$;
		\IF{$\widehat{p}_{1k}=\widehat{p}_{0k}$}
		\STATE{$\widehat{S}_{1n}=\widehat{S}_{1n} \cup  \{i_k\}$;}
		\ELSIF{$\widehat{p}_{2k}=\widehat{p}_{0k}$}
		\STATE{$\widehat{S}_{1n}=\widehat{S}_{1n} \cup \{j_k\}$;}
		\ELSIF{$\widehat{p}_{3k}=\widehat{p}_{0k}$}
		\STATE{$\widehat{S}_{1n}=\widehat{S}_{1n} \cup \{i_k\}\cup\{j_k\}$;}
		\ELSE
		\STATE{$\widehat{S}_{2n}=\widehat{S}_{2n} \cup \{i_k,j_k\}$;}
		\ENDIF
		\ENDFOR
		\STATE return $\widehat{S}_{n}$.
	\end{algorithmic}
\end{algorithm}

\newpage
\section{Additional Material} \label{Appendix_C}

\subsection{Notations} \label{Nota_C}

Throughout this article, we have used the following definitions to present the mathematical results and in related discussions.
\begin{enumerate}
	\item $a_n=o(b_n)$ implies that for every $\epsilon>0$ there exists an $N\in\mathbb{N}$ such that $|a_n/b_n|<\epsilon$ for all $n\geq N$.
	\item $a_n=O(b_n)$ implies that there exist $M>0$ and $N\in\mathbb{N}$ such that $|a_n/b_n|<M$ for all $n\geq N$.
	\item $X_n=o_{\rm P}(a_n)$ implies that the sequence of random variables $X_n/a_n$ converges to 0 in probability as $n\to\infty$. 
	\item $X_n=O_{\rm P}(a_n)$ implies that the sequence of random variables $X_n/a_n$ is stochastically bounded, i.e., for every $\epsilon>0$ there exist a finite $M>0$ and $N\in\mathbb{N}$ such that ${\rm P}[|X_n/a_n|\ge M]<\epsilon$ for all $n\geq N$.
\end{enumerate}

\subsection{Assumption A1: The iid case} \label{iid_C}
\vspace*{-0.1in}

We consider a specific scenario to get a more clear interpretation of the two parts in Assumption A1. Suppose that the components of $\bX \sim \bF$ and $\bY \sim \bG$ are iid with $X_k \sim F$, $Y_k \sim G$ for $k\in S_n$ and $X_k, Y_k \sim H$ for $k\in S_n^c$. Here, $F, G$ and $H$ are univariate absolutely continuous dfs satisfying $F \neq G \neq H$.
Clearly, $\mathcal{E}_{k}=0$ for $k\in S_n^c$, while  $\mathcal{E}_{k}=\mathcal{E}_0>0$ (free of both $k$ as well as $n$) for $k\in S_n$.
This now implies that $\mathcal{E}_{(t_n+1)}=\mathcal{E}_0>0$ while $\max_{(t_n +1) \leq k \leq (d_n-1)} R_k=1$. So, the two parts of assumption A1 simplify to $1/\mathcal{E}_0 = o(n^{\alpha_1})$ with $\alpha_1>0$, which holds trivially as $n \to \infty$.
Further, if we allow the energy distances $\mathcal{E}_{k}$ to depend on $n$, then assumptions A1.1 and A1.2 continue to hold under a restricted~setting.

\subsection{Varying choices of the $\gamma$ function} \label{gamma_C}

We have considered three choices of $\gamma$, namely, $\gamma_1(t)=1-e^{-t}$, $\gamma_2(t)=\log(1+t)$ and $\gamma_3(t)=\sqrt{t}$ for $t \geq 0$. Now, an obvious question that arises from Figures \ref{figsim1} and \ref{figsim2} is a comparative performance among these functions. 
The three functions have non-constant, completely monotone derivatives \citep[see, e.g.,][]{MR0270403, BF10} and they are monotonically increasing. The function $\gamma_1$ is clearly bounded, while the other two functions are unbounded. When dealing with heavy-tailed distributions in  Examples \ref{ex5} and \ref{ex6}, the advantage of using a bounded $\gamma$ is clear. Further, there exists a $C>0$ such that these functions satisfy the ordering $\gamma_1(t) < \gamma_2(t) < \gamma_3(t)$ for all $t>C$. This probably justifies the ordering in their performance for Examples \ref{ex3} and \ref{ex8}.
In general, performances of screening methods based on $\gamma_1$ and $\gamma_2$ are quite similar. 
Overall, we observe that the proposed screening methods for $\gamma_2$ outperform the other two choices.
Further, note that ESP holds for $\gamma_{1}$ (recall Theorem \ref{mainthm_E}). For Gaussian variates, ESP holds for $\gamma_2$ (an $L$-Lipschitz continuous function) as well (see Remark \ref{R_mainthm}). However, we do not have any theoretical results related to ESP for $\gamma_3$ under Gaussianity.

Now, let us look into the relative performance in terms of misclassification rates (see Table \ref{tablesimulation}). 
Example \ref{ex1} (a {\it location} problem) involving light-tailed distributions, $\gamma_3$ performed quite well. Whenever the underlying distributions differ either in scales or shapes, the screening methods associated with $\gamma_1$ performed better than $\gamma_3$. This phenomenon was also observed in Examples \ref{ex4}, \ref{ex7} and \ref{ex8}.
A similar phenomena has been noted by \cite{BF10} for location and scale/shape problems, where the authors were interested in multivariate non-parametric two-sample goodness of fit tests. 

\subsection{Number of features selected by MarS and MixS}

Clearly, the number of features selected by MarS and MixS varies with the choice of $\gamma$. Complete numerical result for the six real and benchmark data sets analyzed in Section \ref{real} is presented below.

\vspace{0.1in}
\begin{table}[htp]
    \caption{Average number of component variables retained by MarS and MixS algorithms with associated standard errors (in italics) in the six real and benchmark data sets.}
    \vspace*{0.25in}
    \renewcommand{\arraystretch}{1.25}
    \centering
    \footnotesize
    \begin{tabular}{|c|ccc|ccc|}
    \hline
    \multirow{2}{*}{Data set} & \multicolumn{3}{c|}{MarS} &\multicolumn{3}{c|}{MixS} \\
 & $\gamma_1$ & $\gamma_2$ & $\gamma_3$ & $\gamma_1$ & $\gamma_2$ & $\gamma_3$\\
\hline
\multirow{2}{*}{Madelon} & 3.310 & 3.310 & 3.310 & 8.333 &	8.389 &	7.778\\
	& {\it 0.148} & {\it 0.148} & {\it 0.148} & {\textit 1.016} &	{\it 1.013} &	{\it 1.128}\\
\hline
\multirow{2}{*}{CorrAL} & 2.454 &	2.464 & 2.464 &	2.680 &	3.140 &	3.150 \\
	& {\it 0.237} & {\it 0.237}& {\it 0.237} &	{\it 0.226} &	{\it 0.238} &	{\it 0.237}\\
\hline
\multirow{2}{*}{Bittner} & 3.640	& 3.610 & 3.510 & 5.610 &	4.870 &	5.740\\
& {\it 0.632}	& {\it 0.348} & {\it 0.348} &	{\it 0.838} &	{\it 0.630} &	{\it 0.678}\\
\hline
\multirow{2}{*}{Shipp} & 4.750 	& 2.980 & 2.520 &	5.200 &	4.33 &	3.66\\
	& {\it 0.511} & {\it 0.372} & {\it 0.251} & {\textit 0.629} &	{\it 0.504} &	{\it 0.291}\\
\hline
\multirow{2}{*}{GSE3726} & 2.650 & 2.640 & 1.010 &	2.850 &	2.740 &	1.030   \\
	& {\it 0.296} & {\it 0.301} & {\it 0.011} &	{\it 0.315} &	{\it 0.356} &	{\it 0.017}\\
\hline
\multirow{2}{*}{GSE967} & 19.850 & 3.720 & 6.140  & 28.700 &	4.680 &	5.220\\
& {\it 1.919}	& {\it 0.578} & {\it 0.526} &	{\it 2.722} &	{\it 0.552} &	{\it 0.689}\\
\hline
    \end{tabular}
    \label{tabcardi}
\end{table}

\end{document}